\documentclass[preprint,flushrt,eqsecnum,10pt]{aastex}
\usepackage{color}
\usepackage{url}
\newcommand{\kk}{\mathbf{k}}
\newcommand{\pp}{\mathbf{p}}

\newcommand{\qq}{\mathbf{q}}

\newcommand{\LL}{\mathbf{L}}
\newcommand{\YY}{\mathbf{\Psi}}
\newcommand{\xx}{\mathbf{x}}

\begin{document}

\title{USING LAGRANGIAN PERTURBATION THEORY FOR PRECISION COSMOLOGY}

\author{Naonori S. Sugiyama}
\affil{Department of Astrophysical Sciences, Peyton Hall, Princeton University, Princeton, NJ 08544-1001, USA}
\affil{Astronomical Institute, Tohoku University, 6-3, Aramakijiaoba, Sendai 980-8578, Japan}
\email{nao.s.sugiyama@gmail.com}

\begin{abstract}
We explore the Lagrangian perturbation theory (LPT) at 1-loop order with Gaussian initial conditions.
We present an expansion method to approximately compute the power spectrum in LPT.
Our approximate solution has good convergence in the series expansion and enables us to compute the power spectrum
in LPT accurately and quickly.
Non-linear corrections in the Lagrangian perturbation theory naturally satisfy the law of conservation of mass
because the relation between matter density and the displacement vector of dark matter corresponds to the conservation of mass.
By matching the 1-loop solution in LPT to the 2-loop solution in standard perturbation theory,
we present an approximate solution of the power spectrum
which has higher order corrections than the 2-loop order in standard perturbation theory with the conservation of mass satisfied.
With this approximation,
we can use LPT to compute a non-linear power spectrum without any free parameters, 
and this solution agrees with numerical simulations at $k=0.2$ $h{\rm Mpc}^{-1}$ and $z=0.35$ to better than 2\%.
\end{abstract}


\maketitle

\section{INTRODUCTION}

Since the first measurement of the baryon acoustic oscillation (BAO) in
the SDSS LRG survey~\citep{Eisenstein:2005su} and the 2dF Galaxy survey~\citep{Cole:2005sx},
various other large-scale structure surveys have measured the galaxy power spectrum
and the position of the baryon acoustic peak with ever increasing precision
\citep{Tegmark:2006az,Percival:2006gs,Percival:2009xn,Kazin:2009cj,Beutler:2011hx,
Blake:2010xz,Blake:2011ep,Blake:2011en}.  In the coming decade,  we anticipate that new ground-based surveys such 
as the Prime Focus Spectrograph and Big BOSS and space-based surveys, such as {\it Euclid} and {\it WFIRST} will make even more
accurate measurements of the galaxy power spectrum. 
Therefore, predicting the precise non-linear behavior of the galaxy power spectrum
using analytical approaches is an essential step in interpreting these data and in elucidating the nature of dark energy.

The past decade has seen the development of a plethora of perturbation approaches to the non-linear matter power spectrum:
standard perturbation theory 
(SPT; \citet{Bernardeau:2001qr,Fry:1983cj,Goroff:1986ep,Suto:1990wf,Makino:1991rp,Jain:1993jh,Scoccimarro:1996se,Scoccimarro:1995if,
Jeong:2006xd,Sugiyama:2013}),
 Lagrangian resummation theory
(LRT; \citet{Matsubara:2007wj,Okamura:2011nu}),
 renormalized perturbation theory
(RPT; \citet{Crocce:2005xy,Crocce:2005xz,Crocce:2007dt}),
 closure theory (\citet{2008ApJ,2009PhRvD}),
 multi-point propagator method
(the $\Gamma$-expansion method; \citet{Bernardeau:2008fa,Bernardeau:2011dp}),
 regularized multi-point propagator method
(RegPT; \citet{Bernardeau:2011dp,Taruya:2012ut,Taruya:2013my}), 
the Wiener Hermite expansion method \citep{Sugiyama:2012pc},
as well as other techniques
~\citep{Pajer:2013jj,Tassev:2012hu,Valageas:2013gba,GilMarin:2012nb,Wang:2012fr,Carlson:2012bu,Tassev:2013pn,Wang:2013hwa}.

In this paper, we explore Lagrangian perturbation theory (LPT).
At the linear order, LPT reduces to the well-studied Zel'dovich approximation (e.g., \citet{Taylor:1996ne}),
but at higher order has not been calculated.
This is because there are numerical difficulties in computing the power spectrum in LPT,
even though some approximate methods in the Lagrangian description have been proposed~\citep{Matsubara:2007wj,Wang:2013hwa,Carlson:2012bu}.
We present an expansion method to approximately compute the LPT power spectrum.
Our approximate solution has good convergence in the series of the expansion
and enables to compute the LPT power spectrum accurately and quickly.
The main goal of the present work is to explore LPT at the 1-loop order
and give higher order correction terms than the 2-loop SPT solution.

The main result of this paper is 
\begin{eqnarray*}
		P(z,k) = D^2 P_{\rm lin}(k) + D^4 P_{\rm 1\mathchar`-loop}(k) + D^6 P_{\rm 2\mathchar`-loop}(k)
		+ \sum_{n=3}^{\infty} P_{\rm n\mathchar`-loop}|_{\rm LPT, 1\mathchar`-loop}(z,k),
\end{eqnarray*}
where $z$ and $D$ are the redshift and the linear growth function,
and $P_{\rm lin}$, $P_{\rm 1\mathchar`-loop}$, and $P_{\rm 2\mathchar`-loop}$ are the SPT solutions at the linear, 1-loop, and 2-loop order,
respectively.
The last term $\sum_{n=3}^{\infty} P_{\rm n\mathchar`-loop}|_{\rm LPT, 1\mathchar`-loop}$
we present in this paper is the correction computed in the 1-loop LPT that have higher order than the 2-loop SPT.
As we will show in Sections~\ref{3loop_and_more} and \ref{Nbody},
this works and agrees very well with the numerical simulations in Figure~\ref{fig:Nbody}.

This paper is organized as follows.
Section~\ref{Review} reviews LPT.
Section~\ref{motivation} gives the motivation for extending LPT to higher order.
Section~\ref{Sec:Sigma} computes correlation functions of the displacement vector.
In Section~\ref{Sec:Review_SPT}, we investigate how the LPT solutions reproduce the SPT solutions.
Section~\ref{Sec:pk_LPT} 
presents an expansion method to approximately compute the LPT power spectrum
and computes the LPT power spectrum in the linear and 1-loop order.
Section~\ref{LPT_RegPT} shows a simple relation between LPT and the $\Gamma$-expansion method.
Section~\ref{3loop_and_more} presents an approximate non-linear power spectrum
which has the 2-loop solution in SPT as well as higher order terms than the 2-loop in SPT computed in the 1-loop LPT.
Section~\ref{Nbody} compares the predicted power spectra in LPT and $N$-body simulation results,
and a final section summarizes our findings.

The cosmological parameters we used are presented by 
the {\it Wilkinson Microwave Anisotropy Probe} five-year release~\citep{Komatsu:2008hk}:
$\Omega_m = 0.279$, $\Omega_{\Lambda} = 0.721$, $\Omega_b = 0.046$, $h = 0.701$, $n_s = 0.96$ and $\sigma_8=0.817$.
We used the publicly available code, RegPT~\citep{Taruya:2012ut}
\footnote{\url{http://www-utap.phys.s.u-tokyo.ac.jp/~ataruya/regpt_code.html}} to compute the 2-loop power spectrum in SPT.

\section{GENERAL FORMULA OF THE LAGRANGIAN PERTURBATION THEORY}
\label{Review}

In the Lagrangian description, the spatial coordinates $\xx$ are transformed as
\begin{eqnarray}
		\xx = \qq_1 + \YY(z,\qq_1),
\end{eqnarray}
where $\YY$ is the displacement vector of dark matter particles.
Conservation of mass implies that the density perturbation $\delta$ can be described as a function of the displacement vector
in real and Fourier spaces, respectively:
\begin{eqnarray}
		\delta(z,\xx) &=&  \int d^3q_1 \delta_{\rm D}(\xx - \qq_1 - \YY(z,\qq_1)) - 1, \nonumber \\
		\delta(z,\kk) &=& \int d^3q_1 e^{-i\kk\cdot\qq_1} \left( e^{-i\kk\cdot\YY(z,\qq_1)}-1 \right) \nonumber \\
		             &=& \sum_{n=1}^{\infty} \frac{(-i)^n}{n!} \int \frac{d^3k_1}{(2\pi)^3} \cdots \frac{d^3k_n}{(2\pi)^3}
					 (2\pi)^3 \delta_{\rm D}(\kk-\kk_{[1,n]})
					 \left[ \kk \cdot \YY (z,\kk_1)\right] \cdots \left[ \kk \cdot \YY(z,\kk_n) \right],
					 \label{relation}
\end{eqnarray}
where $\kk_{[1,n]} \equiv \kk_1 + \dots + \kk_n$.
In LPT, the displacement vector field is expanded in a perturbation series in the linear growth function $D$ in Fourier space
\citep{Bernardeau:2001qr,Rampf:2012up}:
\begin{eqnarray}
		\YY(z,\kk) = \sum_{n=1}^{\infty} D^n\frac{i}{n!} \int \frac{d^3p_1}{(2\pi)^3} \cdots \frac{d^3p_n}{(2\pi)^3}
		(2\pi)^3 \delta_{\rm D}(\kk-\pp_{[1,n]}) \LL_n(\pp_1,\dots,\pp_n) \delta_{\rm lin}(\pp_1) \cdots \delta_{\rm lin}(\pp_n),
		\label{eq:LL}
\end{eqnarray}
where $\delta_{\rm lin}$ is the linearized density perturbation at $z=0$, and
the $n$th order of the kernel function in LPT $\LL_n$ is given by~\cite{Rampf:2012up}.

The linear displacement vector $\YY_{\rm lin}(\pp)
= i\pp \delta_{\rm lin}(\pp)/p^2$, called ``Zel'dovich approximation``, leads to 
\begin{eqnarray}
		\delta(z,\kk) = \sum_{n=1}^{\infty} D^n \int \frac{d^3p_1}{(2\pi)^3} \cdots \int \frac{d^3p_1}{(2\pi)^3}
		(2\pi)^3 \delta_{\rm D}(\kk-\pp_{[1,n]})
		F_{n}|_{\rm ZA}(\pp_1,\cdots,\pp_n) \delta_{\rm lin}(\pp_1) \cdots \delta_{\rm lin}(\pp_n),
\end{eqnarray}
where 
\begin{eqnarray}
		F_{n}|_{\rm ZA}(\pp_1,\cdots,\pp_n) 
		= \frac{1}{n!} \left( \frac{\kk\cdot\pp_1}{p_1^2} \right) \cdots \left( \frac{\kk\cdot\pp_n}{p_n^2} \right).
		\label{F_ZA}
\end{eqnarray}

The power spectrum is given by
\begin{eqnarray}
		P(z,k) &=&  \int d^3q e^{-i\kk\cdot\qq}
		\left\{ 	\left\langle e^{ -i\kk\cdot\left( \YY(z,\qq_1) - \YY(z,\qq_2) \right)  }\right\rangle -1  \right\}\nonumber \\
		&=& \int d^3q e^{-i\kk\cdot\qq}
		\Bigg\{\exp\left[ \sum_{n=1}^{\infty}\frac{(-i)^n}{n!} 
		\left\langle \left( \kk\cdot \YY(z,\qq) - \kk\cdot\YY(z,0)  \right)^n \right\rangle_{\rm c}  \right] - 1 \Bigg\}\nonumber \\
        &=& \int d^3q e^{-i\kk\cdot\qq}\left\{
		e^{ \Sigma(z,\kk,\qq) -\bar{\Sigma}(z,k)} -1 \right\},
		\label{LPT_power_1}
\end{eqnarray}
where the integration variable $\qq$ is the relative coordinate between the initial positions of dark matter particles: $\qq=\qq_1-\qq_2$.
In the second line, we used the translation symmetry in the ensemble average,
and $\langle \cdots \rangle_{\rm c}$ denotes the cumulant.
The functions $\Sigma$ and $\bar{\Sigma}$ are defined as
\begin{eqnarray}
		\Sigma(z,\kk,\qq) &\equiv& \sum_{n=2}^{\infty}\sum_{m=1}^{n-1}
		\frac{(-i)^n(-1)^m}{m!(n-m)!}
		\left\langle \left( \kk\cdot\YY(z,\qq) \right)^{n-m} \left( \kk\cdot\YY(z,0) \right)^m \right\rangle_{\rm c}, \nonumber \\
	\bar{\Sigma}(z,k) &\equiv& 	\Sigma(z,\kk,\qq=0)  =-2\sum_{n=1}^{\infty}\frac{(-1)^n}{(2n)!} 
		 \left\langle \left( \kk\cdot\YY(z,0) \right)^{2n}\right\rangle_{\rm c}.  
		 \label{Sigma_def}
\end{eqnarray}
These functions $\Sigma$ and $\bar{\Sigma}$ are the same as Eqs. (9) and (10) in \citet{Matsubara:2007wj}.
The relation $\bar{\Sigma}(z,k=0) = 0$ recasts Eq.~(\ref{LPT_power_1}) as
\begin{eqnarray}
		P(z,k) = e^{-\bar{\Sigma}(z,k)  } \int d^3q e^{-i\kk\cdot\qq}\left\{ e^{\Sigma(z,\kk,\qq)} -1 \right\},
		\label{power_spectrum}
\end{eqnarray}
where we used $\int d^3q e^{-i\kk\cdot\qq} e^{-\bar{\Sigma}(z,k) } = \int d^3q e^{-i\kk\cdot\qq}$.
Furthermore, we expand $\Sigma$ in Legendre polynomials as
\begin{eqnarray}
		\Sigma(z,\kk,\qq) = \sum_{\ell=0}^{\infty} i^{\ell} \Sigma_{\ell}(z,k,q) {\cal L}_{\ell}(\mu),
\end{eqnarray}
where $\mu = \hat{k} \cdot \hat{q}$.
Note that $\bar{\Sigma}$ comes from the monopole term: $\bar{\Sigma}(z,k) = \Sigma_0(z,k,q=0)$.
In other words, the other $\Sigma_{\ell}$ functions for $\ell \geq 1$ become zero at $q=0$: $\Sigma_{\ell \geq 1}(z,k,q=0)=0$.
For the functions $\Sigma_{\ell}$ to be real, the imaginary number should appear in the Legendre expansion.
Thereby, odd terms in the expansion behave like the changing Lagrangian spatial coordinates $\qq$ in Eq.~(\ref{power_spectrum}).
Finally,
we arrive at the general expression of the power spectrum in LPT:
\begin{eqnarray}
		P(z,k) &=&  e^{-\bar{\Sigma}(z,k)  } \int d^3q 
			   \frac{1}{2}\left(e^{-i\kk\cdot\qq}e^{\Sigma(z,\kk,\qq)} + e^{i\kk\cdot\qq}e^{\Sigma(z,-\kk,\qq)}\right) 
			   - e^{-\bar{\Sigma}(z,k)  } \int d^3q \frac{1}{2}\left(e^{-i\kk\cdot\qq} + e^{i\kk\cdot\qq}\right)  \nonumber \\
			   &=&  2\pi e^{-\bar{\Sigma}(z,k)  } \int_0^{\infty} dq q^2 \int_{-1}^1d\mu
		\Bigg\{\cos\left(  kq {\cal L}_1(\mu) - \sum_{\ell=0}^{\infty} \left( -1 \right)^{\ell}\Sigma_{2\ell+1}(z,k,q) 
		{\cal L}_{2\ell+1}(\mu)\right)
		- \cos\left( kq{\cal L}_1(\mu) \right) \nonumber \\
		&& \hspace{2cm}+
		\cos\left(  kq {\cal L}_1(\mu) - \sum_{\ell=0}^{\infty} \left( -1 \right)^{\ell}\Sigma_{2\ell+1}(z,k,q) {\cal L}_{2\ell+1}(\mu)\right)
		\left(e^{ \sum_{\ell=0}^{\infty} \left( -1 \right)^{\ell} \Sigma_{2\ell}(z,k,q){\cal L}_{2\ell}(\mu)  } -1\right)\Bigg\}, \nonumber \\
		\label{LPT_general}
\end{eqnarray}
where we used ${\cal L}_{2\ell+1}(-\mu) = - {\cal L}_{2\ell+1}(\mu)$ and ${\cal L}_{2\ell}(-\mu) = {\cal L}_{2\ell}(\mu)$.

\section{WHAT IS THE MOTIVATION FOR CONSIDERING LPT?}
\label{motivation}

The relation between the matter density and the displacement vector (Eq.~(\ref{relation})) corresponds to the law of mass conservation.
Therefore, the non-linear solutions in LPT naturally guarantee mass conservation (see Sec.~\ref{3loop_and_more}).
The law is related to various properties of the matter density perturbation.
From the expression in Eq.~(\ref{relation}),
the space-independent displacement vector $\bar{\YY}(z)$ does not yield the matter perturbation:
\begin{eqnarray}
		\delta(z,\xx) \to  \int d^3q \delta_{\rm D}(\xx - \qq - \bar{\YY}(z)) - 1  = 0.
\end{eqnarray}
This implies that 
dark matter particles which globally move in the same way throughout in the universe do not yield the matter density perturbation.
This fact corresponds to the Galilean invariance~\citep{Scoccimarro:1995if,Peloso:2013zw,Kehagias:2013yd,
Bernardeau:2012aq,Sugiyama:2013b,Blas:2013bpa}.
In other words, conservation of mass guarantees Galilean invariance.
In connection with this, in calculating the power spectrum, the integrand in Eq.~(\ref{LPT_power_1}) converges to zero at $\qq=\qq_1-\qq_2=0$,
where $\qq$ is the relative coordinates between the initial positions of dark matter particles,
and the power spectrum has no contribution at this point.
As discussed in Secs.~\ref{IR} and~\ref{Sec:Review_SPT},
this feature is related to the well-known cancellation of the high-$k$ limit solutions and the IR divergence problem in SPT
~\citep{Sugiyama:2013b,Scoccimarro:1995if,Pajer:2013jj,Carrasco:2013sva},
because $q=|\qq_1-\qq_2|\to0$ means the small scale limit.
Furthermore, as shown in Sec.~\ref{LPT_RegPT}, 
the power spectrum in LPT has a simple relation to the $\Gamma$-expansion~\citep{Bernardeau:2008fa,Bernardeau:2011dp}
and RegPT~\citep{Bernardeau:2011dp,Taruya:2012ut,Taruya:2013my}.
Thus, LPT has various interesting properties, and this is the reason we explore LPT.

\section{CORRELATION FUNCTIONS OF THE DISPLACEMENT VECTOR}
\label{Sec:Sigma}
To obtain the power spectrum in LPT, we have to compute the correlation function of the displacement vector $\Sigma$ in Eq.~(\ref{Sigma_def}).
In this section, we investigate the properties of $\Sigma$ at the linear and 1-loop orders,
where the $n$-loop in LPT means $\Sigma_{n\rm \mathchar`-loop} = {\cal O}\left( P_{\rm lin}^{n+1} \right)$:
\begin{eqnarray}
		\Sigma(z,\kk,\qq) &=&  D^2\Sigma_{\rm lin}(\kk,\qq) + D^4\Sigma_{\rm 1\mathchar`-loop}(\kk,\qq), \nonumber \\
		\bar{\Sigma}(z,k) &=& \Sigma_0(z,k,q=0) = D^2\bar{\Sigma}_{\rm lin}(k) + D^4\bar{\Sigma}_{\rm 1\mathchar`-loop}(k).
		\label{sigma_lin2213}
\end{eqnarray}

\subsection{Multipole Expansion of $\Sigma$}
\label{Sec:Sigma_l}

In the Zel'dovich approximation, Eq.~(\ref{Sigma_def}) leads to
\begin{eqnarray}
		\Sigma_{\rm lin}(\kk,\qq) 
		&=& \int \frac{d^3p}{(2\pi)^3}e^{i\pp\cdot\qq} \left[\kk\cdot\LL_1(\pp) \right]^2 P_{\rm lin}(p) \nonumber \\
		&=& \Sigma_{0,\rm lin}(k,q){\cal L}_0(\hat{k}\cdot\hat{q}) - \Sigma_{2, \rm lin}(k,q) {\cal L}_2(\hat{k}\cdot\hat{q}),
\end{eqnarray}
where
\begin{eqnarray}
		\Sigma_{0,\rm lin}(k,q) &=&  \frac{1}{3} k^2 \int_0^{\infty}\frac{dp}{2\pi^2} j_0(pq) P_{\rm lin}(p), \nonumber \\
		\Sigma_{2,\rm lin}(k,q) &=&  \frac{2}{3} k^2 \int_0^{\infty}\frac{dp}{2\pi^2} j_2(pq) P_{\rm lin}(p),
		\label{sigma_lin}
\end{eqnarray}
and $P_{\rm lin}$ denotes the linear power spectrum at the present time.
Then, $\bar{\Sigma}_{\rm lin}(k)$ is given by 
\begin{eqnarray}
		\bar{\Sigma}_{\rm lin}(k) = \Sigma_{0,\rm lin}(k,q=0) =  \frac{1}{3} k^2\int_0^{\infty}\frac{dp}{2\pi^2} P_{\rm lin}(p).
		\label{bar_sigma_lin}
\end{eqnarray}
Thus, the Zel'dovich approximation has monopole and quadrupole terms.
The linear correlation functions of the displacement vector
$\Sigma_{\ell, \rm lin}$ only involve the spherical Bessel functions $j_{\ell}$ in their integrals.
Note that the non-linear scale-dependence of the Zel'dovich solution only comes from the non-linearity of the law of mass conservation,
where at the linear order the mass conservation shows $\YY_{\rm lin}(\kk) = i\kk\delta_{\rm lin}(\kk)/k^2$.
On the other hand, the linear equation of motion of the displacement vector provides the linear growth function $D$.
The factors $\frac{1}{3}$ and $\frac{2}{3}$ in front of the monopole and quadrupole terms
result from isotropy and anisotropy, respectively.
Since $|j_2(pq)| \sim |j_0(pq)|$ is satisfied at large scales,
the quadrupole term has two times greater amplitude than the monopole term at these scales.
The limiting small scale $q = 0$ leads to $\Sigma_{0, \rm lin} \to \bar{\Sigma}_{\rm lin}$ and $\Sigma_{2,\rm lin} \to 0$
due to $j_0(0) = 1$ and $j_2(0) = 0$.

At the 1-loop order in LPT, we decompose $\Sigma_{\rm 1\mathchar`-loop}$ into two parts as in the 1-loop SPT:
\begin{eqnarray}
		\Sigma_{\rm 1\mathchar`-loop}(\kk,\qq) &=& \Sigma_{22}(\kk,\qq) + \Sigma_{13}(\kk,\qq), \nonumber \\
		\bar{\Sigma}_{\rm 1\mathchar`-loop}(k) &=& \bar{\Sigma}_{22}(k) + \bar{\Sigma}_{13}(k).
		\label{sigmas1}
\end{eqnarray}
Equation~(\ref{Sigma_def}) leads to
\begin{eqnarray}
		\Sigma_{22}(\kk,\qq) &=& 
		\frac{1}{2}\int \frac{d^3p_1}{(2\pi)^3}\int \frac{d^3p_2}{(2\pi)^3}
		e^{i\pp_1\cdot\qq} e^{i\pp_2\cdot\qq}\left[ \kk\cdot\LL_2(\pp_1,\pp_2) \right]^2P_{\rm lin}(p_1)P_{\rm lin}(p_2) \nonumber \\
		&& + \int \frac{d^3p_1}{(2\pi)^3}\int \frac{d^3p_2}{(2\pi)^3}
		e^{i\pp_1\cdot\qq} e^{i\pp_2\cdot\qq}\left[ \kk\cdot\LL_1(\pp_1)\kk\cdot\LL_1(\pp_2)\kk\cdot\LL_2(\pp_1,\pp_2) \right]
		P_{\rm lin}(p_1)P_{\rm lin}(p_2) \nonumber \\
		&=& \sum_{\ell=0}^3 i^{\ell} \Sigma_{\ell,22}(k,q) {\cal L}_{\ell}(\hat{k}\cdot\hat{q}),
\end{eqnarray}
and 
\begin{eqnarray}
\Sigma_{13}(\kk,\qq) &=& \int \frac{d^3p_1}{(2\pi)^3}\int \frac{d^3p_2}{(2\pi)^3}
e^{i\pp_1\cdot\qq} \left[ \kk\cdot\LL_1(\pp_1) \kk\cdot\LL_3(\pp_1,\pp_2,-\pp_2) \right]P_{\rm lin}(p_1)P_{\rm lin}(p_2) \nonumber \\
&& -2 \int \frac{d^3p_1}{(2\pi)^3}\int \frac{d^3p_2}{(2\pi)^3}
e^{i\pp_1\cdot\qq}\left[ \kk\cdot\LL_1(\pp_1)\kk\cdot\LL_1(\pp_2)\kk\cdot\LL_2(\pp_1,\pp_2) \right]P_{\rm lin}(p_1)P_{\rm lin}(p_2) \nonumber \\
&=& \sum_{\ell=0}^3 i^{\ell} \Sigma_{\ell,13}(k,q){\cal L}_{\ell}(\hat{k}\cdot\hat{q}).
\end{eqnarray}
Then, the multipole terms in the 1-loop LPT are given by
\begin{eqnarray}
		\Sigma_{\ell, 22}(k,q) &=& 
		 \int_{0}^{\infty}\frac{dp_1p_1^2}{2\pi^2} \int_{0}^{\infty}\frac{dp_2p_2^2}{2\pi^2} \int_{-1}^{1} d\mu
		 j_{\ell}\left( |\pp_1+\pp_2| q\right)	K_{\ell, 22}(k,p_1,p_2,\mu) P_{\rm lin}(p_1)	P_{\rm lin}(p_2), \nonumber \\
		 \bar{\Sigma}_{22}(k) &=&  \int_{0}^{\infty}\frac{dp_1p_1^2}{2\pi^2} \int_{0}^{\infty}\frac{dp_2p_2^2}{2\pi^2} \int_{-1}^{1} d\mu
		 K_{0, 22}(k,p_1,p_2,\mu) P_{\rm lin}(p_1) P_{\rm lin}(p_2), \nonumber \\
		\Sigma_{\ell, 13}(k,q) &=& 
		 \int_{0}^{\infty}\frac{dp_1p_1^2}{2\pi^2}
		 \int_{0}^{\infty}\frac{dp_2p_2^2}{2\pi^2} j_{\ell}\left( p_1 q\right) K_{\ell, 13}(k,p_1,p_2)  P_{\rm lin}(p_1)	P_{\rm lin}(p_2),
		 \nonumber \\
		 \bar{\Sigma}_{13}(k) &=& \int_{0}^{\infty}\frac{dp_1p_1^2}{2\pi^2}
		 \int_{0}^{\infty}\frac{dp_2p_2^2}{2\pi^2} K_{0, 13}(k,p_1,p_2) P_{\rm lin}(p_1) P_{\rm lin}(p_2),
		\label{sigmas2}
\end{eqnarray}
where $\mu \equiv \hat{p}_1\cdot\hat{p}_2$ and $y \equiv p_2/p_1$.
Appendix~\ref{ap:K} summarizes the definitions of the kernel functions $K_{\ell,22}$ and $K_{\ell,13}$.
Note that $\bar{\Sigma}_{22}(k)$ and $\bar{\Sigma}_{13}(k)$ are given by
$\bar{\Sigma}_{22}(k) = \Sigma_{0,22}(k,q=0)$ and $\bar{\Sigma}_{13}(k) = \Sigma_{0,13}(k,q=0)$
and satisfy $\bar{\Sigma}_{22}(k) = \frac{27}{140} \bar{\Sigma}_{13}(k)$.

We find that the 1-loop LPT has monopole, dipole, quadrupole, and octupole terms,
where the dipole and octupole terms come from $\langle \YY \YY \YY \rangle_{\rm c}$ in Eq.~(\ref{Sigma_def}).
The subscripts $13$ and $22$ in $\Sigma_{\ell,13}$ and $\Sigma_{\ell,22}$ mean
that they make the correction terms $P_{13}$ and $P_{22}$ in the 1-loop SPT (for details, see Sec.~\ref{SPT_1loop}).
Unlike the Zel'dovich approximation,
$\Sigma_{\ell,22}$ and $\Sigma_{\ell,13}$ have the kernel functions $K_{\ell,22}$ and $K_{\ell,13}$ in their integrals
which come from the non-linear dynamics of dark matter.
Similarly to the case in the Zel'dovich approximation,
at $q =0$, the dipole, quadrupole, and octupole terms become zero
: $\Sigma_{\ell \geq 1, 22}(k,q=0) = \Sigma_{\ell \geq 1, 13}(k,q=0) = 0$.

\begin{figure}[t]
		\begin{center}
				\plottwo{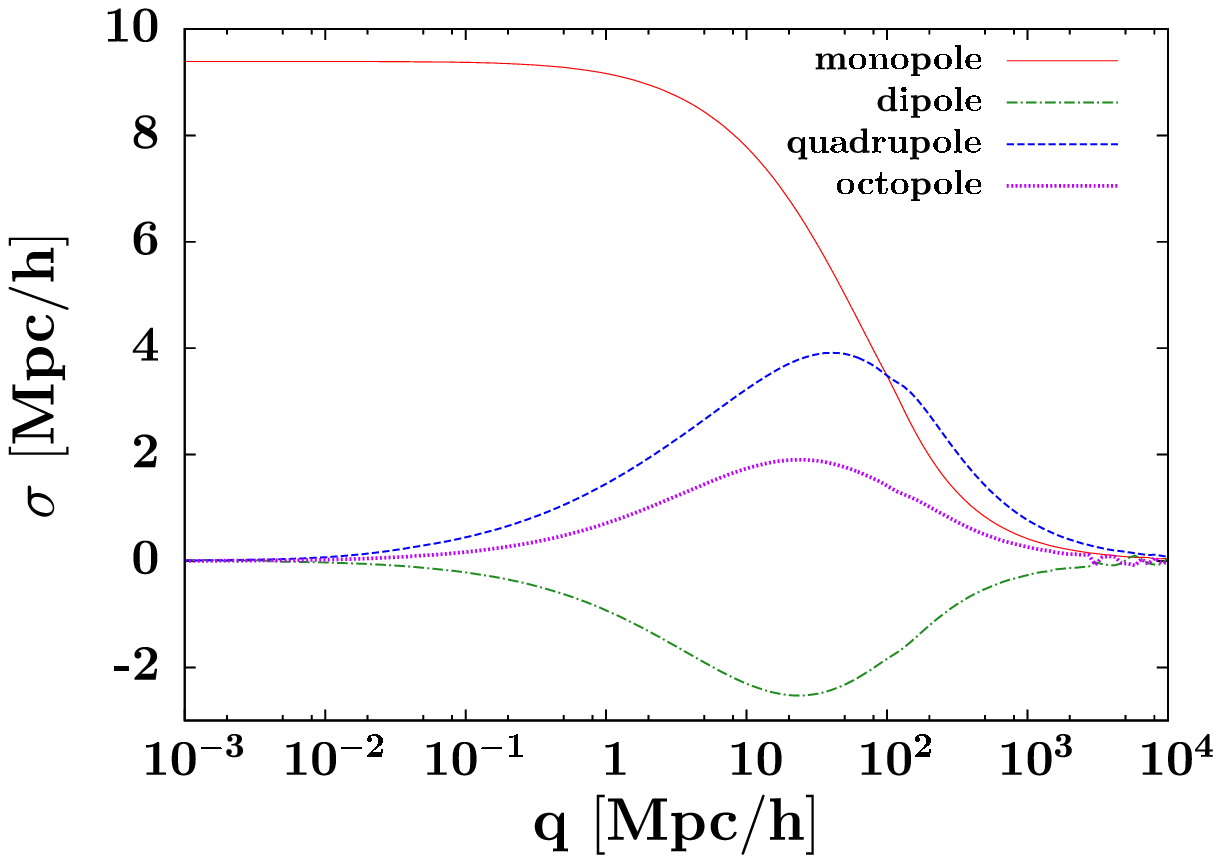}{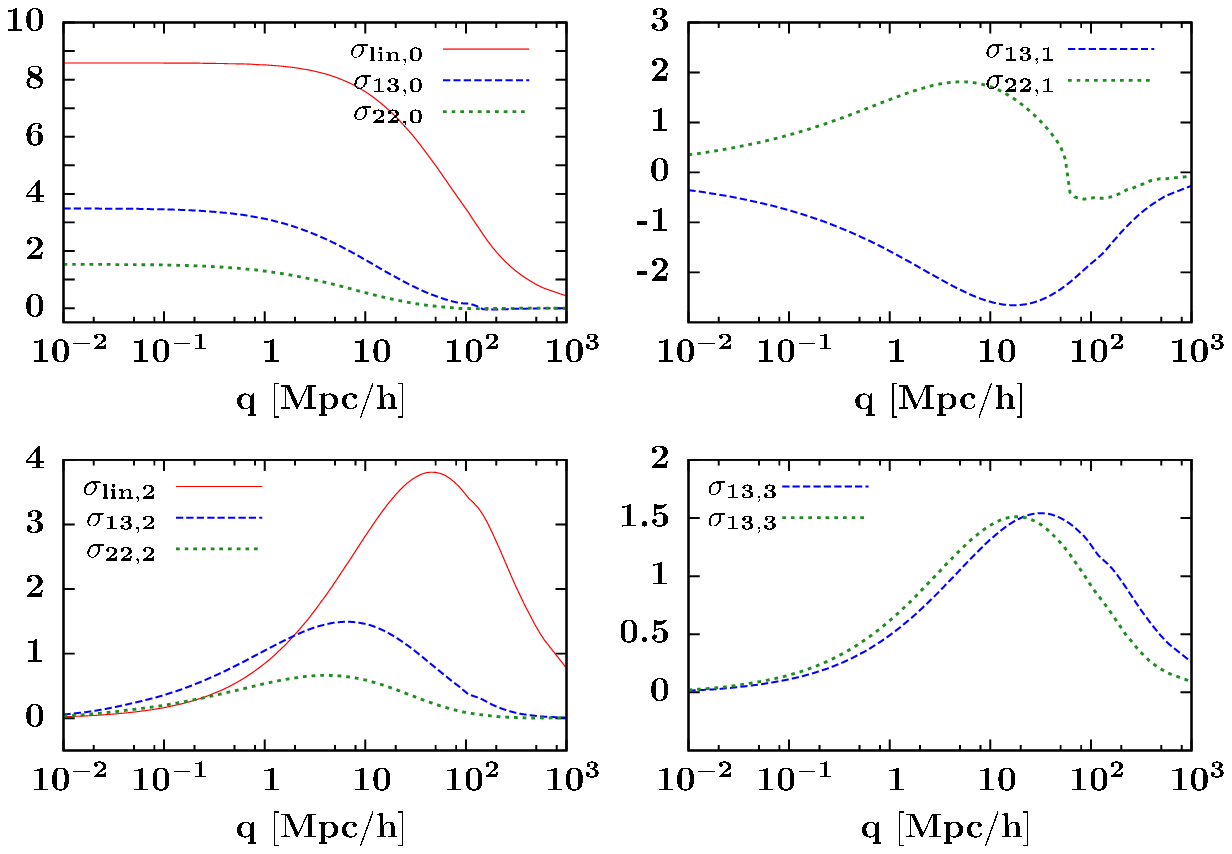}
		\end{center}
		\caption{Left: Functions $\sigma_{\ell}(z,q)$ for $\{\ell = 0,1,2,3\}$ at $z=0$ are plotted.
		The red, green, blue, and violet lines denote the monopole, dipole, quadrupole, and octupole terms 
		defined in Eq.~(\ref{dif:sigma}), respectively.
		Right: $\sigma_{\ell,\rm lin}(q)$, 
		$\sigma_{\ell,22}(q)$, and $\sigma_{\ell,13}(q)$ functions for $\{\ell = 0,1,2,3\}$ defined in Eq.~(\ref{dif:sigma_each})
		are plotted as red, green, and blue lines.}
		\label{fig:sigma}
\end{figure}

Here, we define the quantities $\sigma_{\ell}$ which have the dimension of length $[{\rm Mpc}/h]$:
\begin{eqnarray}
		\Sigma_0(z,k,q) &\equiv& \frac{k^2 \sigma_0^2(z,q)}{2}, \quad
		\Sigma_1(z,k,q) \equiv \frac{k^3 \sigma_1^3(z,q)}{2}, \quad
		\Sigma_2(z,k,q) \equiv \frac{k^2 \sigma_2^2(z,q)}{2}, \quad
		\Sigma_3(z,k,q) \equiv \frac{k^3 \sigma_3^3(z,q)}{2}. \nonumber \\
		\label{dif:sigma}
\end{eqnarray}
and
\begin{eqnarray}
		\Sigma_{0, \rm lin}(k,q) &=& \frac{k^2\sigma_{0, \rm lin}^2(q)}{2}, \quad
		\Sigma_{2, \rm lin}(k,q) = \frac{k^2\sigma_{2, \rm lin}^2(q)}{2}, \nonumber \\
		\Sigma_{0, \rm 13}(k,q) &=&  \frac{k^2\sigma_{0, \rm 13}^2(q)}{2}, \quad
		\Sigma_{1, \rm 13}(k,q) =  \frac{k^3\sigma_{1, \rm 13}^3(q)}{2}, \quad
		\Sigma_{2, \rm 13}(k,q) =  \frac{k^2\sigma_{2, \rm 13}^2(q)}{2}, \quad
		\Sigma_{3, \rm 13}(k,q) =  \frac{k^3\sigma_{3, \rm 13}^3(q)}{2}, \nonumber \\
		\Sigma_{0, \rm 22}(k,q) &=&  \frac{k^2\sigma_{0, \rm 22}^2(q)}{2}, \quad
		\Sigma_{1, \rm 22}(k,q) =  \frac{k^3\sigma_{1, \rm 22}^3(q)}{2}, \quad
		\Sigma_{2, \rm 22}(k,q) =  \frac{k^2\sigma_{2, \rm 22}^2(q)}{2}, \quad
		\Sigma_{3, \rm 22}(k,q) =  \frac{k^3\sigma_{3, \rm 22}^3(q)}{2}. \nonumber \\
		\label{dif:sigma_each}
\end{eqnarray}
The left panel of Figure~\ref{fig:sigma} shows $\sigma_{\ell}$.
At large scales ($q \gtrsim 100\ {\rm Mpc}/h$), 
the linear contributions to the monopole and quadrupole terms are dominant
and the amplitude of the quadrupole is twice larger than that of the monopole.
On the other hand, at small scales the dipole, quadrupole, and octupole terms become zero.
In the right panel of Figure~\ref{fig:sigma}, we find that 
the linear contributions are larger than the non-linear ones at large scales: 
$|\sigma_{\ell, \rm lin}| > |\sigma_{\ell,22}|\ {\rm and}\ |\sigma_{\ell,13}|$.
These features of $\sigma_{\ell}$, $\sigma_{\ell,\rm lin}$, $\sigma_{\ell,\rm 22}$, and $\sigma_{\ell,\rm 13}$ are indeed what we expected.

\subsection{IR Divergence}
\label{IR}

In the Zel'dovich approximation,
$\bar{\Sigma}_{\rm lin}$ has the following integral (Eq.~(\ref{bar_sigma_lin})):
\begin{eqnarray}
		\int_0^{\infty} dp P_{\rm lin}(p).
\end{eqnarray}
For the power-law initial power spectrum $P_{\rm lin}(p) \propto p^n$, this integral has
the IR divergence for $n\leq -1$ and UV divergence for $n\geq -1$~\citep{Scoccimarro:1995if}.
However, computing the power spectrum alleviates the condition of IR divergence,
because the above integral appears as a combination of the monopole terms as follows
\begin{eqnarray}
		\Sigma_{0,\rm lin}(k,q) - \bar{\Sigma}_{\rm lin}(k)
		&\to& - \frac{k^2q^2}{36\pi^2}\int_0^{\infty}dp p^2 P_{\rm lin}(p) \quad \mbox{for $p\to0$}, \nonumber \\
		\Sigma_{2,\rm lin}(k,q)
		&\to& \frac{k^2q^2}{45\pi^2}\int_0^{\infty}dp p^2 P_{\rm lin}(p) \quad \mbox{for $p\to0$},
\end{eqnarray}
where we used $j_0(x) = 1 - x^2/6$ and $j_2(x) = x^2/15$ for $x\ll1$.
Therefore, the Zel'dovich power spectrum has IR divergence for $n\leq -3$~\citep{Taylor:1996ne}.
This is the result of the law of conservation of mass,
because the non-linear $k$-dependence of the power spectrum in the Zel'dovich approximation
results only from the non-linear equation of mass conservation.

At the 1-loop order in LPT, the asymptotic behaviors of the kernel functions $K_{\ell,22}$ and $K_{\ell,13}$
(see Appendix~\ref{ap:K}) lead to those of $\Sigma_{\ell,22}$ and $\Sigma_{\ell,13}$:
for $p_2/p_1\ll1$,
\begin{eqnarray}
		\Sigma_{\ell,22}(k,q)  &\propto& \Sigma_{\ell,13}(k,q) \propto 
		k^2 \int_0^{\infty} dp_1 j_{\ell}(p_1q) P_{\rm lin}(p_1) \int_0^{\infty} dp_2 p_2^2 P_{\rm lin}(p_2), 
		\quad \mbox{for $\ell=0$ and 2},\nonumber \\
		\Sigma_{\ell,22}(k,q)  &\propto& \Sigma_{\ell,13}(k,q) \propto 
		k^3 \int_0^{\infty} dp_1 \frac{j_{\ell}(p_1q)}{p_1} P_{\rm lin}(p_1) \int_0^{\infty} dp_2 p_2^2 P_{\rm lin}(p_2), 
		\quad \mbox{for $\ell=1$ and 3},
		\label{SS_1}
\end{eqnarray}
and for $y=p_2/p_1\gg1$,
\begin{eqnarray}
		\Sigma_{\ell,13}(k,q) &\propto& k^2\int_0^{\infty} dp_1 p_1^2 j_{\ell}(p_1q) P_{\rm lin}(p_1) \int_0^{\infty}dp_2 P_{\rm lin}(p_2)
		\quad \mbox{for $\ell=0$ and 2}, \nonumber \\
		\Sigma_{\ell,13}(k,q)
		&\propto& k^3\int_0^{\infty} dp_1 p_1^2 \frac{j_{\ell}(p_1q)}{p_1} P_{\rm lin}(p_1) \int_0^{\infty}dp_2 P_{\rm lin}(p_2)
		\quad \mbox{for $\ell=1$ and 3},
		\label{SS_2}
\end{eqnarray}
where $\Sigma_{\ell,22}$ for $y=p_2/p_1\gg1$ are given by replacing $p_1$ with $p_2$ in Eq.~(\ref{SS_1}) 
due to the symmetry of $K_{\ell,22}$ about $p_1$ and $p_2$.
For details, see Appendix~\ref{ap:Sigma}.
Thus, we find that for the power-law initial power spectrum $P_{\rm lin}(p) \propto p^n$ with $-3<n<-1$,
$\Sigma_{\ell,22}$ and $\Sigma_{\ell,13}$ have no IR and UV divergences in both cases:
(1) $p_1\to\infty$ and $p_2\to0$, and (2) $p_1\to0$ and $p_2\to\infty$.

\section{POWER SPECTRUM IN SPT}
\label{Sec:Review_SPT}

In this section, we investigate how the solutions of LPT reproduce those of SPT,
where $n$-loop in SPT means $P_{n\rm \mathchar`-loop} = {\cal O}\left( P_{\rm lin}^{n+1} \right)$.

\subsection{At the Linear Order in SPT}
\label{SPT_linear}
Expanding the exponential factor in Eq.~(\ref{LPT_power_1}) and using Eq.~(\ref{sigma_lin}),
the monopole and quadrupole terms yield $\frac{1}{3}P_{\rm lin}$ and $\frac{2}{3}P_{\rm lin}$, respectively:
\begin{eqnarray}
		P_{\rm lin}(k) 
		&=&  \int d^3 q e^{-i\kk\cdot\qq} \left( \Sigma_{\rm lin}(\kk,\qq) - \bar{\Sigma}_{\rm lin}(k)\right)  \nonumber \\
			 &=&  4\pi \int_0^{\infty} d q q^2\left\{ j_0(kq) \Sigma_{0,\rm lin}(k,q) + j_2(kq) \Sigma_{2,\rm lin}(k,q) \right\} 
			 -(2\pi)^3 \delta_{\rm D}(\kk)\bar{\Sigma}_{\rm lin}(k) \nonumber \\
			 &=& \frac{1}{3}P_{\rm lin}(k) + \frac{2}{3}P_{\rm lin}(k),
\end{eqnarray}
where we used the mathematical formula:
$ e^{-i\kk\cdot\qq} =  \sum_{\ell=0}^{\infty}(2\ell+1)(-i)^{\ell} j_{\ell}(kq) {\cal L}_{\ell}( \hat{k} \cdot \hat{q})$,
$\int_{-1}^1d\mu{\cal L}_{\ell}(\mu){\cal L}_{\ell'}(\mu) = 2\delta_{\ell \ell'}/(2\ell+1)$,
and $\int_0^{\infty} dq q^2 j_{\alpha}(kq) j_{\alpha}(pq) =  \frac{\pi}{2k^2} \delta_{\rm D}(k-p)$.
In the second line, the last term $(2\pi)^3\delta_{\rm D}(\kk)\bar{\Sigma}_{\rm lin}(k)$ is zero because $\bar{\Sigma}_{\rm lin}(k=0)=0$.

\subsection{At the One-loop Order in SPT}
\label{SPT_1loop}

Substituting Eq.~(\ref{sigma_lin2213}) into Eq.~(\ref{LPT_power_1})
and expanding the exponential factor in Eq.~(\ref{LPT_power_1}),
we obtain the 1-loop correction term $P_{\rm 1\mathchar`-loop} = P_{22} + P_{13}$ in SPT as follows
\begin{eqnarray}
		P_{\rm 1\mathchar`-loop}(k) 
		&=&  \int d^3q e^{-i\kk\cdot\qq} 
		\Big\{\left( \Sigma_{\rm 1\mathchar`-loop}(\kk,\qq) - \bar{\Sigma}_{\rm 1\mathchar`-loop}(k)  \right)
		+ \frac{1}{2} \left( \Sigma_{\rm lin}(\kk,\qq) - \bar{\Sigma}_{\rm lin}(k) \right)^2\Big\} \nonumber \\
		&=& 4\pi \sum_{\ell=0}^3\int_0^{\infty} dq q^2 j_{\ell}(kq)  \Big\{ \Sigma_{\ell,22}(k,q) + \Sigma_{\ell, 13}(k,q) \Big\}
		\nonumber \\
		&& + \int d^3q e^{-i\kk\cdot\qq} 
		\frac{1}{2} \Big\{ \left(\Sigma_{\rm lin}(\kk,\qq) \right)^2 - 2 \Sigma_{\rm lin}(\kk,\qq) \bar{\Sigma}_{\rm lin}(k) \Big\}
		\nonumber \\
		&=& \sum_{\ell=0}^3 \Big\{ P_{\ell,22}(k) + P_{\ell,13}(k)  \Big\} + P_{\rm 1\mathchar`-loop}|_{\rm ZA}(k), \nonumber \\
		&=& P_{22}(k) + P_{13}(k),
		\label{SPT_LPT_1loop}
\end{eqnarray}
where the terms proportional to $\delta_{\rm D}(\kk)$ become zero due to 
$\bar{\Sigma}_{\rm lin}(k=0) = \bar{\Sigma}_{\rm 1\mathchar`-loop}(k=0)=0$.
Thus, $P_{\rm 1\mathchar`-loop}$ is decomposed into 
the multipole terms $\sum_{\ell=0}^3 \left( P_{\ell,22} + P_{\ell,13} \right)$ and
the contribution from the Zel'dovich approximation $P_{\rm 1\mathchar`-loop}|_{\rm ZA} = P_{22}|_{\rm ZA} + P_{13}|_{\rm ZA}$,
defined as
\begin{eqnarray}
		P_{\ell,22}(k) &\equiv& 4\pi\int_0^{\infty} dq q^2 j_{\ell}(kq) \Sigma_{\ell,22}(k,q), \quad
		P_{\ell,13}(k) \equiv 4\pi\int_0^{\infty} dq q^2 j_{\ell}(kq) \Sigma_{\ell,13}(k,q), \nonumber \\
		P_{22}|_{\rm ZA}(k)  &\equiv&   \int d^3q e^{-i\kk\cdot\qq} 
		\frac{1}{2} \left(\Sigma_{\rm lin}(\kk,\qq) \right)^2, \quad
		P_{13}|_{\rm ZA}(k)  \equiv  -\bar{\Sigma}_{\rm lin}(k) P_{\rm lin}(k).
		\label{def_SPT_LPT}
\end{eqnarray}
In the final line of Eq.~(\ref{SPT_LPT_1loop}), $P_{22}$ and $P_{13}$ are given by
\begin{eqnarray}
		P_{22}(k) =  \sum_{\ell=0}^3P_{\ell,22} + P_{22}|_{\rm ZA}, \quad
		P_{13}(k) = \sum_{\ell=0}^3P_{\ell,13} + P_{13}|_{\rm ZA}.
		\label{SPT1loop}
\end{eqnarray}
The specific expressions of $P_{\ell,22}$, $P_{\ell,13}$, and $P_{\ell,22}|_{\rm ZA}$  are summarized in Appendix~\ref{ap:SPT_1loop}.

For $p/k\ll1$, $P_{22}$ and $P_{13}$ cancel out each other
\citep{Sugiyama:2013b,Scoccimarro:1995if,Pajer:2013jj,Carrasco:2013sva},
\begin{eqnarray}
		P_{22, \rm high\mathchar`-k}(k) &\to&   \bar{\Sigma}_{\rm lin}(k) P_{\rm lin}(k), \nonumber \\
		P_{13, \rm high\mathchar`-k}(k) &\to& - \bar{\Sigma}_{\rm lin}(k) P_{\rm lin}(k), 
\end{eqnarray}
where $P_{22,\rm high\mathchar`-k}$ and $P_{13,\rm high\mathchar`-k}$ are called the high-$k$ solutions of $P_{22}$ and $P_{13}$,
such that $P_{\rm 1\mathchar`-loop}$ is proportional to $\int dp p^2 P_{\rm lin}(p)$ but not to $\int dp P_{\rm lin}(p)$ at $p\to0$ .
In other words, the 1-loop SPT alleviates the IR divergence problem (Sec.~\ref{IR}).
In the context of LPT,
the high-$k$ (small scale) limit corresponds to the limit of $q\to0$,
because $q$ is the relative distance between the initial positions of dark matter particles: $q=|\qq_1-\qq_2|$.
In Eq.~(\ref{SPT_LPT_1loop}),
$\Sigma_{\rm 1\mathchar`-loop}(\kk,\qq)-\bar{\Sigma}_{\rm 1\mathchar`-loop}(k)$ and $(\Sigma_{\rm lin}(\kk,\qq) - \bar{\Sigma}_{\rm lin}(k))^2$
by definition become zero at $q=0$.
Therefore, the cancellation at the high-$k$ limit naturally occurs. 
Specifically, for $p/k\ll1$ ($q\to0$), 
we can show $P_{22}|_{\rm ZA}(k) \to \bar{\Sigma}_{\rm lin}(k) P_{\rm lin}(k)$ 
and $P_{\ell,22} \propto P_{\ell,13}\propto P_{\rm lin}(k)\int dp p^2 P_{\rm lin}(p)$
(for details, see Appendix~\ref{ap:SPT_1loop}.)
Thus, the cancellation of the high-$k$ solutions in the 1-loop SPT comes from the Zel'dovich approximation
and is understood to be the result of the mass conservation,
because the fact that the power spectrum has no contribution at $\qq=0$ is derived only from the law of mass conservation
as mentioned in Sec.~\ref{motivation}.
Furthermore, it is known that the high-$k$ solutions $P_{22,\rm high\mathchar`-k}$ and $P_{13,\rm high\mathchar`-k}$ 
have considerable contributions even at low-$k$ regions in each term of $P_{22}$ and $P_{13}$
despite their complete cancellation~\citep{Sugiyama:2013b}.
As a result, the amplitude of $P_{\rm 1\mathchar`-loop}$ is substantially different from those of $P_{22}$ and $P_{13}$ even at large scales.
Because of all these reasons, we focus on the following quantities
\begin{eqnarray}
		\Delta P_{22}(k) \equiv	
		P_{22}(k) - P_{22, \rm high\mathchar`-k}(k) &=&  \sum_{\ell=0}^3 P_{\ell,22}(k) + P_{\rm 1\mathchar`-loop}|_{\rm ZA}(k), \nonumber \\
		\Delta P_{13}(k) \equiv	
		P_{13}(k) - P_{13, \rm high\mathchar`-k}(k) &=&  \sum_{\ell=0}^3 P_{\ell,13}(k),
		\label{Delta}
\end{eqnarray}
where $P_{\rm 1\mathchar`-loop} = P_{22} + P_{13} = \Delta P_{22} + \Delta P_{13}$.
In our previous work~\citep{Sugiyama:2013b}, we called $\Delta P_{22}$ and $\Delta P_{13}$ short-wavelength terms.
The Zel'dovich approximation only contributes to $\Delta P_{22}$:
$P_{\rm 1\mathchar`-loop}|_{\rm ZA} = \Delta P_{22}|_{\rm ZA}$ and $\Delta P_{13}|_{\rm ZA} = 0$.

\begin{figure}[t]
		\begin{center}
				\plottwo{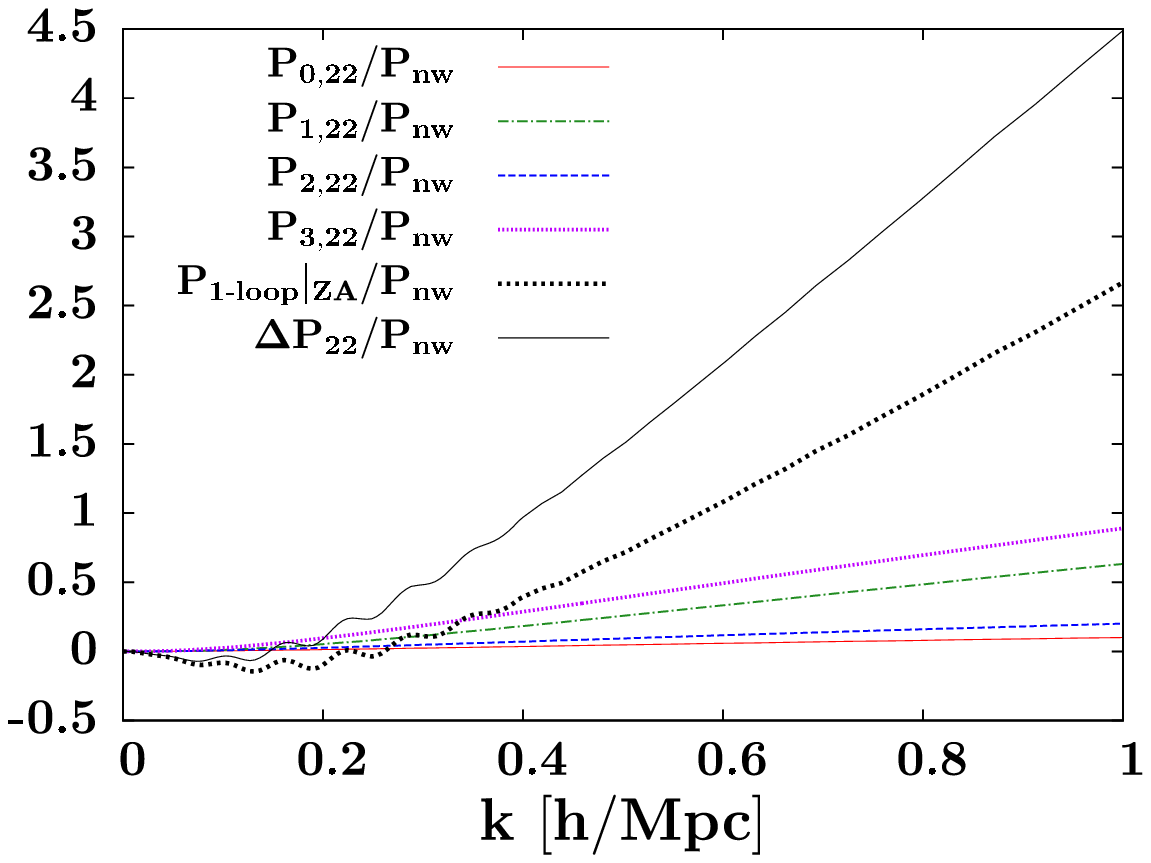}{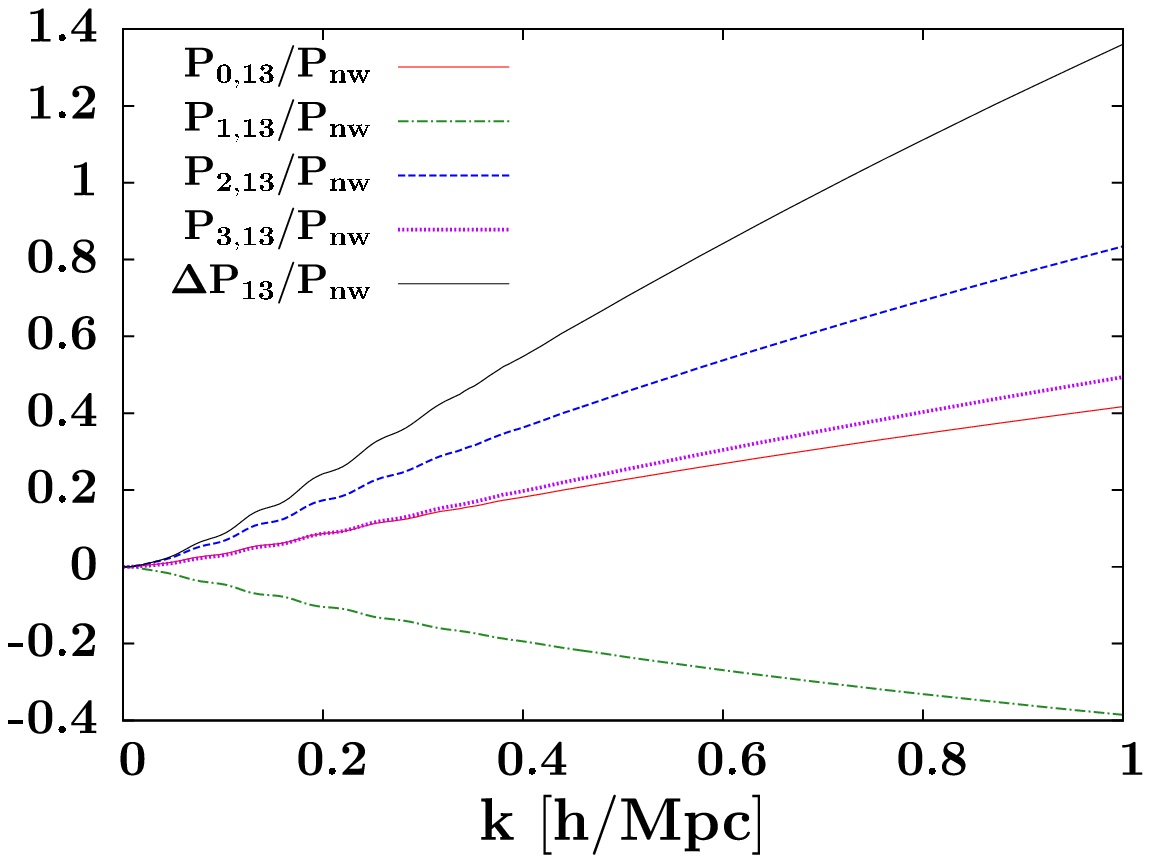}
		\end{center}
		\caption{
		Ratios between the 1-loop correction terms $\Delta P_{22}$ and $\Delta P_{13}$ in SPT, defined in Eq.~(\ref{Delta}),
		and the no-wiggle linear power spectrum $P_{\rm lin}^{\rm nw}$~\citep{Eisenstein:1997ik} are plotted at $z=0$.
		$\Delta P_{22}$ and $\Delta P_{13}$ each satisfies 
		the cancellation at the high-$k$ limit and is proportional to $\int dp p^2 P_{\rm lin}(p)$ at the limit of $p\to0$.
		The total 1-loop correction in SPT is given by $P_{\rm 1\mathchar`-loop}= P_{22} + P_{13} = \Delta P_{22} + \Delta P_{13}$.
		The Zel'dovich solution only contributes to $\Delta P_{22}$:
		$P_{\rm 1\mathchar`-loop}|_{\rm ZA} = \Delta P_{22}|_{\rm ZA} $ and $\Delta P_{13}|_{\rm ZA} = 0$.
		The contributions from the 1-loop LPT are represented as the multipole terms:
		$P_{\ell,22}$ and $P_{\ell,13}$ for $\{\ell = 0,1,2,3\}$.
		This figure shows that $P_{\ell,22}/P_{\rm nw} < 1$ and $P_{\ell,13}/P_{\rm nw} < 1$
		over the range of $k\leq 1.0\ [h{\rm Mpc}^{-1}]$ even at $z=0$, and they are suitable for perturbation quantities.
		}
		\label{fig:LPT_SPT_1loop}
\end{figure}

Figure~\ref{fig:LPT_SPT_1loop} shows how each term of $P_{\ell,22}$, $P_{\ell,13}$, and $P_{\rm 1\mathchar`-loop}|_{\rm ZA}$
contributes to $\Delta P_{22}$ and $\Delta P_{13}$ at $z=0$.
We find that the non-linear effects of the displacement vector are suitable for perturbation quantities
 $P_{\ell,22}/P_{\rm lin}^{\rm nw}<1$ and $P_{\ell,13}/P_{\rm lin}^{\rm nw}<1$ over the range of $k\leq 1.0\ [h{\rm Mpc}^{-1}]$.

\begin{figure}[tpb]
		\begin{center}
				\scalebox{0.9}{\plottwo{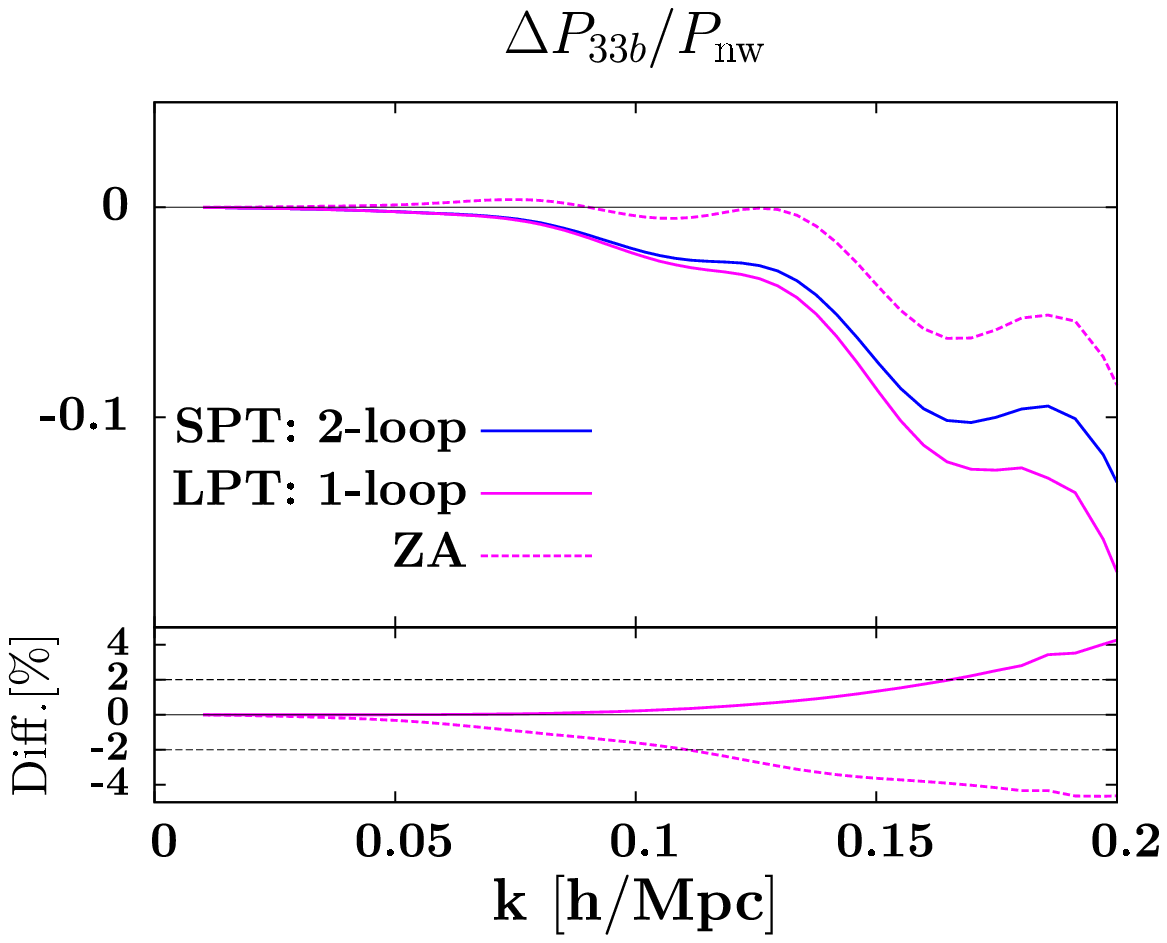}{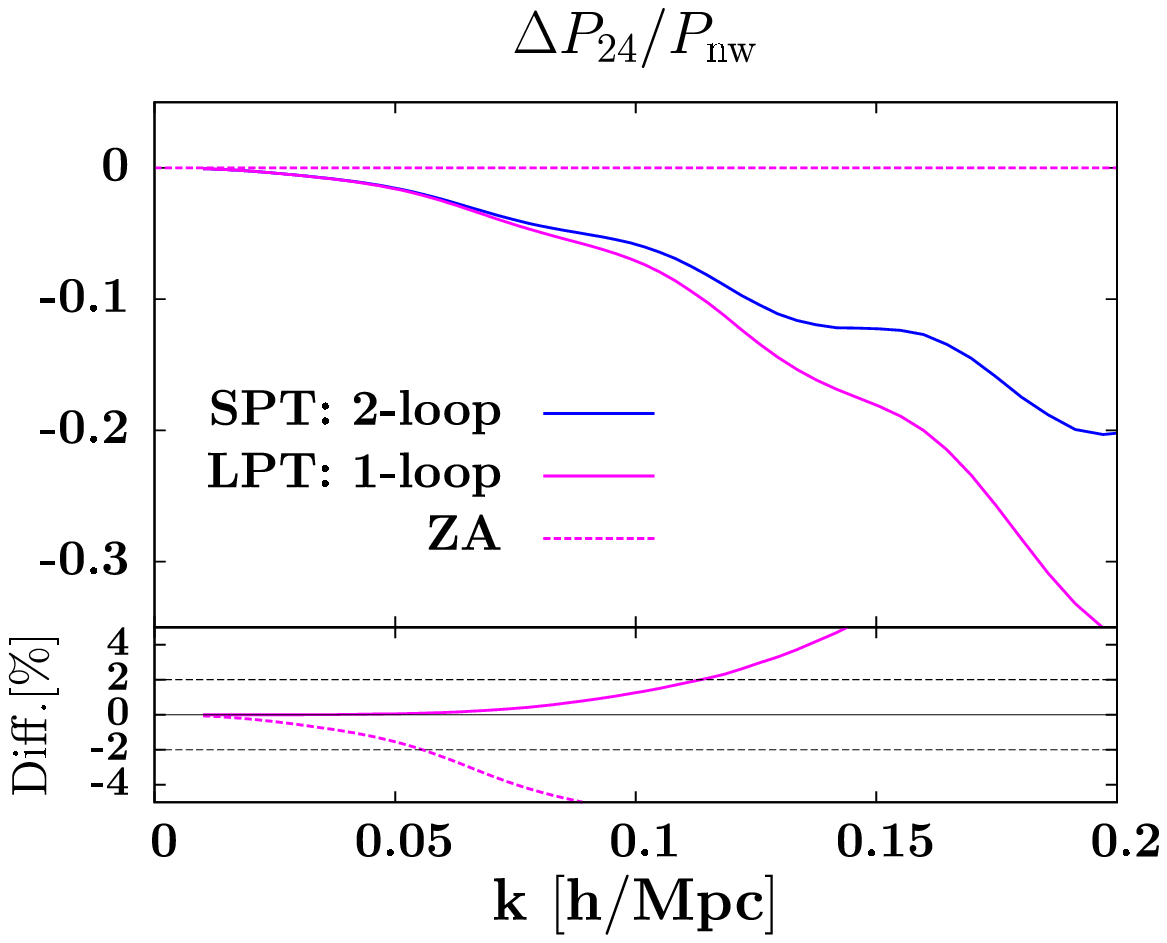}}
		\end{center}
		\begin{center}
				\scalebox{0.9}{\plottwo{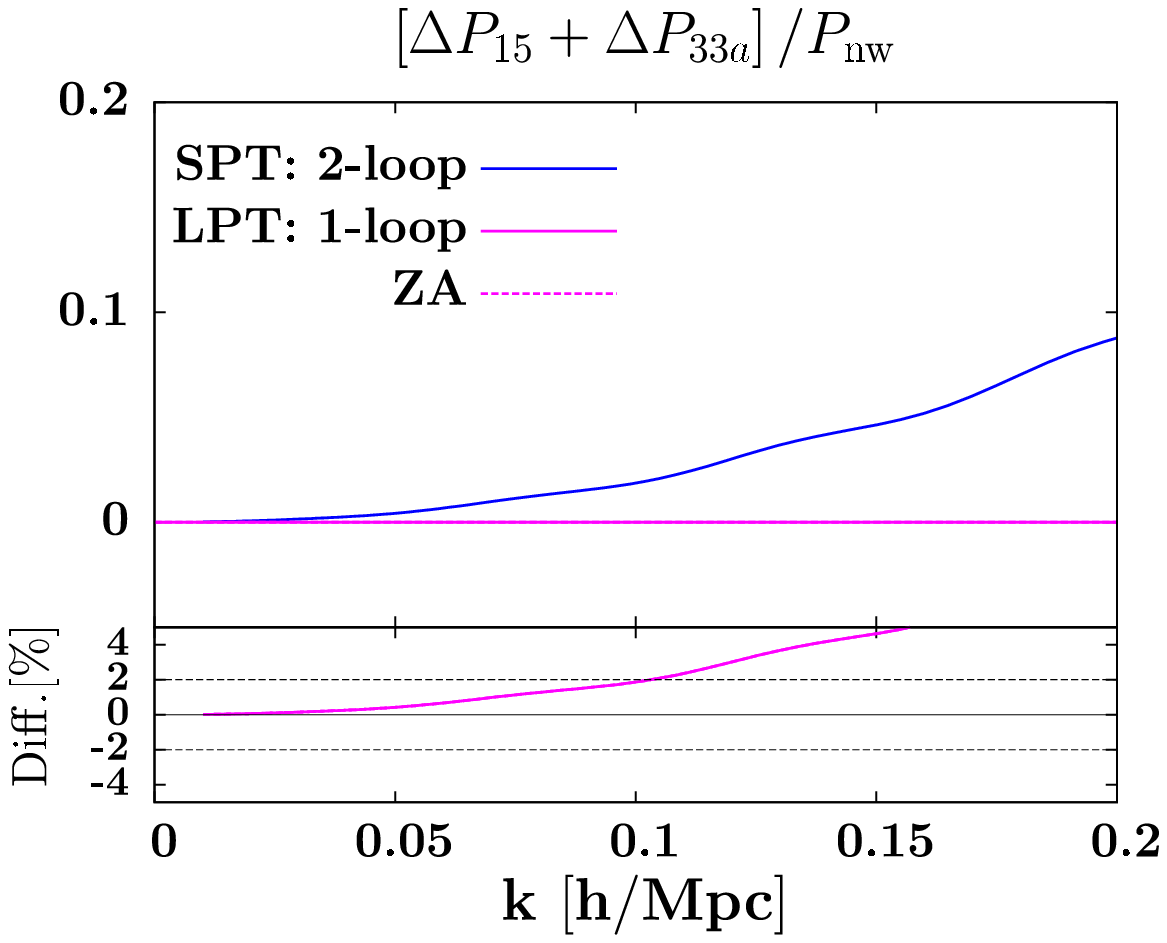}{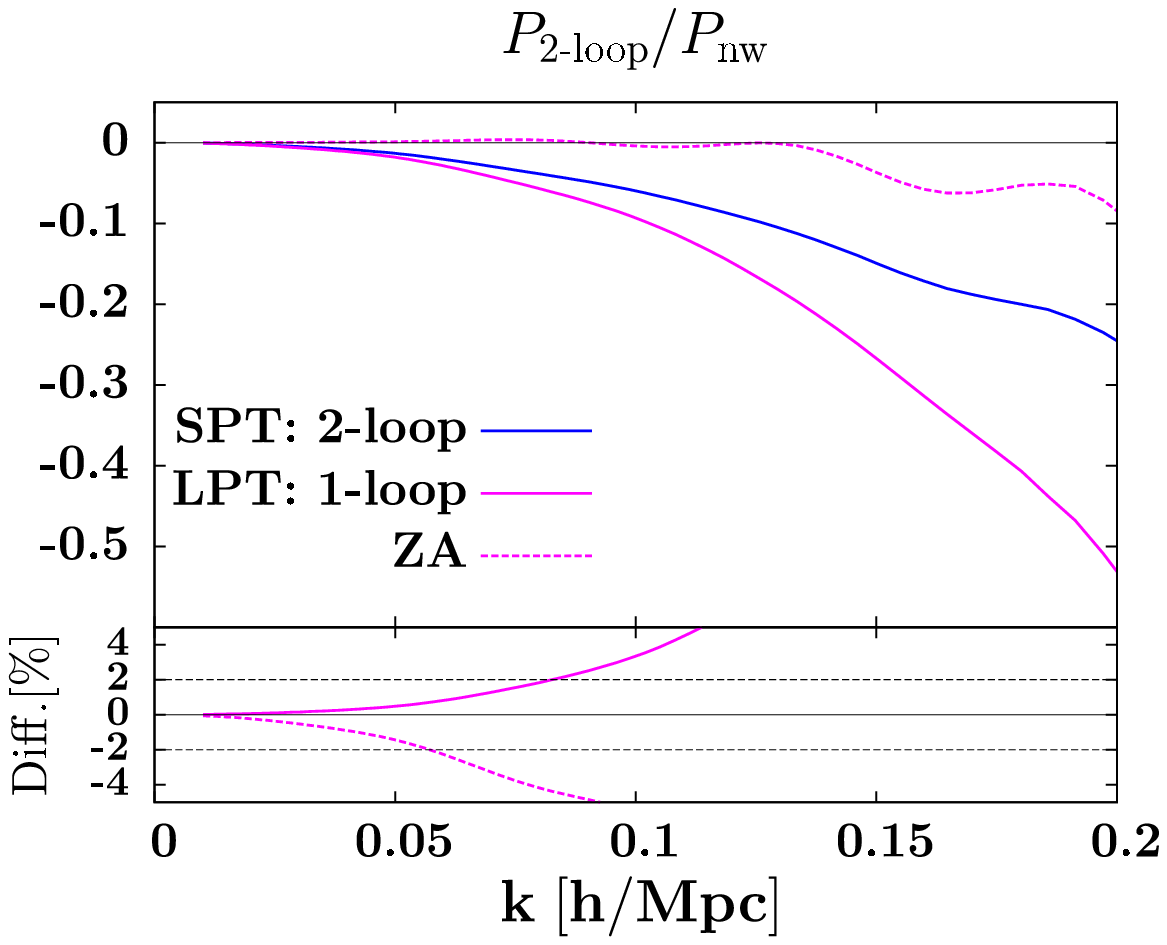}}
		\end{center}
		\caption{
		This figure shows the 2-loop solutions in SPT
		and their approximate solutions computed in the Zel'dovich approximation and the 1-loop LPT at $z=0$.
		Each term of $\Delta P_{33a}$, $\Delta P_{33b}$, $\Delta P_{24}$, and $\Delta P_{15}$
		is defined in Eq.~(\ref{Delta_2loop}).
		The Zel'dovich solution only contributes to $\Delta P_{33b}|_{\rm ZA}$:
		$P_{\rm 2\mathchar`-loop}|_{\rm ZA} = P_{33b}|_{\rm ZA}$ and 
		$\Delta P_{33a}|_{\rm ZA} = \Delta P_{24}|_{\rm ZA} = \Delta P_{15}|_{\rm ZA}  = 0$.
		The 1-loop LPT has contributions from $\Delta P_{33b}|_{\rm LPT,1\mathchar`-loop}$ and $\Delta P_{24}|_{\rm LPT,1\mathchar`-loop}$:
		$P_{\rm 2\mathchar`-loop}|_{\rm LPT,1\mathchar`-loop} = 
		\Delta P_{33b}|_{\rm LPT,1\mathchar`-loop} + \Delta P_{24}|_{\rm LPT,1\mathchar`-loop}$ and 
		$\Delta P_{33a}|_{\rm LPT,1\mathchar`-loop}=\Delta P_{15}|_{\rm LPT,1\mathchar`-loop} =0$.
		Until $k\sim 0.07$ $h{\rm Mpc}^{-1}$,
		the 1-loop LPT has a good agreement with the 2-loop solution of SPT
		at $z=0$ with accuracy of $1\%$
		where Diff [\%] is defined as $[P_{\rm SPT}-P_{\rm LPT}] \times 100/P_{\rm nw}$.
		}
		\label{fig:LPT_SPT_2loop}
\end{figure}

\subsection{At the Two-loop Order in SPT}
\label{SPT2loop}

Because of the non-linearity of the relation between the matter density and the displacement vector,
the 1-loop LPT solution has non-linear correction terms that have the same order as the 2-loop SPT.

The solutions in the 2-loop SPT have the following four terms:
$P_{\rm 2\mathchar`-loop} = P_{15} + P_{24} + P_{33a} + P_{33b}$,
where
\begin{eqnarray}
		P_{15}(k)  &=&
		30P_{\rm lin}(k) \int \frac{d^3p_1}{(2\pi)^3}
		\int \frac{d^3p_2}{(2\pi)^3}F_5(\kk,\pp_1,-\pp_1,\pp_2,-\pp_2)P_{\rm lin}(p_1)P_{\rm lin}(p_2), 
		\nonumber \\
		P_{33a}(k) &=& \frac{(P_{13}(k))^2}{4P_{\rm lin}(k)}, \nonumber \\
		P_{24}(k)  &=&  24	\int \frac{d^3k_1}{(2\pi)^3}\int\frac{d^3k_2}{(2\pi)^3}\int\frac{d^3p}{(2\pi)^3} 
		(2\pi)^3 \delta_{D}(\kk-\kk_{[1,2]}) F_2(\kk_1,\kk_2) F_4(\kk_1,\kk_2,\pp,-\pp)P_{\rm lin}(p) P_{\rm lin}(k_1) P_{\rm lin}(k_2),
		\nonumber \\
		P_{33b}(k) &=& 6	\int \frac{d^3k_1}{(2\pi)^3}\int\frac{d^3k_2}{(2\pi)^3}\int\frac{d^3k_3}{(2\pi)^3} 
		(2\pi)^3 \delta_D(\kk-\kk_{[1,3]}) \left[  F_3(\kk_1,\kk_2,\kk_3)\right]^2 P_{\rm lin}(k_1) P_{\rm lin}(k_2) P_{\rm lin}(k_3).
		\label{SPT_2loop}
\end{eqnarray}
Here, the high-$k$ solutions in the 2-loop SPT are given by~\citep{Sugiyama:2013b}
\begin{eqnarray}
		P_{33a, \rm high\mathchar`-k}(k) &=&  -\frac{1}{2} \bar{\Sigma}_{\rm lin}(k) P_{13}(k) 
		- \frac{1}{4} \left( \bar{\Sigma}_{\rm lin}(k) \right)^2 P_{\rm lin}(k), \nonumber \\
		P_{33b, \rm high\mathchar`-k}(k) &=& 
		\bar{\Sigma}_{\rm lin}(k) P_{22}(k) - \frac{1}{2} \left( \bar{\Sigma}_{\rm lin}(k) \right)^2 P_{\rm lin}(k)
		+ \bar{\Sigma}_{22}(k) P_{\rm lin}(k), \nonumber \\
		P_{24,  \rm high\mathchar`-k}(k) &=&  - \bar{\Sigma}_{\rm lin}(k) P_{22}(k) + \bar{\Sigma}_{\rm lin}(k)P_{13}(k)
  	    + \left( \bar{\Sigma}_{\rm lin}(k) \right)^2 P_{\rm lin}(k) + \bar{\Sigma}_{13}(k) P_{\rm lin}(k) , \nonumber \\ 
		P_{15,  \rm high\mathchar`-k}(k) &=&  - \frac{1}{2}\bar{\Sigma}_{\rm lin}(k) P_{13}(k)
	                                - \frac{1}{4} \left( \bar{\Sigma}_{\rm lin}(k) \right)^2 P_{\rm lin}(k)
                                 	- \bar{\Sigma}_{\rm 1\mathchar`-loop}(k) P_{\rm lin}(k).
\end{eqnarray}
Similarly to the 1-loop SPT, we define the following quantities:
\begin{eqnarray}
		\Delta P_{33a}(k) &\equiv& P_{33a}(k) - P_{33a,\rm high\mathchar`-k}(k), \nonumber \\
		\Delta P_{33b}(k) &\equiv& P_{33b}(k) - P_{33b,\rm high\mathchar`-k}(k), \nonumber \\
 	    \Delta P_{24}(k) &\equiv& P_{24}(k) - P_{24,\rm high\mathchar`-k}(k), \nonumber \\
		\Delta P_{15}(k) &\equiv& P_{15}(k) - P_{15,\rm high\mathchar`-k}(k).
		\label{Delta_2loop}
\end{eqnarray}
Note that $P_{\rm 2\mathchar`-loop} = \Delta P_{15} + \Delta P_{24} + \Delta P_{33a} + \Delta P_{33b}$
and $\Delta P_{33a} = \left( \Delta P_{13} \right)^2/(4 P_{\rm lin})$.

Similarly to the 1-loop SPT, 
$\Delta P_{33a}$, $\Delta P_{33b}$, $\Delta P_{24}$, and $\Delta P_{15}$
are derived from combinations that become zero at $\qq=0$ in Eq.~(\ref{LPT_power_1}).
The Zel'dovich approximation only contributes to $\Delta P_{33b}$:
\begin{eqnarray}
		P_{\rm 2\mathchar`-loop}|_{\rm ZA}(k)
		&=&  \Delta P_{33b}|_{\rm ZA}(k) 
		= \frac{1}{3!}\int d^3q e^{-i\kk\cdot\qq} \left\{ \Sigma_{\rm lin}(\kk,\qq) - \bar{\Sigma}_{\rm lin}(k) \right\}^3,
		\nonumber \\
		&=& P_{33b}|_{\rm ZA}(k) - \bar{\Sigma}_{\rm lin}(k) P_{22}|_{\rm ZA}(k) 
		+ \frac{1}{2} \left( \bar{\Sigma}_{\rm lin}(k) \right)^2 P_{\rm lin}(k),
\end{eqnarray}
and $\Delta P_{33a}|_{\rm ZA} = \Delta P_{24}|_{\rm ZA} = \Delta P_{15}|_{\rm ZA} = 0$.
On the other hand, 
we derive the following expressions corresponding to the SPT solutions from the 1-loop LPT:
\begin{eqnarray}
	\Delta P_{33b}|_{\rm LPT, 1\mathchar`-loop}(k) &=& \Delta P_{33b}|_{\rm ZA}(k)
	+ \int d^3q e^{-i\kk\cdot\qq} \left\{ \left( \Sigma_{22}(\kk,\qq) - \bar{\Sigma}_{22}(k)  \right)
	\left( \Sigma_{\rm lin}(\kk,\qq) - \bar{\Sigma}_{\rm lin}(k) \right)\right\},  \nonumber \\
	&=& P_{33b}|_{\rm LPT,1\mathchar`-loop}(k) - P_{33b,\rm high\mathchar`-k}(k) \nonumber \\
	\Delta P_{24}|_{\rm LPT, 1\mathchar`-loop}(k) &=& 
	\int d^3q e^{-i\kk\cdot\qq} \left\{ \left( \Sigma_{13}(\kk,\qq) - \bar{\Sigma}_{13}(k) \right) 
	\left( \Sigma_{\rm lin}(\kk,\qq) - \bar{\Sigma}_{\rm lin}(k) \right)\right\}, \nonumber \\
	&=& P_{24}|_{\rm LPT, 1\mathchar`-loop}(k) - P_{24,\rm high\mathchar`-k}(k), 
	\label{SPT2loop_in_LPT}
\end{eqnarray}
and $\Delta P_{33a}|_{\rm LPT, 1\mathchar`-loop}= \Delta P_{15}|_{\rm LPT, 1\mathchar`-loop} = 0$.
The specific expressions of $P_{33b}|_{\rm ZA}$, $P_{33b}|_{\rm LPT,1\mathchar`-loop}$, and $P_{24}|_{\rm LPT,1\mathchar`-loop}$ 
are given in Appendix~\ref{ap:SPT_2loop}.
Figure~\ref{fig:LPT_SPT_2loop} 
compares the 2-loop solutions in SPT with the approximate ones computed in the linear and 1-loop LPT at $z=0$.
Around $k\simeq0.2\ [h{\rm Mpc}^{-1}]$,
the validity of the approximate solutions in the 1-loop LPT is violated.
This is because of lack of non-linear dynamics.
The 1-loop LPT has the third order displacement vector in the perturbation series,
but we need the fifth order displacement vector to completely reproduce the 2-loop SPT solutions.
The limitation scale of the validity of the 1-loop LPT is estimated 
as $|P_{\rm 2\mathchar`-loop} - P_{\rm 2\mathchar`-loop}|_{\rm LPT, 1\mathchar`-loop}| > |P_{\rm 2\mathchar`-loop}|$.
The scale where this relation is satisfied is $k \gtrsim 0.2\ [h{\rm Mpc}^{-1}]$.
In other words, at these scales, the 1-loop SPT solution is better than the 1-loop LPT solution.
This behavior of the 1-loop LPT solution is independent of redshifts.
Thus, we can theoretically check the validity of the 1-loop LPT solution without using $N$-body simulations.

\section{POWER SPECTRUM IN LPT}
\label{Sec:pk_LPT}

Generally, the power spectrum is represented as~\citep{Crocce:2007dt}
\begin{eqnarray}
		P(z,k) = G^2(z,k) P_{\rm lin}(k) + P_{\rm MC}(z,k),
		\label{GMC}
\end{eqnarray}
where $G$ and $P_{\rm MC}$ are referred to as ``propagator''  and ``mode-coupling term'' in the context of RPT.
While we can compute the propagator with relative ease,
it is difficult to explicitly compute the mode-coupling term in LPT, even for the Zel'dovich approximation.
In this section, we decompose the LPT power spectrum into these two parts
and present an expansion method to approximately compute the mode-coupling term in LPT.
Our approximate solution has good convergence in the series expansion
and enables a computation of the LPT power spectrum accurately and quickly.

\subsection{At the Linear Order (Zel'dovich approximation)}
\label{Sec:Zel'dovich}
We derive the Zel'dovich power spectrum from Eqs.~(\ref{power_spectrum}) and (\ref{sigma_lin}) as follows
\begin{eqnarray}
		P(z,k) &=& 
		e^{-D^2\bar{\Sigma}_{\rm lin}(k)}\int d^3q e^{-i\kk\cdot\qq}
		\Big\{ D^2 \Sigma_{\rm lin}(\kk,\qq) + 
		\left( e^{D^2 \Sigma_{\rm lin}(\kk,\qq)} -1 - D^2 \Sigma_{\rm lin}(\kk,\qq) \right)\Big\} \nonumber \\
		&=& G^2(z,k) P_{\rm lin}(k) + P_{\rm MC}(z,k),
		\label{LPT}
\end{eqnarray}
where 
\begin{eqnarray}
		G^2(z,k) &=& e^{-D^2\bar{\Sigma}_{\rm lin}(k)} D^2, \nonumber \\
		P_{\rm MC}(z,k) &=& 2\pi e^{-D^2\bar{\Sigma}_{\rm lin}(k)}\int_0^{\infty} dq q^2 \int_{-1}^1 d\mu 
		\cos\left(kq \mu \right) \nonumber \\
		& \times &
		\left\{ e^{D^2 \Sigma_{0,\rm lin}(k,q)  -  D^2\Sigma_{2,\rm lin}(k,q){\cal L}_2(\mu)} -1 
		-\left(  D^2 \Sigma_{0,\rm lin}(k,q)  -  D^2\Sigma_{2,\rm lin}(k,q){\cal L}_2(\mu)  \right) \right\}
		\label{G-mode}
\end{eqnarray}
with $\mu = \hat{k}\cdot\hat{q}$, and we used $\int d^3q e^{-i\kk\cdot\qq} \Sigma_{\rm lin}(\kk,\qq) = P_{\rm lin}(k)$.

We naturally find the exponential damping behavior of the propagator in the Zel'dovich approximation,
even though the damping behavior in the high-$k$ limit have been obtained in several previous works
(for one of the latest works, see \citet{Bernardeau:2012ux}).
The exponential damping behavior of the propagator is the result of mass conservation,
because the non-linear scale-dependence of the Zel'dovich power spectrum comes only from the non-linearity of the law of mass conservation.

It is difficult to numerically compute the mode-coupling term in the Zel'dovich approximation,
because the integrand in the mode-coupling term has complicated oscillatory behavior caused by $\cos\left( kq \mu \right)$.
Therefore, here we present an approximation method to reproduce well the Zel'dovich power spectrum.
Note that the first term $G^2 P_{\rm lin}$
mainly contributes at large scales because of the contribution from the linear order in SPT $P_{\rm lin}$,
while the mode-coupling term is dominant at small scales.
Since Figure~\ref{fig:sigma} shows $\Sigma_0 \gg \Sigma_2$ at small scales,
we expand the exponential factor in the mode-coupling term (Eq.~(\ref{G-mode})) provided that $\Sigma_0 \gg \Sigma_2$,
obtaining the following approximate mode-coupling term:
\begin{eqnarray}
		P_{\rm MC}(z,k) = \sum_{n=0}^{\infty} P_{\rm MC}^{(n)}(z,k),
		\label{P_MC_EX}
\end{eqnarray}
where 
\begin{eqnarray}
		P_{\rm MC}^{(0)} &\equiv& 4\pi e^{-D^2\bar{\Sigma}_{\rm lin}(k)}
		\int_0^{\infty} dq q^2 j_{0}(kq)\left( e^{D^2 \Sigma_{0,\rm lin}(k,q) } -1 -  D^2 \Sigma_{0,\rm lin}(k,q) \right), \nonumber \\
        P_{\rm MC}^{(1)} &\equiv&   4\pi   e^{-D^2\bar{\Sigma}_{\rm lin}(k)}
		\int_0^{\infty} dq q^2 j_2(kq) D^2\Sigma_{2,\rm lin}(k,q) \left( e^{D^2 \Sigma_{0,\rm lin}(k,q)}-1 \right), \nonumber \\
		P_{\rm MC}^{(n)} &\equiv&  4\pi   e^{-D^2\bar{\Sigma}_{\rm lin}(k)} 
		\int_0^{\infty} dq q^2 J^{(n)}(z,k,q) e^{D^2 \Sigma_{0,\rm lin}(k,q)} \quad \mbox{for $n\geq 2$},
\end{eqnarray}
with
\begin{eqnarray}
		J^{(n)}(z,k,q) \equiv 
		\frac{\left(  D^2\Sigma_{2,\rm lin}(k,q) \right)^n}{n!}	
		\sum_{\ell=0}^{2n}(-i)^{\ell} 
		j_{\ell}(kq) \left( \frac{2\ell+1}{2}  \right) \int_{-1}^{1} d \mu {\cal L}_{\ell}(\mu) \left( - {\cal L}_2(\mu) \right)^n.
\end{eqnarray}
Analytical calculations of the $\mu$ integral in $J^{(n)}$ (Eq.~(\ref{J_Z}))
enable computation of the mode-coupling term quickly and safely.

We have another theoretical reason for our approximation method (Eq.~(\ref{P_MC_EX})).
As mentioned in Sec~\ref{motivation},
we should keep a combination of $\Sigma_0(z,k,q) - \bar{\Sigma}(z,k)$
to satisfy the fact that the power spectrum has no contribution at $\qq=0$ and to respect the law of mass conservation.
This is also related to the IR divergence problem and the cancellation of the high-$k$ solutions in SPT
(see Secs.~\ref{IR} and \ref{Sec:Review_SPT}).
Here, note that the propagator and the mode-coupling term each have the integral $\int dp P_{\rm lin}(p)$ in the limit of $p\to0$.
However, the total Zel'dovich power spectrum does not have $\int dp P_{\rm lin}(p)$,
but $\int dp p^2 P_{\rm lin}(p)$ in the limit as shown in Sec.~\ref{IR}.
The same thing also occurs in the 1-loop SPT (Sec.~\ref{SPT_1loop}).
The propagator and mode-coupling term are described in the 1-loop SPT as
$G^2(z,k)P_{\rm lin}(k) = D^2P_{\rm lin}(k) + D^4P_{13}(k)$ and $P_{\rm MC}(z,k) = D^4 P_{22}(k)$.
Each of their terms is proportional to $\int dp P_{\rm lin}(p)$ in the limit of $p\to0$ (in the high-$k$ limit),
but completely cancels out each other.
Thus, to satisfy this cancellation at all orders in SPT, 
we should not expand the exponential factor for the monopole term $e^{\Sigma_0}$ 
when we do not expand the exponential damping factor $e^{-\bar{\Sigma}}$ in the mode-coupling term.
This idea is the first main result of this paper.

\subsection{At the One-loop Order}

\begin{figure}[tpb]
		\begin{center}
				\scalebox{1.0}{\plottwo{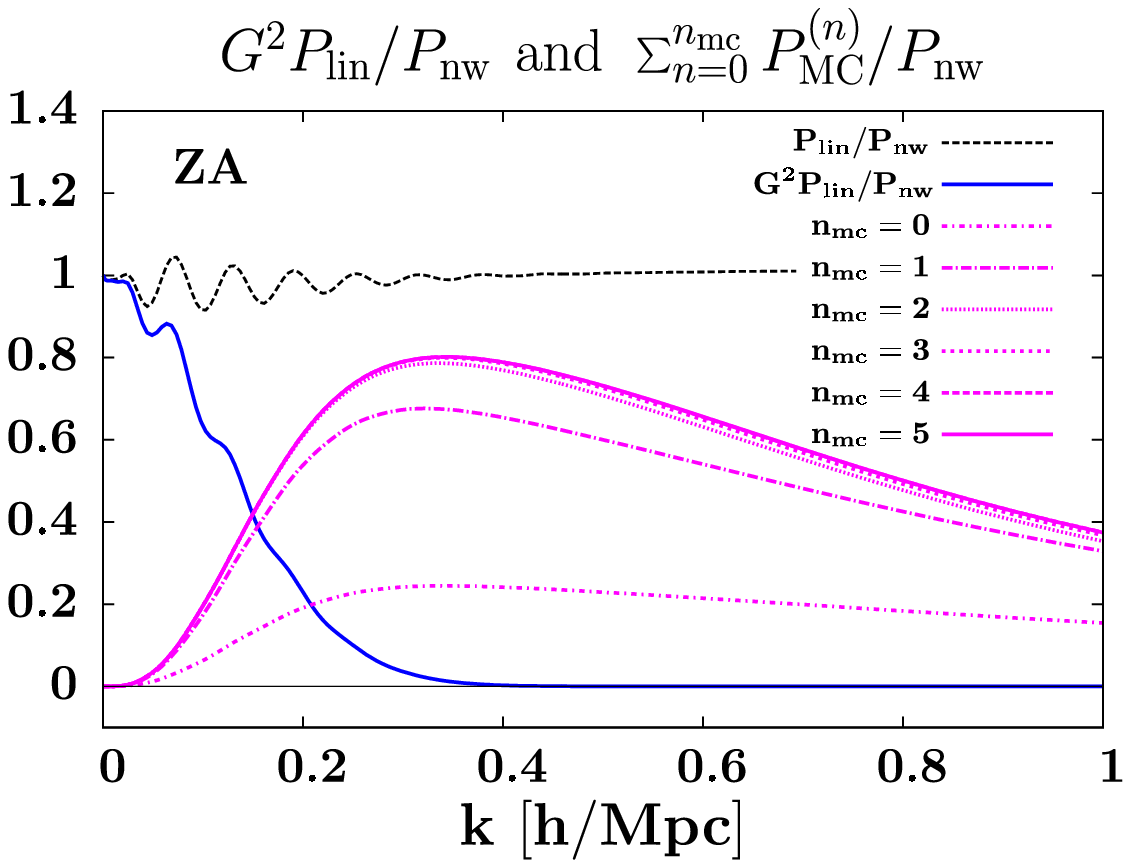}{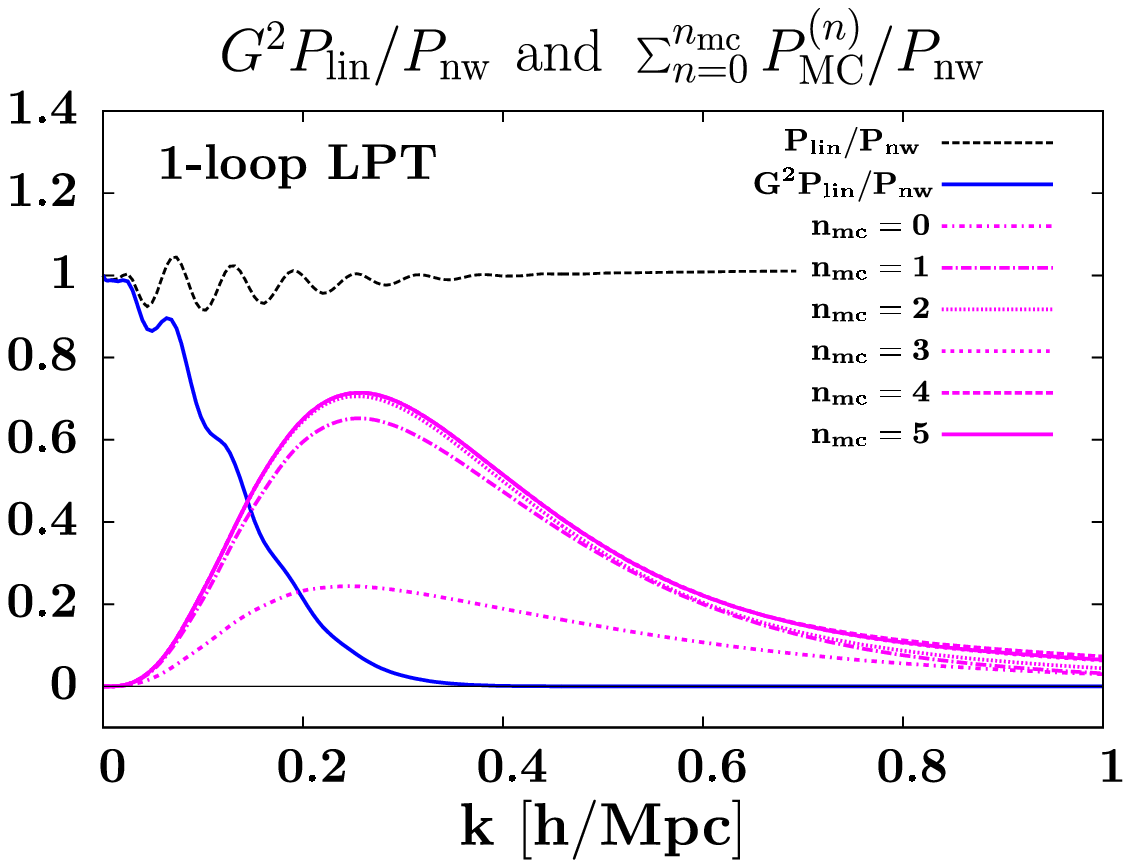}}
		\end{center}
		\begin{center}
				\scalebox{1.0}{\plottwo{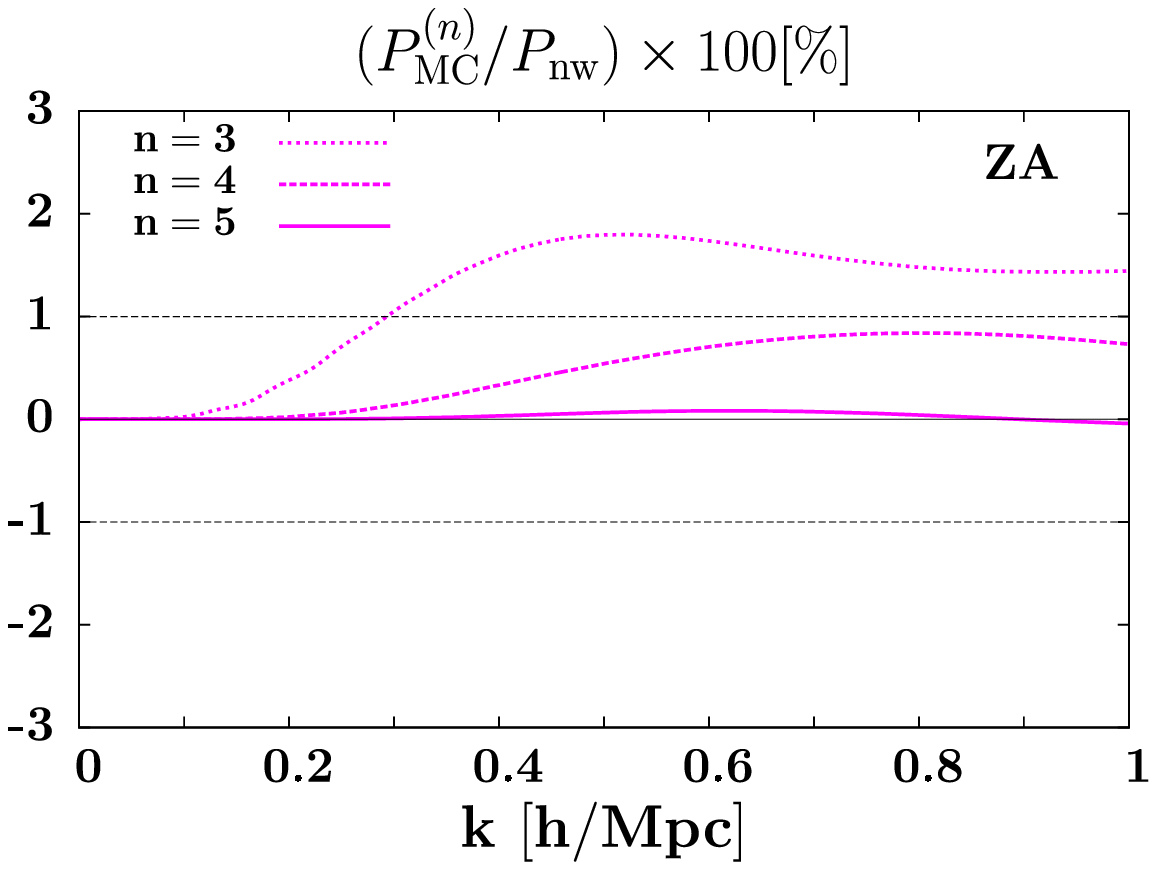}{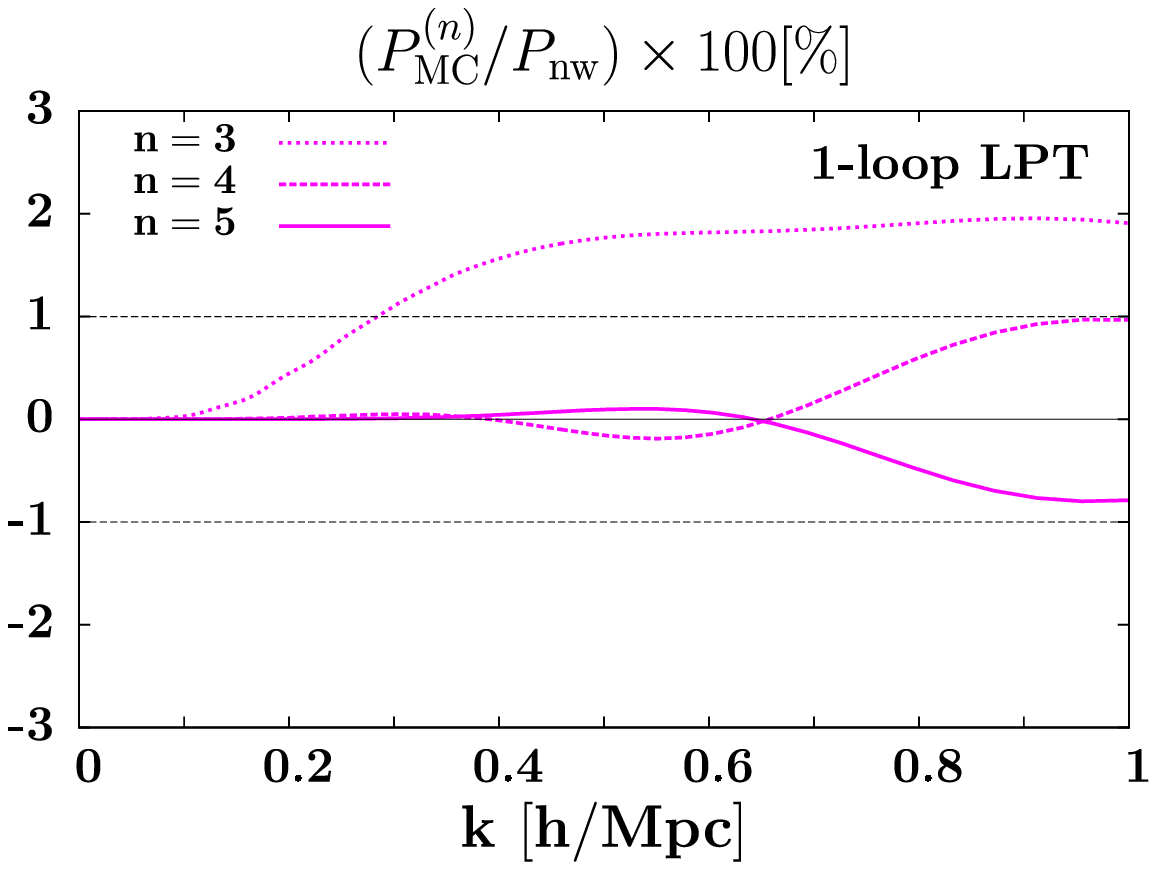}}
		\end{center}
		\caption{
		Two contributions to the power spectrum 
		$G^2(z,k)P_{\rm lin}(k)/P_{\rm nw}(k)$ and $\sum_{n=0}^{n_{\rm mc}}P_{\rm MC}^{(n)}(z,k)/P_{\rm nw}(k)$
		for $\{n_{\rm mc}=0,1,2,3,4,5\}$ are plotted in the Zel'dovich approximation and the 1-loop LPT at $z=0$.
		This figure shows the performance of our method of expanding the mode-coupling term presented 
		in Eqs.~(\ref{P_MC_EX}) and (\ref{P_MC_EX_LPT}).
		In particular, the bottom panels imply that 
		the fourth and fifth orders in the expansion ($P_{\rm MC}^{(4)}$ and $P_{\rm MC}^{(5)}$)
		contribute less than $1\%$ compared to the linear power spectrum.
		In other words, the approximate mode-coupling term has good convergence in the series of the expansion,
		and we only have to compute up to the third order of the expansion $P_{\rm MC} = \sum_{n=0}^3P_{\rm MC}^{(n)}$
		with an accuracy of $<1\%$ until $k=1\ [h{\rm Mpc}^{-1}]$.
		At large scales $k\lesssim 0.2\ [h{\rm Mpc}^{-1}]$,
		$P_{\rm MC} = \sum_{n=0}^2P_{\rm MC}^{(n)}$ presents a good approximate solution with an accuracy of $<1\%$.
		}
		\label{fig:Zel'dovich_and_LPT}
\end{figure}

\begin{figure}[tpb]
		\begin{center}
				\scalebox{1.0}{\plottwo{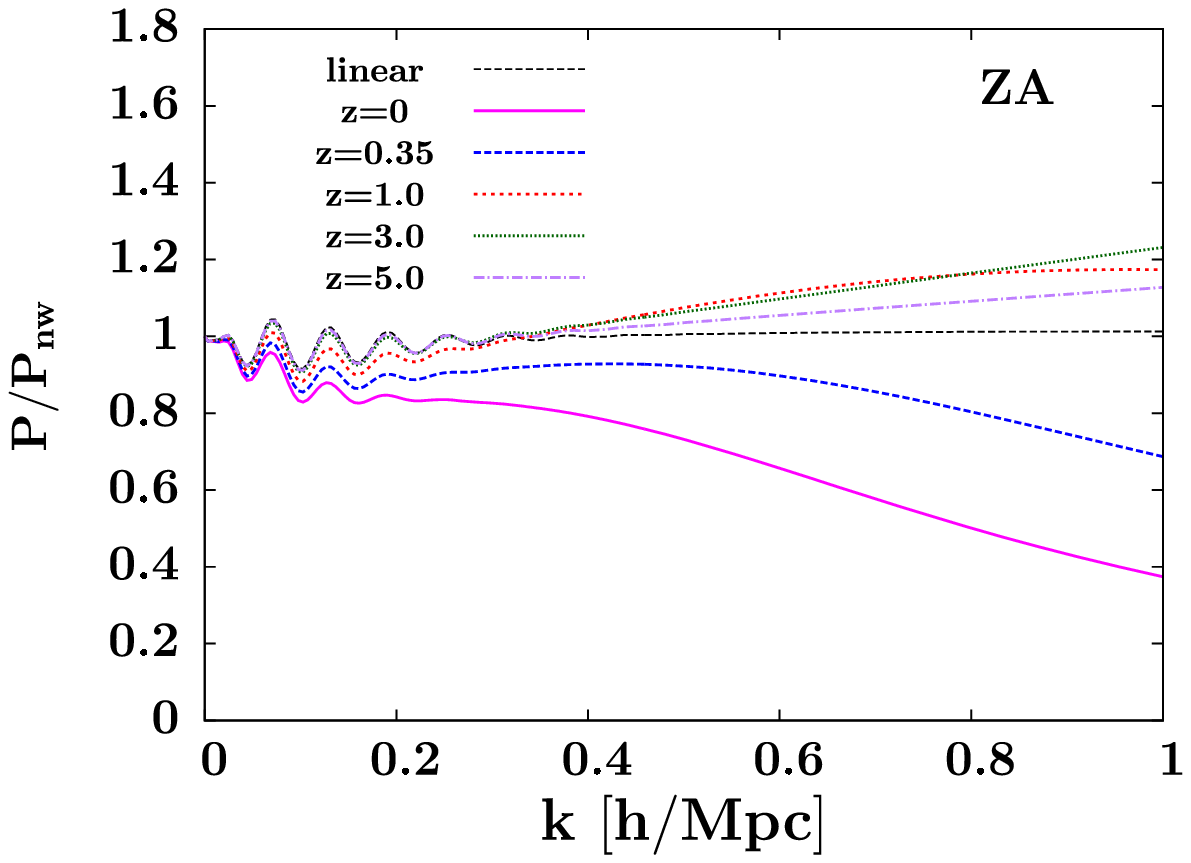}{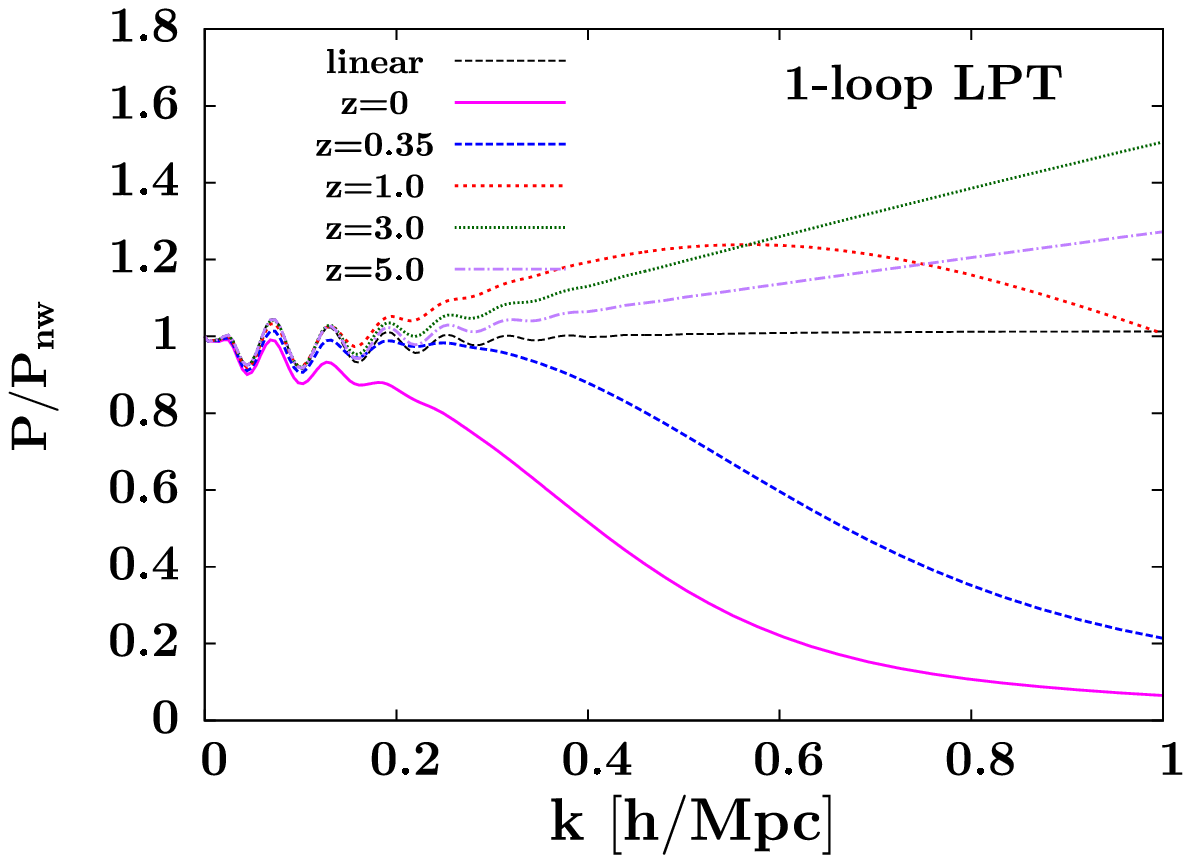}}
		\end{center}
		\caption{
		Zel'dovich and 1-loop LPT power spectra are shown at various redshifts ($z=0$, $0.35$, $1.0$, $3.0$, and $5.0$).
		The LPT solution has full non-linear effects from the law of conservation of mass,
		but its non-linear equation of the motion of dark matter
		(the equation of motion of the displacement vector) is solved in the perturbation series.
		Therefore, at high-$z$, the LPT solution can well describe the non-linear evolution of dark matter,
		and the ratio $P_{\rm LPT}/P_{\rm nw}$ becomes larger than unity.
		On the other hand, at low-$z$,
		because of lack of non-linear dynamics of dark matter, the ratio $P_{\rm LPT}/P_{\rm nw}$ becomes less than unity.
		\label{fig:ZA_and_LPT_red}
		}
\end{figure}

From Eqs~(\ref{LPT_general}), (\ref{sigma_lin2213}), and (\ref{Delta}), the propagator term is given by
\begin{eqnarray}
		G^{2}(z,k) P_{\rm lin}(k)
		&=&  e^{-\bar{\Sigma}(z,k)}
		\int d^3q e^{-i\kk\cdot\qq} \Big\{ D^2 \Sigma_{\rm lin}(\kk,\qq) + D^4 \Sigma_{13}(\kk,\qq) \Big\} \nonumber \\
		&=& 
		e^{-D^2\bar{\Sigma}_{\rm lin}(k) - D^4\bar{\Sigma}_{\rm 1\mathchar`-loop}(k)}
		\left( 1 + D^2 \frac{\Delta P_{13}(k)}{P_{\rm lin}(k)}   \right)D^2 P_{\rm lin}(k).
		\label{1-loop_propagator}
\end{eqnarray}
Compared to the Zel'dovich solution of the propagator,
the additional factors $e^{-D^4\bar{\Sigma}_{\rm 1\mathchar`-loop}(k)}$ and $\left( 1 + D^2 \Delta P_{13}/P_{\rm lin} \right)$
appear in the 1-loop LPT.
They come from the non-linear equation of the motion of the displacement vector,
the kernel functions $\LL_2$ and $\LL_3$ in Eq.~(\ref{eq:LL}).

We obtain the mode-coupling term from
\begin{eqnarray}
		P_{\rm MC}(z,k) &=&  e^{-\bar{\Sigma}(z,k)}
		\int d^3q e^{-i\kk\cdot\qq} \Big\{ e^{\Sigma(z,\kk,\qq)}
		- 1 - D^2 \Sigma_{\rm lin}(\kk,\qq) - D^4 \Sigma_{13}(\kk,\qq) \Big\} \nonumber \\
		&=& e^{-\bar{\Sigma}(z,k)} D^4\left( P_{22}(k) - P_{22}|_{\rm ZA}(k) \right) + \tilde{P}_{\rm MC}(z,k),
		\label{1-loop_modecoupling}
\end{eqnarray}
where
\begin{eqnarray}
		\tilde{P}_{\rm MC}(z,k) &=&   e^{-\bar{\Sigma}(z,k)}
		\int d^3q e^{-i\kk\cdot\qq} \Big\{ e^{\Sigma(z,\kk,\qq)} - 1 - \Sigma(z,\kk,\qq)\Big\}.
\end{eqnarray}
In computing the mode-coupling term, we can use the same analysis as the Zel'dovich approximation.
Provided $\Sigma_0 \gg \Sigma_1,\ \Sigma_2,\ {\rm and}\ \Sigma_3$,
the mode-coupling term $P_{\rm MC}$ is approximated as follows
\begin{eqnarray}
		P_{\rm MC}(z,k) =  \sum_{n=0}^{\infty} P_{\rm MC}^{(n)}(z,k)
		= e^{-\bar{\Sigma}(z,k)}\left( D^4P_{22}(k) - D^4P_{22}|_{\rm Z}(k) \right)
		+ \sum_{n=0}^{\infty} \tilde{P}_{\rm MC}^{(n)}(z,k),
		\label{P_MC_EX_LPT}
\end{eqnarray}
where
\begin{eqnarray}
		\tilde{P}_{\rm MC}^{(0)}(z,k) &=& 
		4\pi e^{-\bar{\Sigma}(z,k)} \int_0^{\infty} dq  q^2 j_0(kq) \big(  e^{\Sigma_0(z,k,q)} - 1 - \Sigma_0(z,k,q)  \big), \nonumber \\
        \tilde{P}_{\rm MC}^{(1)}(z,k) &=& 
		4\pi e^{-\bar{\Sigma}(z,k)} \int_0^{\infty} dq  q^2	
		\sum_{\ell=1}^3 j_{\ell}(kq) \Sigma_{\ell}(z,k,q)\left(  e^{\Sigma_0(z,k,q)} - 1  \right), \nonumber \\
		 \tilde{P}_{\rm MC}^{(n)}(z,k) &=& 
		 4\pi e^{-\bar{\Sigma}(z,k)} \int_0^{\infty} dq q^2 J^{(n)}(z,k,q) e^{\Sigma_0(z,k,q)} \quad \mbox{for $n\geq 2$},
\end{eqnarray}
with
\begin{eqnarray}
		J^{(n)}(z,k,q) \equiv
		\frac{1}{n!}\sum_{\ell=0}^{3n} (-i)^{\ell} j_{\ell}(kq)
		\frac{2\ell+1}{2} \int_{-1}^{1} d\mu 
		{\cal L}_{\ell}(\mu) \left(  \sum_{\ell'=1}^3 i^{\ell'} \Sigma_{\ell'}(z,k,q)  {\cal L}_{\ell'}(\mu) \right)^n.\nonumber \\
\end{eqnarray}
Note that $\Sigma_{\ell}(z,k,q) = D^2 \Sigma_{\ell,\rm lin}(k,q) + D^4 \Sigma_{\ell,\rm 22}(k,q) + D^4 \Sigma_{\ell,\rm 13}(k,q)$.
The analytical calculation of the $\mu$ integral in $J^{(2)}$ is given in Eq.~(\ref{J_2}).
	
Figure~\ref{fig:Zel'dovich_and_LPT} shows the performance of the our method for expanding the mode-coupling term.
The fourth and fifth orders in the expansion of the mode-coupling term, $P_{\rm MC}^{(4)}$ and $P_{\rm MC}^{(5)}$,
contribute less than $1\%$ over the range of $k \leq 1\ [h{\rm Mpc}^{-1}]$ at $z=0$.
Therefore, the approximate mode-coupling term has good convergence in the series expansion,
and we only have to compute up to the third order of the expansion $P_{\rm MC} = \sum_{n=0}^3 P_{\rm MC}^{(n)}$
to compute the mode-coupling term with an accuracy of  $<1\%$ until $k=1\ [h{\rm Mpc}^{-1}]$.
The approximate solution $P = G^2 P_{\rm lin} + \sum_{n=0}^{3} P_{\rm MC}^{(n)}$ works well at any redshift
because $P_{\rm MC}^{(n)}$ $(n\geq 4)$ are non-linear effects and become progressively smaller at high $z$.
When we focus only on large scales $k\lesssim 0.2\ [h{\rm Mpc}^{-1}]$,
$P_{\rm MC} = \sum_{n=0}^2P_{\rm MC}^{(n)}$ is enough to reproduce the LPT power spectrum with an accuracy of $<1\%$.

Figure~\ref{fig:ZA_and_LPT_red} shows 
the Zel'dovich and 1-loop LPT power spectra at various redshifts ($z=0$, 0.35, 0.5, 1.0, 3.0, and 5.0).
The LPT solution has full non-linear effects from the law of conservation of mass,
but its non-linear equation of motion of dark matter
(the equation of motion of the displacement vector) is solved in the perturbation series.
Therefore, at high $z$, the LPT solution can describe well the non-linear evolution of dark matter,
and the ratio $P_{\rm LPT}/P_{\rm nw}$ becomes larger than unity.
On the other hand, at low $z$,
the third order of the displacement vector in the perturbation series
is not enough to accurately describe the non-linear growth of dark matter, and
the ratio $P_{\rm LPT}/P_{\rm nw}$ becomes less than unity.

\section{COMPARISON WITH THE $\Gamma$-EXPANSION METHOD}
\label{LPT_RegPT}

To clarify the relation between LPT and existing works,
in this section we will show that the expansion method used in LRT~\citep{Matsubara:2007wj} corresponds to the $\Gamma$-expansion
\citep{Bernardeau:2008fa,Bernardeau:2011dp,Taruya:2012ut,Sugiyama:2012pc}, leading to the solution of RegPT.
The LRT solution is derived from expanding $e^{\Sigma(z,\kk,\qq)}$ in Eq.~(\ref{power_spectrum}) as
\begin{eqnarray}
		P(z,k) = e^{-\bar{\Sigma}(z,k)} \sum_{n=1}^{\infty} \frac{1}{n!} \int d^3q e^{-i\kk\cdot\qq} \Big\{ \Sigma(z,\kk,\qq)\Big\}^n,
\end{eqnarray}
and truncating at a finite order of $n$.
Unlike our expansion method (Sec.~\ref{Sec:pk_LPT}),
LRT (the $\Gamma$-expansion method) expands the exponential factor including the monopole term $e^{\Sigma_{0}}$.

\subsection{Review of the $\Gamma$-Expansion}

The $\Gamma$-expansion is only used to obtain information on the power spectrum at large scale regions.
The higher order terms of the $\Gamma$-expansion have information on smaller scales.
In the $\Gamma$-expansion method, the full non-linear power spectrum is described as
\begin{eqnarray}
		P(z,k) = G^2(z,k) P_{\rm lin}(k) + \sum_{r=2}^{\infty} P_{\rm \Gamma}^{(r)}(z,k).
		\label{Gamma}
\end{eqnarray}
Therefore, the mode-coupling term is $P_{\rm MC} = \sum_{r=2}^{\infty} P_{\rm \Gamma}^{(r)}$,
where $P_{\rm \Gamma}^{(r)}$ is the $r$th-order contribution to the power spectrum in the $\Gamma$-expansion, defined as
\begin{equation}
		P_{\rm \Gamma}^{(r)}(z,k) \equiv  r! \int \frac{d^3k_1}{(2\pi)^3} \cdots  \int \frac{d^3k_r}{(2\pi)^3}
		(2\pi)^3 \delta_D(\kk-\kk_{[1,r]}) \left[\Gamma^{(r)}(z,\kk_1,\dots,\kk_r)  \right]^2P_{\rm lin}(k_1) \cdots P_{\rm lin}(k_r)
	\label{power-r}
\end{equation}
with
\begin{eqnarray}
		\Gamma^{(r)}(z,\kk_1,\dots,\kk_r)
		&\equiv& \sum_{n=0}^{\infty} D^{r+2n} 
		\frac{(r+2n)!}{2^n n!r!} 
          \int \frac{d^3p_1}{(2\pi)^3} \cdots  \int \frac{d^3p_n}{(2\pi)^3} \nonumber \\
		&& F_{r+2n}(\kk_1,\dots,\kk_r,\pp_1,-\pp_1,\dots,\pp_n,-\pp_n)P_{\rm lin}(p_1)\cdots P_{\rm lin}(p_n). \nonumber \\
\end{eqnarray}
The propagator is defined as
\begin{eqnarray}
		G(z,k) = P_{\rm \Gamma}^{(1)}(z,k) \equiv \frac{\langle \delta(z, \kk) \delta_{\rm lin}(z=0,\kk') \rangle }
								          {\langle \delta_{\rm lin}(z=0,\kk) \delta_{\rm lin}(z=0,\kk') \rangle}
										  = \left( 1 + \sum_{n=1}^{\infty}D^{2n}\frac{P_{1(2n+1)}(k)}{2P_{\rm lin}(k)} \right) D.
		\label{pro}
\end{eqnarray}

\subsection{Zel'dovich Approximation}
In the Zel'dovich approximation, 
the LPT expansion method provides $P_{\rm \Gamma}^{(r)}$ as
\begin{eqnarray}
		P_{\rm \Gamma}^{(r)}(z,k)
		&=&  e^{-D^2\Sigma_{\rm lin}(k)} D^{2r} \frac{1}{r!} \int d^3q e^{-i\kk\cdot\qq} 
		\Big\{ \Sigma_{\rm lin}(\kk,\qq)\Big\}^{r} \nonumber \\
		&=&  e^{-D^2\Sigma_{\rm lin}(k)} D^{2r} \frac{1}{r!} \int d^3q e^{-i\kk\cdot\qq} 
		\Bigg\{ \int \frac{d^3p}{(2\pi)^3} e^{i\pp\cdot\qq}\left( \frac{\kk\cdot\pp}{p^2} \right)^2 P_{\rm lin}(p)\Bigg\}^{r} \nonumber \\
		&=& e^{-D^2\bar{\Sigma}_{\rm lin}(k)} D^{2r} r! \int \frac{d^3k_1}{(2\pi)^3} \cdots  \int \frac{d^3k_r}{(2\pi)^3}
		(2\pi)^3 \delta_D(\kk-\kk_{[1,r]}) \big[F_r|_{\rm ZA}(\kk_1,\cdots,\kk_r)  \big]^2P_{\rm lin}(k_1) \cdots P_{\rm lin}(k_r). \nonumber \\
\end{eqnarray}
It is worth noting that while we need $3(r-1)$-dimensional integral to compute the expression using $F_r|_{\rm ZA}$ in the final line,
in the first line we need only need a two-dimensional integral for any $r$ as follows
\begin{eqnarray}
		P_{\rm \Gamma}^{(r)}(z,k)
		&=&  4\pi e^{- D^2 \bar{\Sigma}_{\rm lin}(k)} \frac{1}{r!}\int_0^{\infty} dq q^2
		\sum_{\ell=0}^{\infty} (-i)^{\ell}j_{\ell}(kq) \frac{2\ell+1}{2}
		\int_{-1}^1 d\mu {\cal L}_{\ell}(\mu) \left(  D^2 \Sigma_{0, \rm lin}(k,q) 
		- {\cal L}_2(\mu) D^2  \Sigma_{2, \rm lin}(k,q) \right)^r. \nonumber \\
\end{eqnarray}

\subsection{LPT at the One-loop Order}

\begin{figure}[t]
		\begin{center}
				\plottwo{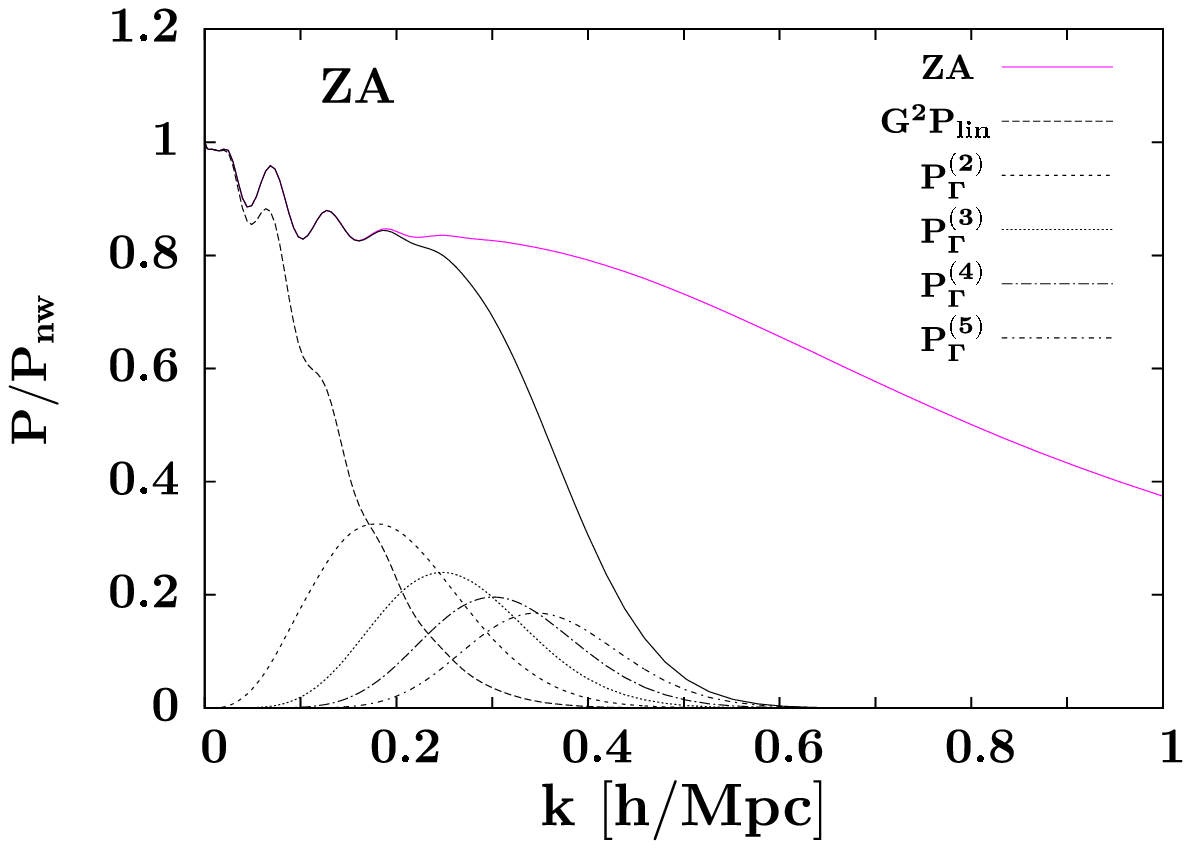}{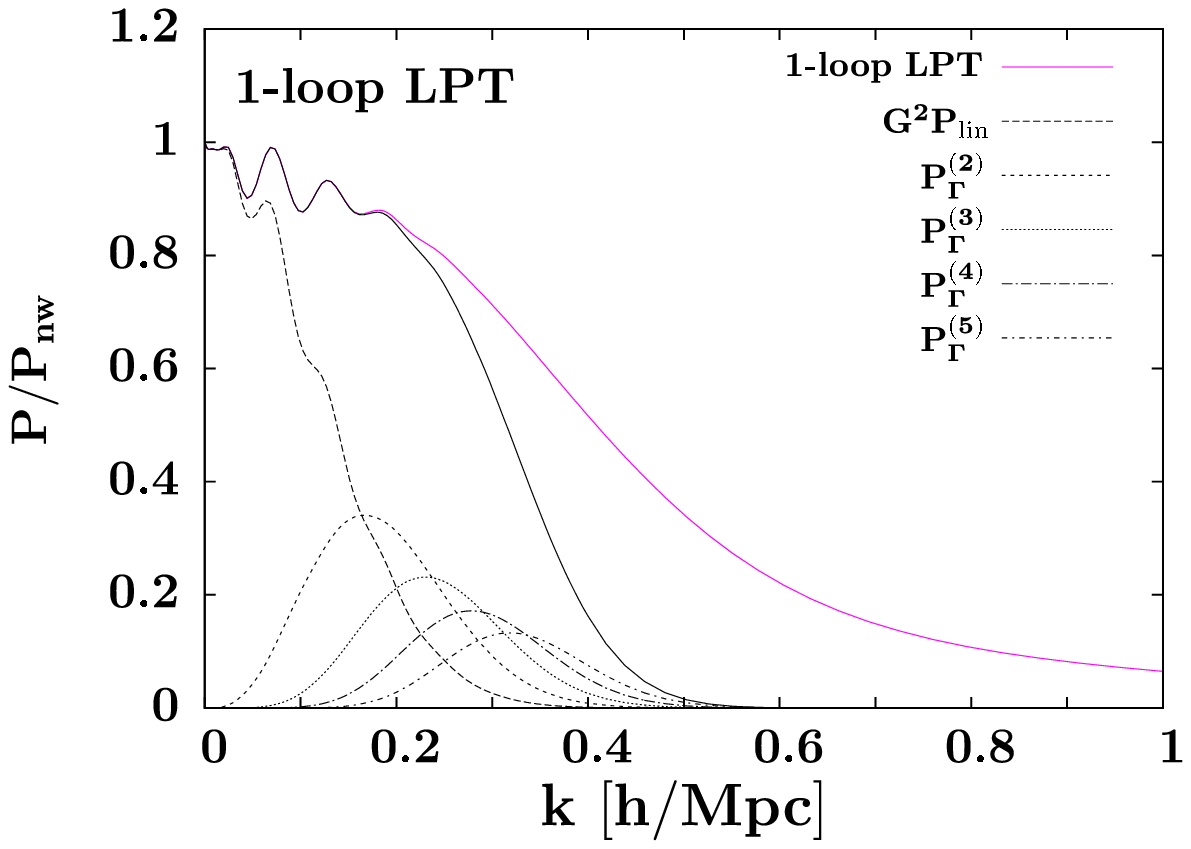}
		\end{center}
		\caption{
		LPT power spectrum in the $\Gamma$-expansion is shown.
		$G^2P_{\rm lin}/P_{\rm nw}$, $P_{\Gamma}^{(r)}/P_{\rm lin}^{\rm nw}$ for $r=\{2,3,4,5\}$ 
		and $G^2P_{\rm lin}/P_{\rm nw} + \sum_{r=2}^{5}P_{\Gamma}^{(r)}/P_{\rm nw}$ (black solid)
		are plotted at $z=0$ in the Zel'dovich approximation and the 1-loop LPT.
		Thus, the LPT solution enables a computation of any order in the $\Gamma$-expansion.
		}
		\label{fig:Gamma_expansion}
\end{figure}

Similarly to the case of the Zel'dovich approximation, we obtain $P_{\rm \Gamma}^{(r)}$ in the 1-loop LPT:
\begin{eqnarray}
		P_{\rm MC}(z,k) =\sum_{r=2}^{\infty}P_{\rm \Gamma}^{(r)}(z,k) = 
		\sum_{n=2}^{\infty}P_{\rm \Gamma\mathchar`-1}^{(n)}(z,k)
		+\sum_{n=1}^{\infty}P_{\rm \Gamma\mathchar`-2}^{(2n)}(z,k)
		+\sum_{n=2}^{\infty}\sum_{m=1}^{n-1}	P_{\rm \Gamma\mathchar`-3}^{(n+m)}(z,k),
\end{eqnarray}
where 
\begin{eqnarray}
		P_{\rm \Gamma\mathchar`-1}^{(n)}(z,k) &=& 4\pi e^{- \bar{\Sigma}(z,k)} \frac{1}{n!}\int_0^{\infty} dq q^2
		\sum_{\ell=0}^{\infty} (-i)^{\ell}j_{\ell}(kq) \frac{2\ell+1}{2}
		\int_{-1}^1 d\mu {\cal L}_{\ell}(\mu)  \left[ D^2 \Sigma_{\rm lin}(\kk,\qq) +  D^4 \Sigma_{13}(\kk,\qq)  \right]^n. \nonumber \\
		P_{\rm \Gamma\mathchar`-2}^{(2n)}(z,k) &=& 4\pi e^{- \bar{\Sigma}(z,k)} \frac{1}{n!}\int_0^{\infty} dq q^2
		\sum_{\ell=0}^{\infty} (-i)^{\ell}j_{\ell}(kq) \frac{2\ell+1}{2}
		\int_{-1}^1 d\mu {\cal L}_{\ell}(\mu)  [D^4 \Sigma_{22}(\kk,\qq)]^n. \nonumber \\
		P_{\rm \Gamma\mathchar`-3}^{(n+m)}(z,k)&=& 4\pi e^{- \bar{\Sigma}(z,k)} \frac{1}{n!}\int_0^{\infty} dq q^2
		\sum_{\ell=0}^{\infty} (-i)^{\ell}j_{\ell}(kq) \frac{2\ell+1}{2}
		\int_{-1}^1 d\mu {\cal L}_{\ell}(\mu) \left( \begin{array}{c} n \\ m \end{array} \right) \nonumber \\
				&&\hspace{2.5cm} \times	
		\left[ D^2 \Sigma_{\rm lin}(\kk,\qq) +  D^4 \Sigma_{13}(\kk,\qq) \right]^{n-m} \left[ D^4 \Sigma_{22}(\kk,\qq) \right]^m.
\end{eqnarray}
Specifically, we have the following expressions up to the third order of the $\Gamma$-expansion:
\begin{eqnarray}
		P_{\rm \Gamma}^{(2)}(z,k)
		&=& e^{- D^2\bar{\Sigma}_{\rm lin}(k) - D^4\bar{\Sigma}_{\rm 1\mathchar`-loop}(k)}
		 \left( D^4P_{22}(k) 
		 + D^6\left( P_{24}|_{\rm LPT,1\mathchar`-loop}(k) + \bar{\Sigma}_{\rm lin}(k)P_{22}(k) \right) + \cdots\right),\nonumber \\
		P_{\rm \Gamma}^{(3)}(z,k)
		 &=&  e^{- D^2\bar{\Sigma}_{\rm lin}(k) - D^4\bar{\Sigma}_{\rm 1\mathchar`-loop}(k)}
		 \left( D^6 P_{33b}|_{\rm LPT, 1\mathchar`-loop}(k) + \cdots \right).
		 \label{Gamma_fifth}
\end{eqnarray}
Note that $P_{\rm \Gamma}^{(2)}$ and $P_{\rm \Gamma}^{(3)}$ correspond to those in the original 2-loop RegPT,
even though the approximate solutions $P_{24}|_{\rm LPT,1\mathchar`-loop}$ and $P_{33b}|_{\rm LPT,1\mathchar`-loop}$ are used
in the 1-loop LPT.
Figure~\ref{fig:Gamma_expansion} gives a demonstration of how the $\Gamma$-expansion reproduces the LPT power spectrum,
where we computed up to the fifth order of the $\Gamma$-expansion.

Truncating the $\Gamma$-expansion at the second order and ignoring some non-linear effects in the 1-loop LPT (Eq.~(\ref{Gamma_fifth})),
we have the 1-loop LRT solution:
\begin{eqnarray}
		P|_{\rm LRT, 1\mathchar`-loop}(z,k) &=&  P_{\rm \Gamma}^{(1)}(z,k) + P_{\rm \Gamma}^{(2)}(z,k) \nonumber \\
	&=& e^{-D^2 \bar{\Sigma}_{\rm lin}(k)} \left( 1 +D^2\frac{\Delta P_{13}(k)}{P_{\rm lin}(k)} \right)D^2 P_{\rm lin}(k)
	+ e^{-D^2 \bar{\Sigma}_{\rm lin}(k)}D^4 P_{22}(k) \nonumber \\
	&=&  e^{-D^2 \bar{\Sigma}_{\rm lin}(k)} \left( D^2 P_{\rm lin}(k) 
	+ D^4 \left( P_{\rm 1\mathchar`-loop} + \bar{\Sigma}_{\rm lin}(k)P_{\rm lin}(k)  \right)\right),
\end{eqnarray}
where we ignored $\bar{\Sigma}_{13}$, $\bar{\Sigma}_{22}$, $P_{24}|_{\rm LPT, 1\mathchar`-loop}$, and so on.
On the other hand, the original 1-loop RegPT solution is given by
\begin{eqnarray}
	P|_{\rm RegPT, 1\mathchar`-loop}(z,k) &=&  P_{\rm \Gamma}^{(1)}(z,k) + P_{\rm \Gamma}^{(2)}(z,k) \nonumber \\
	&=& e^{-D^2 \bar{\Sigma}_{\rm lin}(k)} \left( 1 +D^2\frac{\Delta P_{13}(k)}{2 P_{\rm lin}(k)} \right)^2D^2 P_{\rm lin}(k)
	+ e^{-D^2 \bar{\Sigma}_{\rm lin}(k)}D^4 P_{22}(k) \nonumber \\
	&=&	P|_{\rm LRT, 1\mathchar`-loop}(z,k)  +  e^{-D^2 \bar{\Sigma}_{\rm lin}(k)}D^6 \Delta P_{33a}(k).
	\label{RegPT1loop}
\end{eqnarray}
Since the 1-loop LPT solution does not have the 2-loop correction term $\Delta P_{33a}$ in SPT which comes from the 2-loop LPT solution 
(see Sec.~\ref{SPT2loop}), the 1-loop LPT solution does not completely reproduce the 1-loop RegPT solution.
However, the term $e^{-D^2 \bar{\Sigma}_{\rm lin}(k)}D^6 \Delta P_{33a}$ is small enough to be ignored
and we can actually regard as $P|_{\rm RegPT, 1\mathchar`-loop} \simeq P|_{\rm LRT, 1\mathchar`-loop}$.
Clearly, the 2-loop LPT includes the 1-loop RegPT.

Finally, we present the general expression of the propagator in LPT:
\begin{eqnarray}
		\langle\delta(z,\kk) \delta_{\rm lin}(z=0,\kk')\rangle
		&=&  
		\int d^3q_1 \int d^3q_2  e^{-i\kk\cdot\qq_1} e^{-i\kk'\cdot\qq_2}
		\left\langle e^{-i\kk\cdot\YY(z,\qq_1)}\left( -i\kk'\cdot\YY_{\rm lin}(z=0,\qq_2) \right) \right\rangle \nonumber \\
		&=& (2\pi)^3 \delta_{\rm D}(\kk+\kk') \left\langle e^{-i\kk\cdot\YY(z,0)} \right\rangle 
		\int d^3qe^{-i\kk\cdot\qq}\left\langle e^{-i\kk\cdot\YY(z,\qq)} \left( i\kk\cdot\YY_{\rm lin}(z=0,0) \right)\right\rangle_{\rm c},
		\nonumber \\
\end{eqnarray}
where $\qq = \qq_1-\qq_2$.
This implies 
\begin{eqnarray}
		G(z,k) = \exp\left(-\frac{\bar{\Sigma}(z,k)}{2} \right)
		\left( 1 + \sum_{n=1}^{\infty} D^{2n}\frac{\Delta P_{1(2n+1)}(k)}{2P_{\rm lin}(k)} \right)D,
		\label{G_gen}
\end{eqnarray}
where we used 
$\left\langle \left( -i\kk\cdot\YY(z,0)\right)^n \right \rangle_{\rm c}
=\left\langle \left( i\kk\cdot\YY(z,0) \right)^n \right \rangle_{\rm c}$,
and 
\begin{eqnarray}
		\int d^3qe^{-i\kk\cdot\qq}\left\langle e^{-i\kk\cdot\YY(z,\qq)} \left( i\kk\cdot\YY_{\rm lin}(z=0,0) \right)\right\rangle_{\rm c}
		=\left( 1 + \sum_{n=1}^{\infty} D^{2n}\frac{\Delta P_{1(2n+1)}(k)}{2P_{\rm lin}(k)} \right)DP_{\rm lin}(k).
		\label{Ps}
\end{eqnarray}
In other words, the above relation is the definition of $\Delta P_{1(2n+1)}$.
For example, 
Eq.~(\ref{Ps}) leads to
\begin{eqnarray}
		\Delta P_{13}(k) 
		&=&  P_{\rm lin}(k)\int \frac{d^3p}{(2\pi)^3} 
		\left[ \kk\cdot\LL_3(\kk,\pp,-\pp) - \kk\cdot\LL_2(\kk,\pp) \kk\cdot\LL_1(\pp)  \right] P_{\rm lin}(p) \nonumber \\
		&=& P_{13}(k) + \bar{\Sigma}_{\rm lin}(k) P_{\rm lin}(k).
\end{eqnarray}
This expression is the same as Eq.~(\ref{Delta}).
Furthermore, at the 1-loop order, the square of Eq.~(\ref{G_gen}) leads to Eq.~(\ref{1-loop_propagator}) with the term $\Delta P_{33a}$ ignored.

\section{BEYOND THE 2-LOOP SOLUTION IN SPT}
\label{3loop_and_more}

Our main goal is to obtain non-linear information on the matter perturbation beyond the 2-loop SPT.
While the exact 2-loop solution in SPT has been well studied (Sec.~\ref{SPT2loop}),
it is too computationally expensive to compute solutions of higher order than the 2-loop in SPT.
Therefore, we want approximate information on the 3- and higher loop orders in SPT.
For that purpose, 
we have so far solved the 1-loop LPT solution, which is described in the standard perturbation series as follows
\begin{eqnarray}
		P(z,k)|_{\rm LPT,1\mathchar`-loop}(k)
		= D^2P_{\rm lin}(k) + D^4P_{\rm 1\mathchar`-loop}(k) + D^6P_{\rm 2\mathchar`-loop}|_{\rm LPT,1\mathchar`-loop}(k) 
		+ \sum_{n=3}^{\infty} D^{2n+2}P_{n\rm \mathchar`-loop}|_{\rm LPT, 1\mathchar`-loop}(k).
\end{eqnarray}
Note that 
solutions of higher order than the 2-loop order in SPT come from the non-linearity of the conservation of mass.
The 1-loop LPT solution has the exact 1-loop correction in SPT
because of the third order of the displacement vector in the perturbation expansion.
In this section, we focus the 3- and more order terms in SPT computed in the 1-loop LPT
$\sum_{n=3}^{\infty} D^{2n+2}P_{n\rm \mathchar`-loop}|_{\rm LPT, 1\mathchar`-loop}$ and investigate how they behave.

\begin{table}[tbp]
		\begin{center}
		\begin{tabular}{|c||c|c|c|c|c|c|} \hline
		Redshift	                & $z=0$ & $z=0.35$ & $z=0.5$ & $z=1.0$ & $z=2.0$ & $z=3.0$ \\ \hline \hline
		SPT: 1-loop	[$h{\rm Mpc}^{-1}$] 
		& $\lesssim 0.04$ & $\lesssim 0.05$ & $\lesssim 0.06$ & $\lesssim 0.1$ & $\lesssim 0.25$ & $\lesssim 0.5$ \\ \hline
		SPT: 2-loop [$h{\rm Mpc}^{-1}$]
		& $\lesssim 0.1$ & $\lesssim 0.12$ & $\lesssim 0.15$ & $\lesssim 0.3$ & $\lesssim 0.4$ & $\lesssim 0.6$ \\ \hline
		\end{tabular}
		\end{center}
		\caption{Limitation of the Validity of the Solution in SPT at the 1- and 2-loop Order
		with an accuracy of $<1\%$ are Estimated from Figure~\ref{fig:higher_loop}.}
		\label{limit_scale}
\end{table}
\begin{figure}[htbp]
		\begin{center}
				\scalebox{0.7}{\plotone{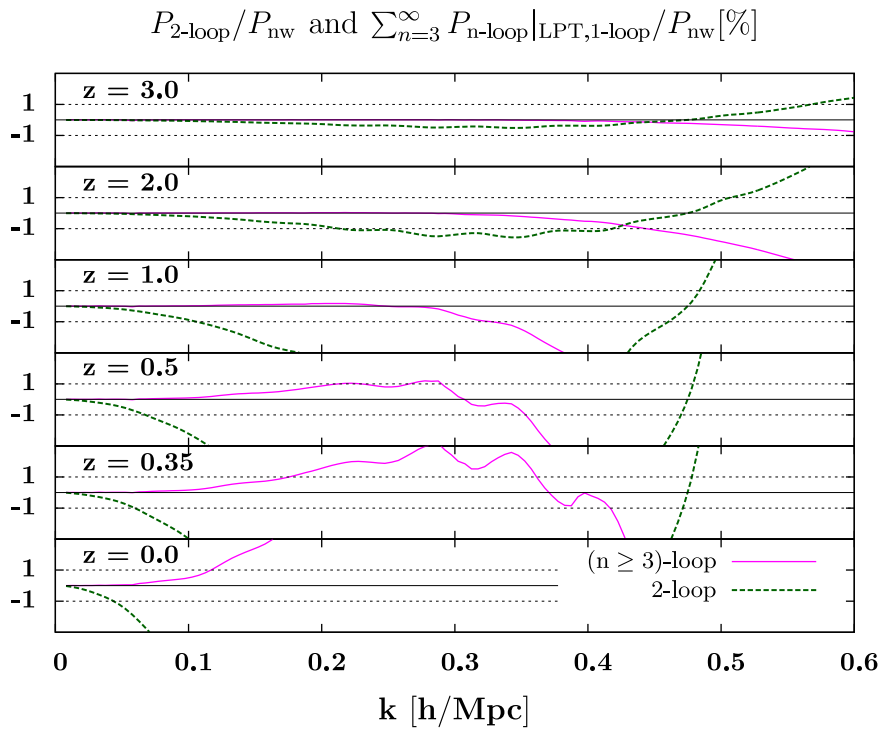}}
		\end{center}
		\caption{
		We show how the contributions of the exact 2-loop solution in SPT
		and the approximate loop solution higher than the 2-loop computed in the 1-loop LPT affect the non-linear power spectrum.
		The ratios $P_{\rm 2\mathchar`-loop}\times 100/P_{\rm lin}^{\rm nw} [\%]$ (green) and 
		$\sum_{n=3}^{\infty} P_{n\rm \mathchar`-loop}|_{\rm LPT, 1\mathchar`-loop}\times 100/P_{\rm lin}^{\rm nw} [\%]$
		(magenta) are plotted at $z=0,\ 0.35,\ 0.5,\ 1.0,\ 2.0,\ {\rm and},\ 3.0$.} 
		This figure can be used to estimate the limitation of the validity of the SPT solutions.
		For example, at $z=3.0$, the 2-loop and higher order solutions are too small to be considered,
		and the 1-loop solution in SPT, therefore, can describe the precise non-linear power spectrum until $k\sim 0.5$ $h{\rm Mpc}^{-1}$.
		At $z=1.0$, $\sum_{n=3}^{\infty} P_{n\rm \mathchar`-loop}|_{\rm LPT, 1\mathchar`-loop}$ can be ignored 
		until $k\sim 0.3$ $h{\rm Mpc}^{-1}$. This implies that the 2-loop SPT solution works well until this scale.
		At $z=0.35$, 
		the fact that $\sum_{n=3}^{\infty} P_{n\rm \mathchar`-loop}|_{\rm LPT, 1\mathchar`-loop}$ is too large to be ignored
		at $k=0.2$ $h{\rm Mpc}^{-1}$ shows that the validity of the 2-loop SPT is violated at this scale.
		The limitation of the validity of the SPT solutions is summarized in Table~\ref{limit_scale}.
		\label{fig:higher_loop}
\end{figure}

Figure~\ref{fig:higher_loop} shows the behavior of $P_{\rm 2\mathchar`-loop}$ and 
$\sum_{n=3}^{\infty} D^{2n+2}P_{n\rm \mathchar`-loop}|_{\rm LPT, 1\mathchar`-loop}$
at various redshifts ($z=0,\ 0.35,\ 0.5,\ 1.0,\ 2.0,\ {\rm and},\ 3.0$).
This figure implies the limitation of the validity of the solutions in SPT at the 1- and 2-loop order.
For example, at $z=3.0$ the 2-loop correction in SPT is small enough to be ignored until $k\simeq0.5\ [h{\rm Mpc}^{-1}]$ 
with an accuracy of $<1\%$,
and the 1-loop solution in SPT, therefore, works well until this scale.
On the other hand,
at $z=0$ the validity of the 2-loop solution in SPT violates around $k\simeq0.1\ [h{\rm Mpc}^{-1}]$
because the approximate higher loop solutions have a considerable contribution of more than 1\% around these scales.
In other words, we expect that around $k\simeq0.1\ [h{\rm Mpc}^{-1}]$ and $z=0$
the 2-loop SPT solution will be too small to predict the precise non-linear power spectrum.
A rough estimate of the scales where the 1- and 2-loop solutions in SPT are valid with an accuracy of $<1\%$
is summarized in Table~\ref{limit_scale}.
These predictions of the behavior of the SPT solutions are confirmed by comparing $N$-body simulations in Sec.~\ref{Nbody}.

Let us mention the exact 3-loop solution in SPT recently computed by~\citet{Blas:2013aba}.
The converging properties of the 3- and higher loop corrections computed in the 1-loop LPT differ from the exact 3-loop results.
The origin of this difference is higher order of the displacement vector than the third order,
because we need up to the seventh order of the displacement vector in the perturbation series to reproduce the exact 3-loop SPT solutions.
We leave the investigation how their non-linear corrections affect the power spectrum for our future work.
At least, we find that the third order displacement vector and the full non-linear law of conservation of mass yield good converging properties.

We end this section by presenting the following approximate solution of the non-linear power spectrum:
\begin{eqnarray}
		P(z,k) = D^2 P_{\rm lin}(k) + D^4 P_{\rm 1\mathchar`-loop}(k) + D^6 P_{\rm 2\mathchar`-loop}(k)
		+ \sum_{n=3}^{\infty} D^{2n+2}P_{\rm n\mathchar`-loop}|_{\rm LPT, 1\mathchar`-loop}(k).
		\label{main_result}
\end{eqnarray}
This is the second main result of this paper.
Since we already have the exact 2-loop solution in SPT,
we do not need to use the approximate 2-loop solution in the 1-loop LPT $P_{\rm 2\mathchar`-loop}|_{\rm LPT,1\mathchar`-loop}$.
Therefore, by matching the solutions at the 3- and higher loop orders in SPT computed in the 1-loop LPT to the 2-loop SPT,
we can obtain more information on the non-linearity of the law of conservation of mass 
and a better approximate non-linear power spectrum than the 2-loop SPT solution.

\section{COMPARISON WITH N-BODY SIMULATION: POWER SPECTRUM}
\label{Nbody}

\begin{table}[t]
		\begin{center}
		\begin{tabular}{|c||c|c|c|c|} \hline
		Name              & $L_{\rm box}$            & Particles & $z_{\rm ini}$ & Runs  \\ \hline \hline
		Low               &  1,000 $h^{-1}{\rm Mpc}$ & $512^3$      & 31            & 30       \\ \hline
		High (L11-N11)    &  2,048 $h^{-1}{\rm Mpc}$ & $2,048^3$     & 99            &  1       \\
	\ \ \ \ \ \ \ (L12-N11)    &  4,096 $h^{-1}{\rm Mpc}$ & $2,048^3$     & 99            &  1       \\ \hline
		\end{tabular}
		\end{center}
		\caption{Sets of $N$-body Simulations we used are Summarized.}
		\label{table:Nbody}
\end{table}

In this section, we compare the analytical predicted power spectra and $N$-body simulation results.
We use two $N$-body simulation results 
created by the public $N$-body codes {\it GADGET2} and {\it 2LPT}~\citep{Springel:2005mi,Crocce:2006ve}
with low and high resolutions presented in~\citet{2009PhRvD} and ~\citet{Valageas:2010yw}, respectively.
The high-resolution $N$-body simulations are
computed by combining the results with different box sizes, called $L11$-$N11$ and $L12$-$N11$. 
We summarize our sets of $N$-body simulation in Table~\ref{table:Nbody}.

\subsection{One-loop Order}

\begin{figure}[tpb]
		\begin{center}
				\scalebox{1.0}{\plottwo{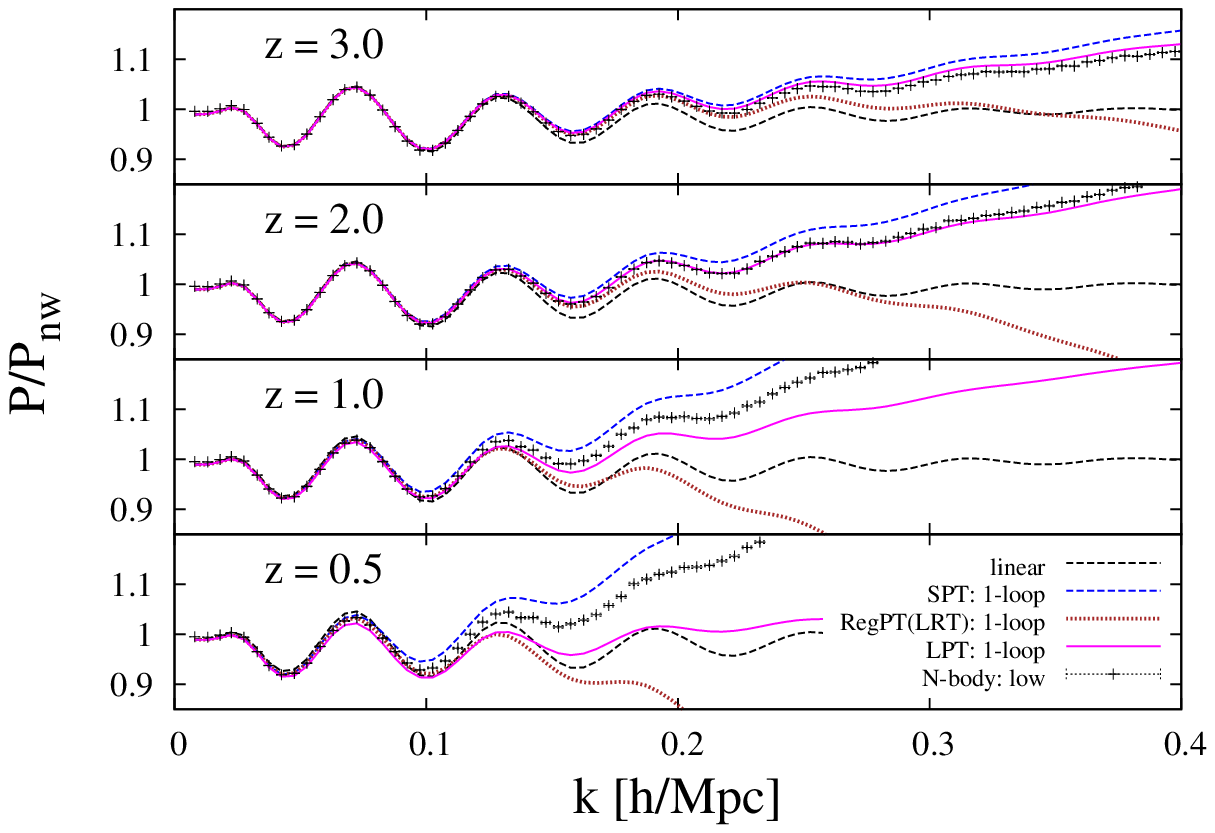}{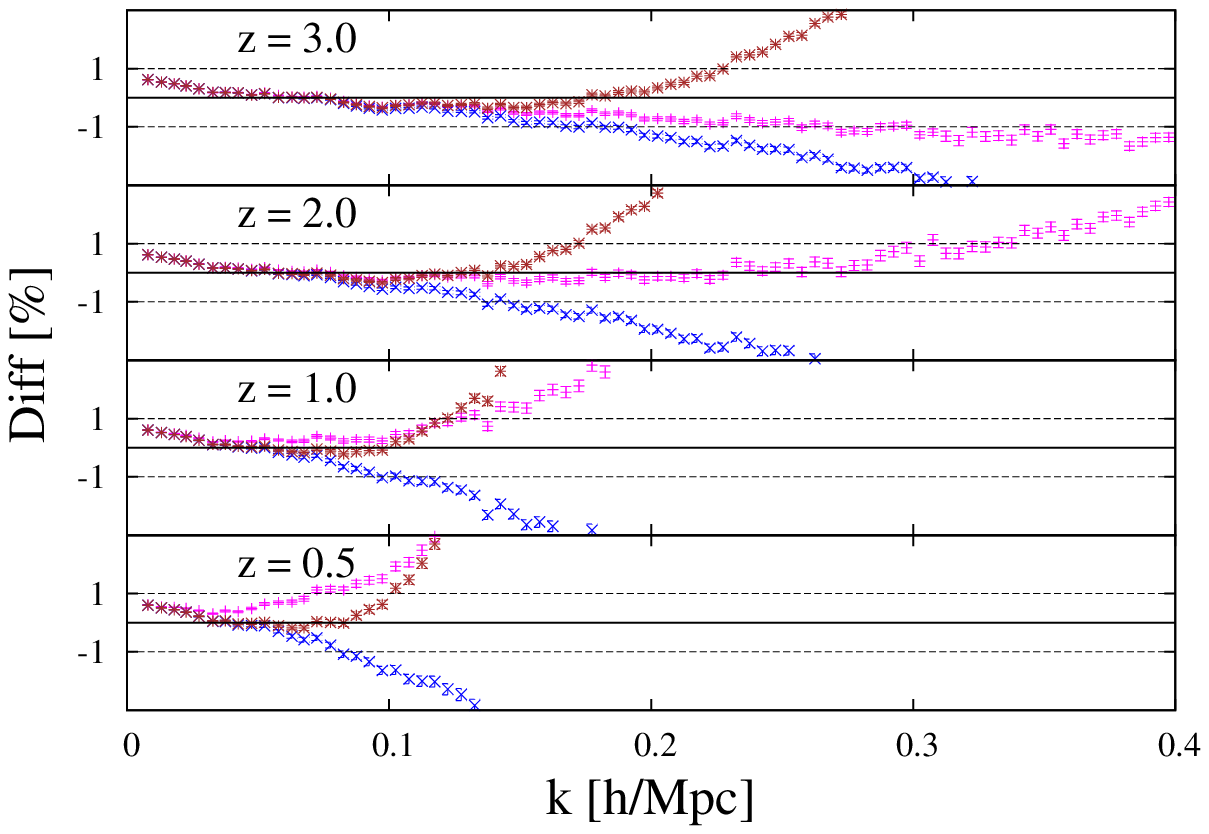}}
		\end{center}
		\begin{center}
				\scalebox{1.0}{\plottwo{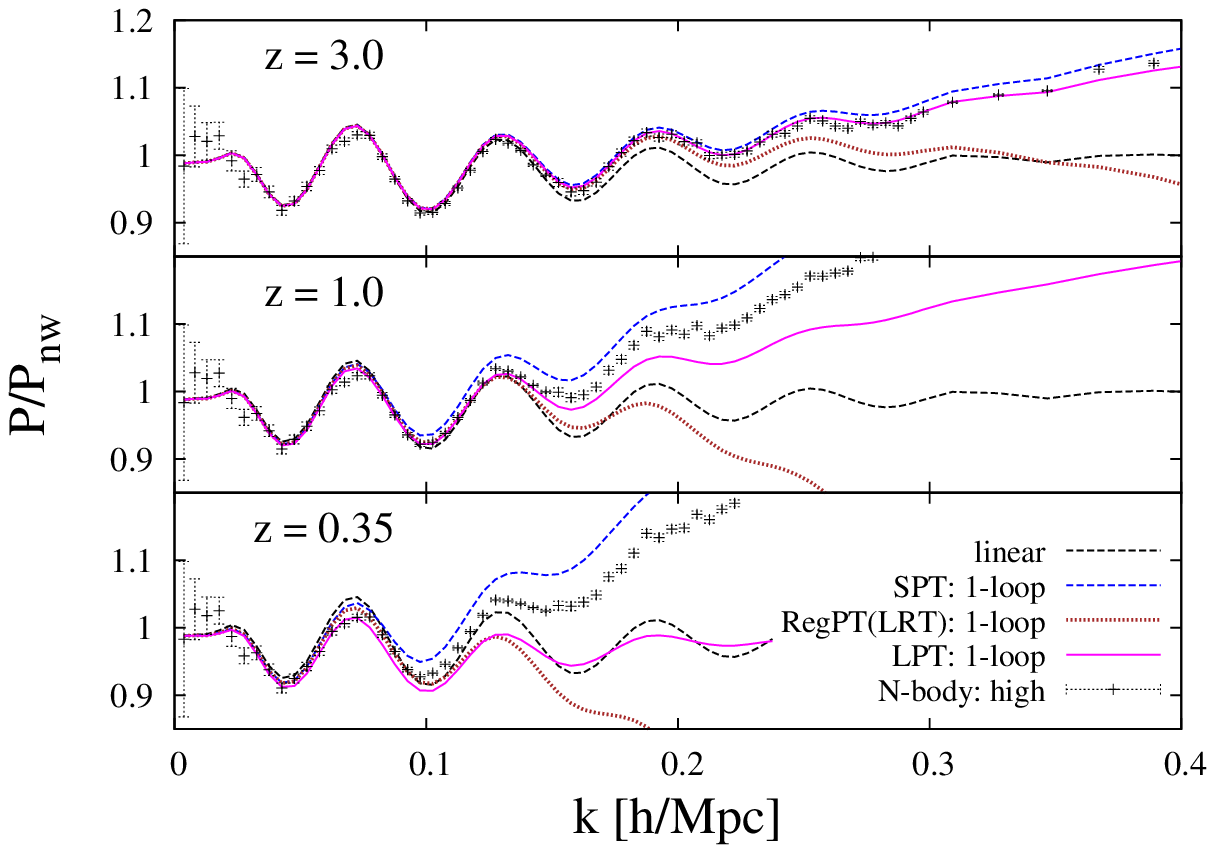}{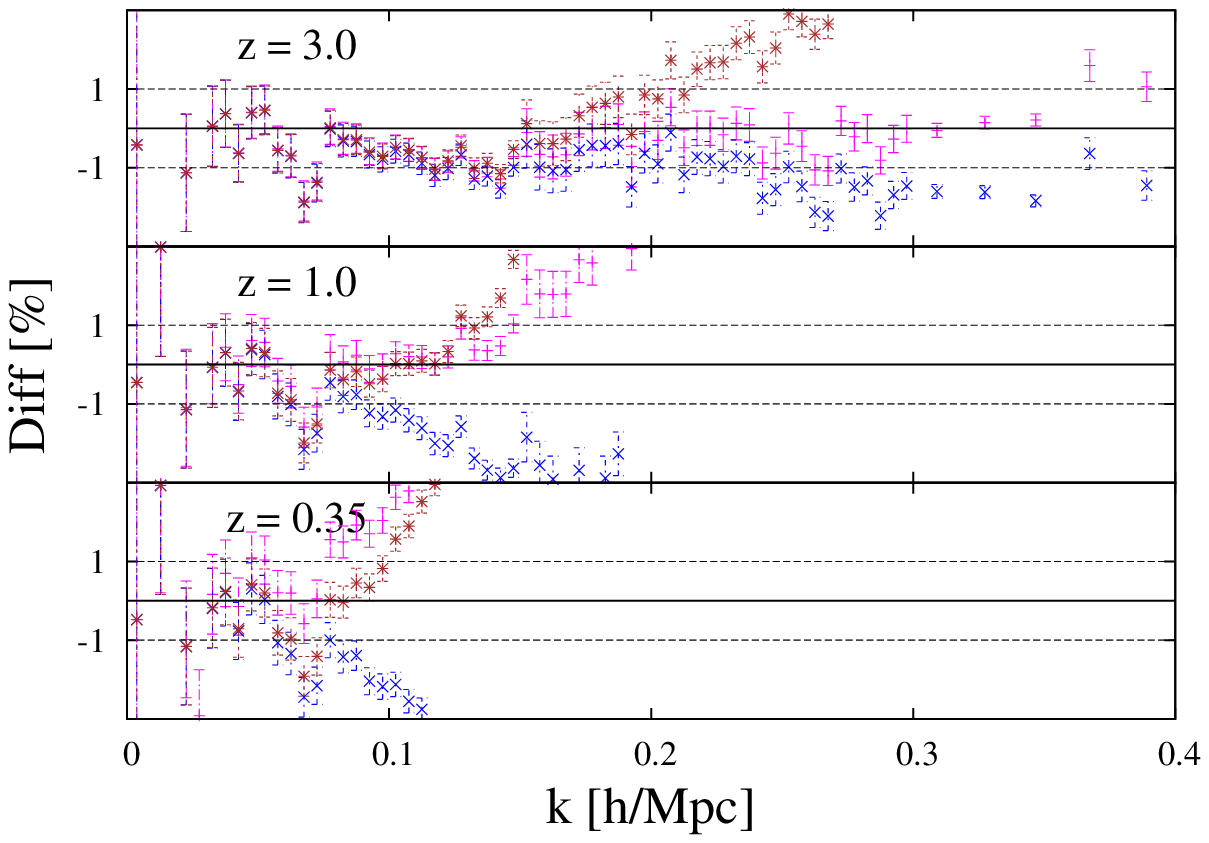}}
		\end{center}
		\caption{
		Comparison between the $N$-body simulation results with the low- and high resolutions 
		and various analytical predictions at the 1-loop order are shown.
		The top panels and bottom panels plot the $N$-body simulations with the low- and high resolutions, respectively,
		even though the analytical predictions are the same.
		Left panels: ratios of the predicted non-linear power spectra and the no-wiggle linear power spectrum
		$P/P_{\rm lin}^{\rm nw}$ are plotted:
		1-loop SPT (blue), 1-loop RegPT (brown), 1-loop LPT (magenta), and $N$-body simulations (black symbols).
		Right panels: 
		Fractional differences ${\rm Diff}[\%]\equiv [P_{\rm N\mathchar`-body}-P]\times100/P_{\rm lin}^{\rm nw}$ are plotted.
		}
		\label{fig:Nbody1loop}
\end{figure}

In Figure~\ref{fig:Nbody1loop},
we plot the analytically predicted power spectra at the 1-loop order 
(SPT in Eqs.~(\ref{SPT1loop}), RegPT in Eq.~(\ref{RegPT1loop}), 
and LPT in Eqs.~(\ref{1-loop_propagator}) and (\ref{1-loop_modecoupling}) ) and the $N$-body simulations.
The top and bottom panels show the $N$-body simulations with the low and high resolutions, respectively,
while the analytical predictions are the same.
First, let us recall that the 1-loop SPT solution should be correct until $k\simeq0.5\ [h{\rm Mpc}^{-1}]$ at $z=3$
within accuracy less than 1\%
and until $k\simeq0.4\ [h{\rm Mpc}^{-1}]$ at $z=2$ within accuracy less than 2\% (Figure~\ref{fig:higher_loop}).
Nevertheless, 
the top panels in Figure~\ref{fig:Nbody1loop} show that the low-resolution $N$-body simulations do not agree with the 1-loop SPT result.
This inconsistency implies that the low-resolution $N$-body simulations underestimate true values at $z=2.0$ and $z=3.0$.
This fact is not surprising.
It is well known that
this underestimation happens due to difficulty of describing small fluctuations of dark matter at high-$z$.
In fact, the $N$-body simulations with the high-resolutions are in excellent agreement with the 1-loop SPT result at $z=3.0$ in the bottom panels.
Second, as expected, 
the 1-loop LPT solution is better than the 1-loop SPT solution at relatively low-$z$: $z=1.0$, $z=0.5$, and $z=0.35$.
This is because the 2-loop contribution becomes large enough to not be ignored at $k\lesssim 0.2\ [h{\rm Mpc}^{-1}]$ at these redshifts.

\begin{figure}[tpb]
		\begin{center}
				\scalebox{1.0}{\plottwo{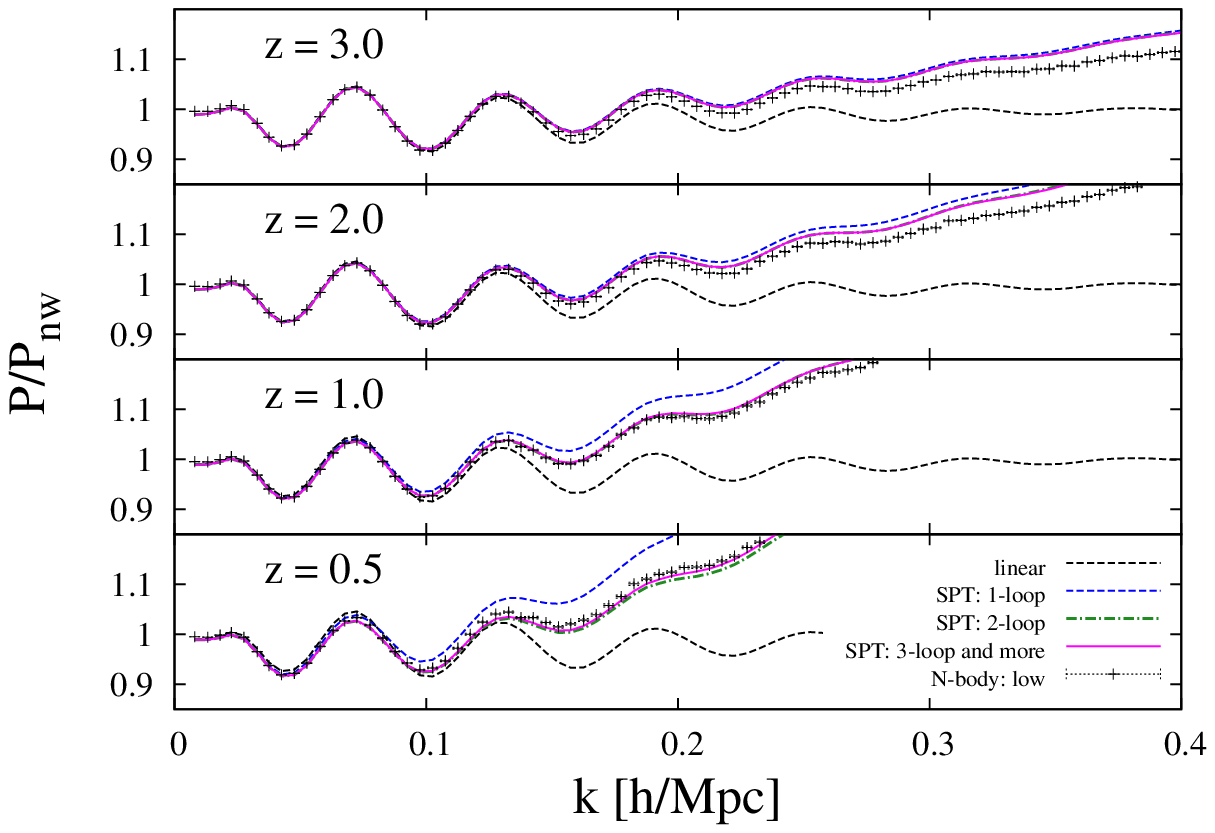}{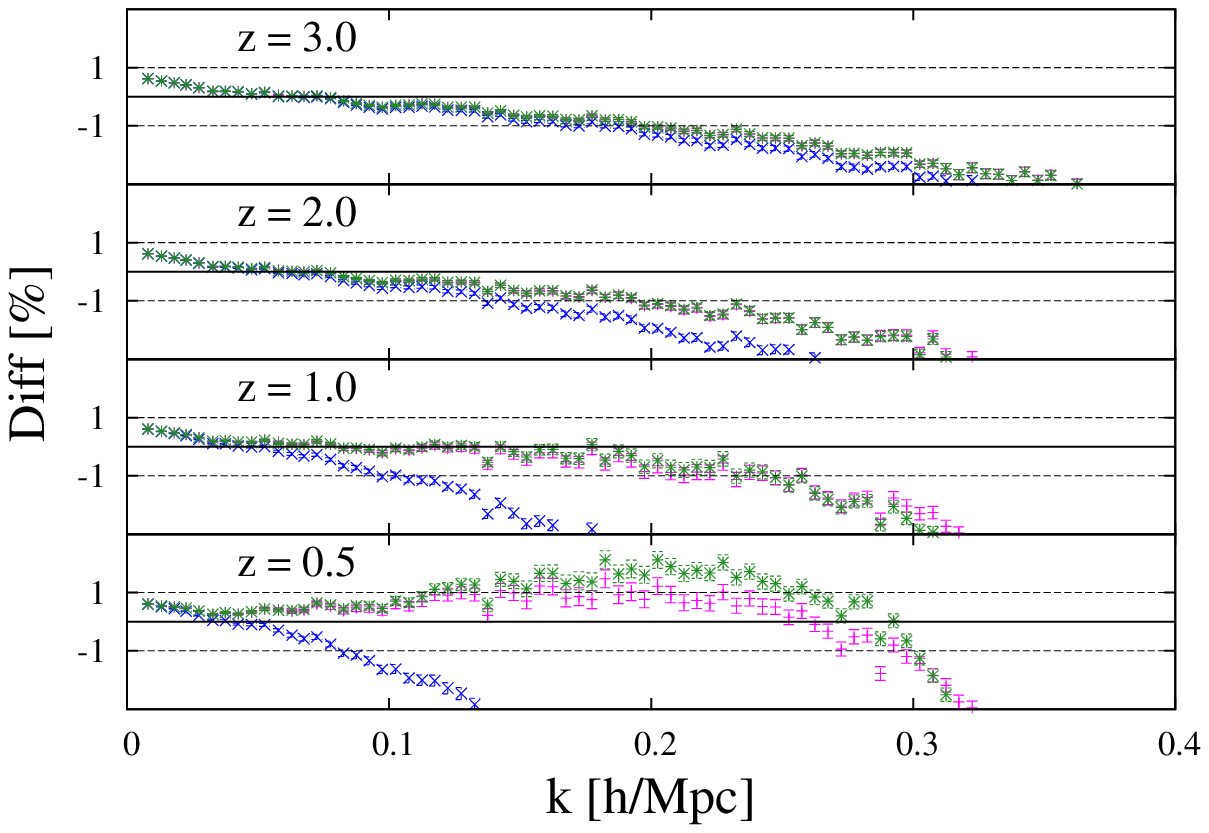}}
		\end{center}
		\begin{center}
				\scalebox{1.0}{\plottwo{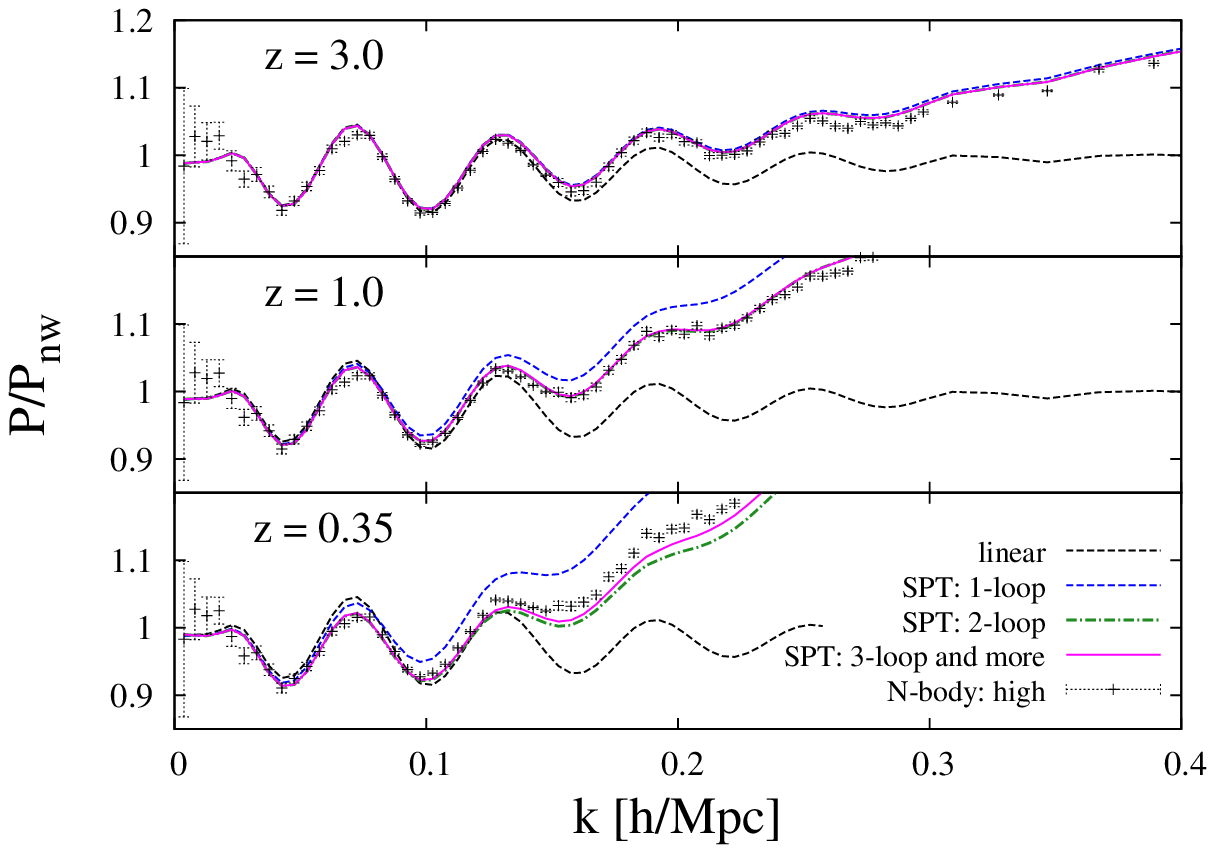}{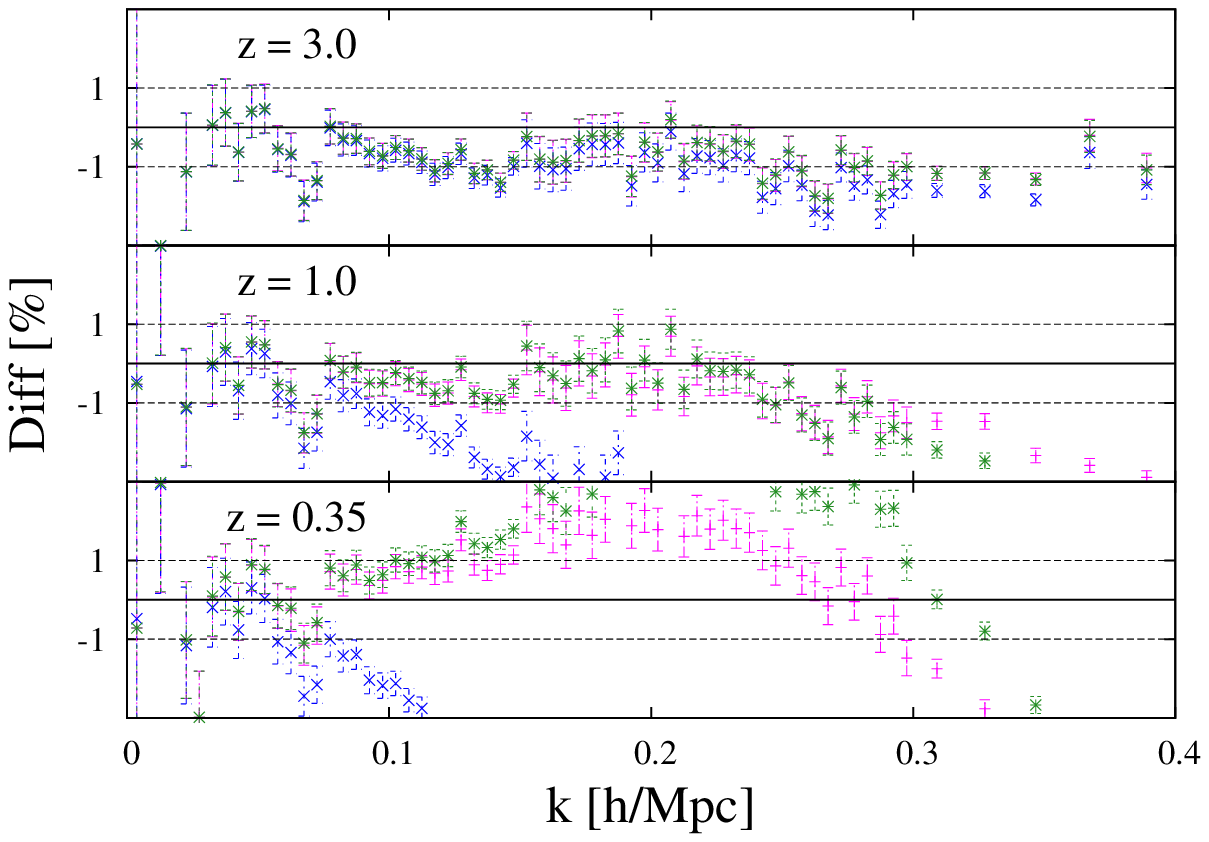}}
		\end{center}
		\caption{Same as Figure~\ref{fig:Nbody1loop}:
		predicted power spectra
	 	(the 1-loop SPT in Eq.~(\ref{SPT1loop}), the 2-loop SPT in Eq.~(\ref{SPT_2loop}),
		and our main result in Eq.~(\ref{main_result}) ) are plotted as blue, green and magenta lines.
		As expected from Figure~\ref{fig:higher_loop},
		at $z=3.0$ the 1-loop SPT solution works well until $k=0.4$ $h{\rm Mpc}^{-1}$, and 
		at $z=1.0$ the 2-loop SPT solution is in extremely agreement with the $N$-body simulation result until $k=0.3$ $h{\rm Mpc}^{-1}$.
		At $z=0.35$, the 2-loop SPT solution is not enough to describe the non-linear power spectrum at $k=0.2$ $h{\rm Mpc}^{-1}$
		and our main result is indeed better than the 2-loop SPT solution at the scale.
		Our result agrees with numerical simulations at $k=0.2$ $h{\rm Mpc}^{-1}$ and $z=0.35$ to better than 2\%.
		}
		\label{fig:Nbody}
\end{figure}

\subsection{Two-loop Order and More}

In Figure~\ref{fig:Nbody},
we compare the 1- and 2-loop SPT solutions 
(Eqs.~(\ref{SPT1loop}) and (\ref{SPT_2loop}))
and the 2-loop SPT solution in addition to the approximate solutions at 3- and higher loop orders
computed in the 1-loop LPT (Eq.~(\ref{main_result})).
Similarly to the case in the last subsection,
there is the disagreement between the analytical results and the low-resolution $N$-body simulations at $z=2.0$ and $z=3.0$,
but it is no problem.
At $z=1.0$, the 2-loop SPT result agrees well with the $N$-body result
as expected from Figure~\ref{fig:higher_loop}.
Furthermore, 
higher order contributions than the 2-loop in SPT which come from the non-linearity of the law of conservation of mass
$\sum_{n=3}^{\infty} P_{n\rm \mathchar`-loop}|_{\rm LPT,1\mathchar`-loop}$
indeed improve the 2-loop SPT solution at $z=0.35$ and $z=0.5$.
(see magenta and green symbols in the right panel of Figure~\ref{fig:Nbody}).
Although the 2-loop LPT would give better solutions, the calculations are left to our future works.

\section{CONCLUSION}
\label{Conclusion}

We calculated the LPT power spectrum at the 1-loop order.
In LPT, the full non-linear law of conservation of mass is naturally satisfied 
by the relation between the matter density and the displacement vector.
The conservation of mass relates various properties of the matter density perturbation:
Galilean invariance, cancellation of high-$k$ solutions in SPT, and IR divergence problem.
Furthermore, the LPT solution has a simple relation to the $\Gamma$-expansion method.

Although it is difficult to explicitly compute the LPT power spectrum even using the Zel'dovich approximation,
we presented an expansion method to approximately compute the LPT power spectrum.
Our approximate solution has good convergence in the series expansion
and enables to compute the LPT power spectrum accurately and quickly.

The 1-loop LPT solution has full non-linear information on the conservation of mass.
Therefore, 
by matching the 1-loop LPT solution to the 2-loop SPT solution,
we can obtain a better approximate solution of the power spectrum than the 2-loo SPT without any free parameter.
This solution agrees with the $N$-body simulation at $k=0.2$ $[h{\rm Mpc}^{-1}]$ and $z=0.35$ to better than 2\%.

\acknowledgments
We would like to express our deepest gratitude to D. N. Spergel who provided carefully considered feedback and valuable comments.
We also owe a very important debt to T. Nishimich and A. Taruya for providing the numerical simulation data and useful comments.
We also would like to thank E. Komatsu whose opinions have helped us in this study.
This work is supported in part by a Grant-in-Aid for Scientific Research from JSPS (No. 24-3849).
Finally, N.S.S. gratefully appreciates the Department of Astrophysical Science at Princeton University for providing a good environment for research.

\appendix

\section{NON-LINEAR CORRECTION TERMS IN LPT}
The LPT power spectrum is described as
\begin{eqnarray}
		P(z,k) = \int d^3q e^{-i\kk\cdot\qq} \Big\{e^{\Sigma(z,\kk,\qq) - \bar{\Sigma}(z,k)} - 1 \Big\},
\end{eqnarray}
where in the 1-loop LPT, $\Sigma$ is given by
\begin{eqnarray}
		\Sigma(z,\kk,\qq) &=&  \sum_{\ell=0}^{3}i^{\ell}
		\left( D^2\Sigma_{\ell,\rm lin}(k,q) + D^4\Sigma_{\ell, 22}(k,q) + D^4\Sigma_{\ell, 13}(k,q) \right)
	{\cal L}_{\ell}(\hat{k}\cdot\hat{q}), \nonumber \\
	\bar{\Sigma}(z,k) &=& D^2\bar{\Sigma}_{\rm lin}(k) +D^4\bar{\Sigma}_{22}(k) + D^4\bar{\Sigma}_{13}(k),
\end{eqnarray}
with 
\begin{eqnarray}
		\Sigma_{0,\rm lin}(k,q) &=&  \frac{1}{3}k^2 \int \frac{dp}{2\pi^2} j_{0}(pq)P_{\rm lin}(p), \quad
		\Sigma_{2,\rm lin}(k,q) =  \frac{2}{3}k^2 \int \frac{dp}{2\pi^2} j_{2}(pq)P_{\rm lin}(p), \nonumber \\
		\Sigma_{\ell,\rm 22}(k,q) &=&  \int_0^{\infty} \frac{dp_1p_1^2}{2\pi^2}\int_0^{\infty}\frac{dp_2p_2^2}{2\pi^2} 
		\int_{-1}^1 d\mu j_{\ell}\left( |\pp_1+\pp_2|q \right)K_{\ell,22}(k,p_1,p_2,\mu) P_{\rm lin}(p_1) P_{\rm lin}(p_2), \nonumber \\
		\Sigma_{\ell,\rm 13}(k,q) &=&  \int_0^{\infty} \frac{dp_1p_1^2}{2\pi^2}\int_0^{\infty}\frac{dp_2p_2^2}{2\pi^2} 
		j_{\ell}\left( p_1q \right)K_{\ell,13}(k,p_1,p_2) P_{\rm lin}(p_1) P_{\rm lin}(p_2),
		\label{ap_sigma}
\end{eqnarray}
and $\bar{\Sigma}_{\rm lin}(k) = \Sigma_{0,\rm lin}(k,q=0)$,
$\bar{\Sigma}_{22}(k) = \Sigma_{0,22}(k,q=0)$, and $\bar{\Sigma}_{13}(k) = \Sigma_{0,13}(k,q=0)$.

\subsection{Kernel Functions $K_{\ell,22}$ and $K_{\ell,13}$}
\label{ap:K}

In Eq.~(\ref{ap_sigma}), the kernel functions $K_{\ell,13}$ and $K_{\ell,22}$ are given by
\begin{eqnarray}
		K_{0,22}(k,p_1,p_2,\mu) &=& k^2\frac{3}{196} \frac{\left( 1-\mu^2 \right)^2}{|\pp_1+\pp_2|^2}, \nonumber \\
		K_{1,22}(k,p_1,p_2,\mu) &=& k^3\frac{3}{70} \frac{\left( 1 - \mu^2 \right)}{|\pp_1+\pp_2|^3}
		\left( 3\mu\left( \frac{p_1^2 + p_2^2}{p_1p_2} \right)  + 4\mu^2 + 2 \right),\nonumber \\
		K_{2,22}(k,p_1,p_2,\mu) &=& 2 K_{0,22}(k,p_1,p_2,\mu), \nonumber \\
		K_{3,22}(k,p_1,p_2,\mu) &=& k^3\frac{3}{70}\frac{\left(  1 - \mu^2 \right)}{|\pp_1+\pp_2|^3}
		\left(  2\mu\left( \frac{p_1^2 + p_2^2}{p_1p_2} \right) + \mu^2 + 3\right),
		\label{K_l_22}
\end{eqnarray}
and
\begin{eqnarray}
		K_{0,13}(k,p_1,p_2) &=&  k^2\frac{5}{1008}\frac{1}{p_1^2} 
		\frac{1}{y^5} \left( \left( y^2-1 \right)^4 \ln\Bigg| \frac{1+y}{1-y}\Bigg|
		-\frac{2}{3}y\left( 3y^6 - 11y^4 -11y^2 + 3 \right)\right) , \nonumber \\
		K_{1,13}(k,p_1,p_2) &=&  k^3\frac{3}{560} \frac{1}{p_1^3} 
		\frac{1}{y^5} \left( \left( y^2-1 \right)^3(2y^2+4)  \ln\Bigg| \frac{1+y}{1-y}\Bigg|
		- \frac{2}{3}y\left( 6y^6 - 4 y^4 + 26 y^2 -12 \right)\right) , \nonumber \\
		K_{2,13}(k,p_1,p_2) &=& 2 K_{0,13}(k,p_1,p_2) ,\nonumber \\	
		K_{3,13}(k,p_1,p_2) &=&  k^3\frac{3}{560}\frac{1}{p_1^3}
		 \frac{1}{y^5}\left( \left( y^2-1 \right)^3(3y^2+1)  \ln\Bigg| \frac{1+y}{1-y}\Bigg|
		- \frac{2}{3}y\left( 9y^6 - 21 y^4 - y^2 - 3 \right)\right),
		\label{K_l_13}
\end{eqnarray}
where $y = p_2/p_1$ and $\mu = \hat{p}_1 \cdot \hat{p}_2$.

For $y=p_2/p_1\ll1$, $K_{\ell,22}$ and $K_{\ell,13}$ become
\begin{eqnarray}
		K_{0,22}(k,p_1,p_2,\mu) &\to& k^2 \frac{3}{196} \frac{(1-\mu^2)^2}{p_1^2}, \nonumber \\
		K_{1,22}(k,p_1,p_2,\mu) &\to& k^3 \frac{3}{70} \frac{(1-\mu^2)}{p_1^3}\left( 3\mu\frac{p_1}{p_2} + 4\mu^2+2 \right), \nonumber \\
		K_{0,22}(k,p_1,p_2,\mu) &\to& k^2 \frac{6}{196} \frac{(1-\mu^2)^2}{p_1^2}, \nonumber \\
		K_{0,22}(k,p_1,p_2,\mu) &\to& k^3 \frac{3}{70} \frac{(1-\mu^2)}{p_1^3}\left( 2\mu\frac{p_1}{p_2} + \mu^2 + 3 \right),
		\label{K22_1}
\end{eqnarray}
and 
\begin{eqnarray}
		K_{0,13}(k,p_1,p_2) &\to& \frac{16k^2}{189p_1^2} - \frac{16k^2}{441p_1^2}y^2 + \frac{16k^2}{3969p_1^2}y^4, \nonumber \\ 
		K_{1,13}(k,p_1,p_2) &\to& -\frac{4k^3}{175p_1^3} - \frac{12k^3}{245p_1^3}y^2 + \frac{44k^3}{3675p_1^3}y^4, \nonumber \\
		K_{2,13}(k,p_1,p_2) &\to& \frac{32k^2}{189p_1^2} - \frac{32k^2}{441p_1^2}y^2 + \frac{32k^2}{3969p_1^2}y^4, \nonumber \\
		K_{3,13}(k,p_1,p_2) &\to& \frac{24k^3}{175p_1^3} - \frac{24k^3}{245p_1^3}y^2 + \frac{8k^3}{525p_1^3}y^4.
		\label{K13_1}
\end{eqnarray}
On the other hand, for $y = p_2/p_1\gg1$, 
$K_{\ell,22}$ are given by replacing $p_1$ with $p_2$ in Eq.~(\ref{K22_1}) due to the symmetry of $K_{\ell,22}$ about $p_1$ and $p_2$,
and $K_{\ell,13}$ are given by
\begin{eqnarray}
		K_{0,13}(k,p_1,p_2) &\to& \quad \frac{16}{189}\frac{k^2}{p_1^2}\frac{1}{y^2} - \frac{16k^2}{441p_1^2}\frac{1}{y^4}, \nonumber \\
		K_{1,13}(k,p_1,p_2) &\to& -\frac{4k^3}{25p_1^3}\frac{1}{y^2} + \frac{156k^3}{1225p_1^3}\frac{1}{y^4}, \nonumber \\
		K_{2,13}(k,p_1,p_2) &\to& \quad \frac{32}{189}\frac{k^2}{p_1^2}\frac{1}{y^2} - \frac{32k^2}{441p_1^2}\frac{1}{y^4}, \nonumber \\
		K_{3,13}(k,p_1,p_2) &\to& \frac{8k^3}{175p_1^3}\frac{1}{y^2} + \frac{24k^3}{1225p_1^3}\frac{1}{y^4}.
		\label{K13_2}
\end{eqnarray}

\subsection{Asymptotic Expressions of $\Sigma_{\ell,22}$ and $\Sigma_{\ell,13}$}
\label{ap:Sigma}

The asymptotic behaviors of $K_{\ell,22}$ and $K_{\ell,13}$ (Eqs.~(\ref{K22_1}), (\ref{K13_1}), and (\ref{K13_2}))
lead to those of $\Sigma_{\ell,22}$ and $\Sigma_{\ell,13}$:
for $p_2/p_1\ll1$, 
\begin{eqnarray}
		\Sigma_{0,22}(k,q) - \bar{\Sigma}_{22}(k)
		&\to& \frac{k^2}{245\pi^4}\int_0^{\infty} dp_1 \left[ j_0(p_1q) - 1\right]
		P_{\rm lin}(p_1) \int_0^{\infty} dp_2 p_2^2 P_{\rm lin}(p_2), \nonumber \\
		\Sigma_{1,22}(k,q) &\to& \frac{k^3}{25\pi^4}
		\int_0^{\infty} dp_1 \frac{j_1(p_1q)}{p_1} P_{\rm lin}(p_1) \int_0^{\infty} dp_2 p_2^2 P_{\rm lin}(p_2), \nonumber \\
		\Sigma_{2,22}(k,q) &\to& \frac{2k^2}{245\pi^4}
		\int_0^{\infty} dp_1 j_2(p_1q) P_{\rm lin}(p_1) \int_0^{\infty} dp_2 p_2^2 P_{\rm lin}(p_2), \nonumber \\
		\Sigma_{1,22}(k,q) &\to& \frac{8k^3}{175\pi^4}
		\int_0^{\infty} dp_1 \frac{j_3(p_1q)}{p_1} P_{\rm lin}(p_1) \int_0^{\infty} dp_2 p_2^2 P_{\rm lin}(p_2),
		\label{22_k_large}
\end{eqnarray}
and
\begin{eqnarray}
		\Sigma_{0,13}(k,q) - \bar{\Sigma}_{13}(k) 
		&\to& \frac{4k^2}{189\pi^4}\int_0^{\infty} dp_1 \left[ j_0(p_1q) -1 \right] P_{\rm lin}(p_1) 
		\int_0^{\infty} dp_2 p_2^2 P_{\rm lin}(p_2), \nonumber \\
		\Sigma_{1,13}(k,q) 
		&\to& - \frac{k^3}{175 \pi^4} \int_0^{\infty} dp_1 \frac{j_1(p_1q)}{p_1} P_{\rm lin}(p_1) 
		\int_0^{\infty} dp_2 p_2^2 P_{\rm lin}(p_2), \nonumber \\
		\Sigma_{2,13}(k,q) 
		&\to& \frac{8k^2}{189\pi^4}\int_0^{\infty} dp_1 j_2(p_1q)  P_{\rm lin}(p_1)
		\int_0^{\infty} dp_2 p_2^2 P_{\rm lin}(p_2), \nonumber \\
		\Sigma_{3,13}(k,q) 
		&\to& \frac{6k^3}{175\pi^4} \int_0^{\infty} dp_1 \frac{j_3(p_1q)}{p_1} P_{\rm lin}(p_1) 
		\int_0^{\infty} dp_2 p_2^2 P_{\rm lin}(p_2),
		\label{13_k_large}
\end{eqnarray}
and for $y=p_2/p_1\gg1$,
\begin{eqnarray}
		\Sigma_{0,22}(k,q) - \bar{\Sigma}_{22}(k)
		&\to& \frac{k^2}{245\pi^4} \int_0^{\infty} dp_1 p_1^2 P_{\rm lin}(p_1)
		\int_0^{\infty} dp_2 \left[ j_0(p_2q) - 1\right]P_{\rm lin}(p_2), \nonumber \\
		\Sigma_{1,22}(k,q) &\to& \frac{k^3}{25\pi^4}
		 \int_0^{\infty} dp_1 p_1^2 P_{\rm lin}(p_1)\int_0^{\infty} dp_2 \frac{j_1(p_2q)}{p_2} P_{\rm lin}(p_2), \nonumber \\
		\Sigma_{2,22}(k,q) &\to& \frac{2k^2}{245\pi^4}
		\int_0^{\infty} dp_1 p_1^2 P_{\rm lin}(p_1)\int_0^{\infty} dp_2 j_2(p_2q) P_{\rm lin}(p_2) , \nonumber \\
		\Sigma_{1,22}(k,q) &\to& \frac{8k^3}{175\pi^4}
		\int_0^{\infty} dp_1 p_1^2 P_{\rm lin}(p_1)\int_0^{\infty} dp_2 \frac{j_3(p_2q)}{p_2} P_{\rm lin}(p_2) ,
\end{eqnarray}
and
\begin{eqnarray}
		\Sigma_{0,13}(k,q) - \bar{\Sigma}_{13}(k) 
		&\to& \frac{4k^2}{189\pi^4} \int_{0}^{\infty}dp_1 p_1^2
		\left[ j_0(p_1q) -1 \right]P_{\rm lin}(p_1) \int_0^{\infty} dp_2 P_{\rm lin}(p_2), \nonumber \\
		\Sigma_{1,13}(k,q) 
		&\to& - \frac{k^3}{25\pi^4} \int_{0}^{\infty}dp_1 p_1^2 \frac{j_1(p_1q)}{p_1}
		P_{\rm lin}(p_1) \int_0^{\infty} dp_2 P_{\rm lin}(p_2), \nonumber \\
		\Sigma_{2,13}(k,q) 
		&\to& \frac{8k^2}{189\pi^4} \int_{0}^{\infty}dp_1 p_1^2 j_2(p_1q) P_{\rm lin}(p_1) \int_0^{\infty} dp_2 P_{\rm lin}(p_2), \nonumber \\
		\Sigma_{3,13}(k,q) 
		&\to& \frac{2k^3}{175} \int_{0}^{\infty}dp_1 p_1^2 \frac{j_3(p_1q)}{p_1}
		P_{\rm lin}(p_1) \int_0^{\infty} dp_2 P_{\rm lin}(p_2).
		\label{13_k_small}
\end{eqnarray}

\section{NON-LINEAR CORRECTIONS IN THE ONE-LOOP SPT}
\label{ap:SPT_1loop}

In LPT, the 1-loop correction term in SPT, $P_{\rm 1\mathchar`-loop} = P_{22} + P_{13}$, is described as
\begin{eqnarray}
		P_{22}(k) =  \sum_{\ell=0}^3 P_{\ell,22}(k) + P_{22}|_{\rm ZA}(k) \quad \mbox{and} \quad
		P_{13}(k) =  \sum_{\ell=0}^3 P_{\ell,13}(k) + P_{13}|_{\rm ZA}(k).
\end{eqnarray}

The contribution from the Zel'dovich approximation, $P_{\rm 1\mathchar`-loop}|_{\rm ZA}=P_{22}|_{\rm ZA} + P_{13}|_{\rm ZA}$,
is given by
\begin{eqnarray}
		P_{\rm 1\mathchar`-loop}|_{\rm ZA}(k)
		&=& \frac{1}{2} \int d^3q e^{-i\kk\cdot\qq}
		\left\{ \sum_{\ell=0}^2 i^{\ell}\Sigma_{\ell,\rm lin}(k,q) {\cal L}_{\ell}(\hat{k}\cdot\hat{q}) - \bar{\Sigma}_{\rm lin}(k) \right\}^2,
		\nonumber \\
		P_{22}|_{\rm ZA}(k)
		&=&  \frac{1}{2} \int d^3q e^{-i\kk\cdot\qq}
		\left\{ \sum_{\ell=0}^2 i^{\ell} \Sigma_{\ell,\rm lin}(k,q) {\cal L}_{\ell}(\hat{k}\cdot\hat{q}) \right\}^2 \nonumber \\
		&=&  4\pi \int_0^{\infty} dq q^2 \Bigg\{
		j_0(kq) \frac{\left( \Sigma_{0, \rm lin}(k,q) \right)^2}{2} 
		+ j_2(kq) \Sigma_{0, \rm lin}(k,q)\Sigma_{2, \rm lin}(k,q) \nonumber \\
		&& \hspace{2cm} + 		\Bigg(\frac{18}{35}j_4(kq) - \frac{2}{7}j_2(kq) + \frac{1}{5}j_0(kq) \Bigg)
		\frac{\left( \Sigma_{2, \rm lin}(k,q) \right)^2}{2} \Bigg\}, \nonumber \\
		P_{13}|_{\rm ZA}(k)
		&=& -\bar{\Sigma}_{\rm lin}(k)\int d^3q e^{-i\kk\cdot\qq}
		\left\{ \sum_{\ell=0}^2 i^{\ell} \Sigma_{\ell,\rm lin}(k,q) {\cal L}_{\ell}(\hat{k}\cdot\hat{q}) \right\} \nonumber \\
		&=& -\bar{\Sigma}_{\rm lin}(k)P_{\rm lin}(k).
\end{eqnarray}
Here, note that $P_{22}|_{\rm ZA}$ is also represented as
\begin{eqnarray}
		P_{22}|_{\rm ZA}(k) &=&	\frac{1}{4}\int_0^{\infty} \frac{dpp^2}{2\pi^2} \int_{-1}^{1}
								d \mu \frac{\mu^2\left( 1-r\mu \right)^2}{r^2\left( 1-2r\mu+r^2 \right)^2}
								P_{\rm lin}(p) P_{\rm lin}(|\kk-\pp|).
	\label{ap:P22ZA}
\end{eqnarray}
where $r\equiv p/k$ and $\mu \equiv \hat{k}\cdot\hat{p}$.

The non-linear correlation functions of the displacement vector $\Sigma_{\ell,22}$ and $\Sigma_{\ell,13}$
(Eq.~(\ref{ap_sigma})) yield $P_{\ell,22}$ and $P_{\ell,13}$ as follows
\begin{eqnarray}
		P_{\ell,22}(k) = 4\pi\int_0^{\infty} dq q^2 j_{\ell}(kq)\Sigma_{\ell,22}(k,q) \quad \mbox{and} \quad
		P_{\ell,13}(k) = 4\pi\int_0^{\infty} dq q^2 j_{\ell}(kq)\Sigma_{\ell,13}(k,q),
\end{eqnarray}
where
\begin{eqnarray}
		P_{0, 22}(k) &=&  \frac{3}{196}\int_0^{\infty}\frac{dpp^2}{2\pi^2} \int_{-1}^{1} d\mu 
                          \frac{(1-\mu^2)^2}{\left( 1 - 2 r\mu +r^2 \right)^2} P_{\rm lin}(p) P_{\rm lin}(|\kk-\pp|), \nonumber \\
		P_{1, 22}(k) &=&  \frac{3}{70}\int_0^{\infty}\frac{dpp^2}{2\pi^2} \int_{-1}^{1} d\mu 
 						  \frac{(1-\mu^2)\left( 3r\mu - r^2 - 2r^2\mu^2 \right)}
						  {r^2\left( 1 - 2 r\mu +r^2 \right)^2}P_{\rm lin}(p) P_{\rm lin}(|\kk-\pp|), \nonumber \\
		P_{2, 22}(k) &=& 2 	P_{0, 22}(k), \nonumber \\
		P_{3, 22}(k) &=&  \frac{3}{70}\int_0^{\infty}\frac{dpp^2}{2\pi^2} \int_{-1}^{1} d\mu 
		\frac{(1-\mu^2)\left( 2r\mu + r^2 - 3r^2\mu^2 \right)}{r^2\left( 1 - 2 r\mu +r^2 \right)^2}P_{\rm lin}(p) P_{\rm lin}(|\kk-\pp|),
\end{eqnarray}
and
\begin{eqnarray}
		P_{0,13}(k) &=& \frac{5}{1008} P_{\rm lin}(k) \int_{0}^{\infty} \frac{dpp^2}{2\pi^2}
		\frac{1}{r^5} \left( \left( r^2-1 \right)^4 \ln\Bigg| \frac{1+r}{1-r}\Bigg|
		-\frac{2}{3}r\left( 3r^6 - 11r^4 -11r^2 + 3 \right)\right)  P_{\rm lin}(p), \nonumber \\
		P_{1,13}(k) &=&  \frac{3}{560} P_{\rm lin}(k) \int_{0}^{\infty} \frac{dpp^2}{2\pi^2}
		\frac{1}{r^5} \left( \left( r^2-1 \right)^3(2r^2+4)  \ln\Bigg| \frac{1+r}{1-r}\Bigg|
		- \frac{2}{3}r\left( 6r^6 - 4 r^4 + 26 r^2 -12 \right)\right)  P_{\rm lin}(p), \nonumber \\
		P_{2,13}(k) &=&  2 P_{0,13}(k), \nonumber \\
		P_{3,13}(k) &=&  \frac{3}{560} P_{\rm lin}(k) \int_{0}^{\infty} \frac{dpp^2}{2\pi^2} 
		 \frac{1}{r^5}\left( \left( r^2-1 \right)^3(3r^2+1)  \ln\Bigg| \frac{1+r}{1-r}\Bigg|
		- \frac{2}{3}r\left( 9r^6 - 21 r^4 - r^2 - 3 \right)\right) P_{\rm lin}(p). \nonumber \\
\end{eqnarray}

Finally, we can show the following well known results:
\begin{eqnarray}
		P_{22}(k) &=&  \sum_{\ell=0}^3P_{\ell,22} + P_{22}|_{\rm ZA} \nonumber \\
		&=&  \frac{1}{196}\int_0^{\infty} \frac{dpp^2 }{2\pi^2} \int_{-1}^{1}d\mu \frac{(3r + 7\mu-  10r \mu^2)^2}{r^2(1-2r\mu+r^2)^2}
		P_{\rm lin}(|\kk-\pp|) P_{\rm lin}(p), \nonumber \\
		P_{13}(k) &=& \sum_{\ell=0}^3P_{\ell,13} + P_{13}|_{\rm ZA} \nonumber \\
		&=& \frac{1}{504} P_{\rm lin}(k)
		\int_0^{\infty} \frac{dpp^2}{2\pi^2}
		\frac{1}{r^2}\left( \frac{12}{r^2}- 158 + 100r^2 - 42 r^4 
		+ \frac{3}{r^3}\left( r^2-1 \right)^3(7r^2+2)\ln \Bigg| \frac{1+r}{1-r}\Bigg| \right) P_{\rm lin}(p). \nonumber \\
		\label{ap:SPT1loop_correction}
\end{eqnarray}

\subsection{Asymptotic Behaviors of $P_{\ell,22}$ and $P_{\ell,13}$}

For $r = p/k\ll1$,
$P_{22}|_{\rm ZA}$ (Eq.~(\ref{ap:P22ZA})) becomes
\begin{eqnarray}
		P_{22}|_{\rm ZA}(k) &\to& 
		P_{22,\rm high\mathchar`-k}|_{\rm ZA}(k) = 2 \times \frac{1}{4}P_{\rm lin}(k) \int_0^{\infty} \frac{dpp^2}{2\pi^2} \int_{-1}^{1} 
		d \mu \frac{\mu^2}{r^2} P_{\rm lin}(p)  \nonumber \\
		&=& \bar{\Sigma}_{\rm lin}(k) P_{\rm lin}(k),
\end{eqnarray}
Thus, at the high-$k$ limit, the 1-loop contributions in SPT from the Zel'dovich approximation cancel out each other,
\begin{eqnarray}
		P_{\rm 1\mathchar`-loop, high\mathchar`-k}|_{\rm ZA}(k) 
		= P_{22, \rm high\mathchar`-k}|_{\rm ZA}(k) + P_{13}|_{\rm ZA}(k)  = 0 \quad \mbox{for $p/k \to0$}.
\end{eqnarray}

The asymptotic behaviors of $\Sigma_{\ell,22}$ and $\Sigma_{\ell,13}$ (Eqs.~(\ref{22_k_large}) and (\ref{13_k_large}))
lead to those of $P_{\ell,22}$ and $P_{\ell,13}$:
for $r = p/k \ll1$,
\begin{eqnarray}
		P_{0,22}(k) &\to& \frac{4}{245\pi^2}  P_{\rm lin}(k) \int_0^{\infty} dp p^2 P_{\rm lin}(k), \nonumber \\
		P_{1,22}(k) &\to& \frac{4}{25\pi^2}   P_{\rm lin}(k) \int_0^{\infty} dp p^2 P_{\rm lin}(k), \nonumber \\
		P_{2,22}(k) &\to& \frac{8}{245\pi^2}  P_{\rm lin}(k) \int_0^{\infty} dp p^2 P_{\rm lin}(k), \nonumber \\
		P_{3,22}(k) &\to& \frac{32}{175\pi^2} P_{\rm lin}(k) \int_0^{\infty} dp p^2 P_{\rm lin}(k), 
\end{eqnarray}
\begin{eqnarray}
		P_{0,13}(k) &\to&  \frac{8}{189\pi^2}P_{\rm lin}(k) \int_0^{\infty} dp p^2 P_{\rm lin}(k), \nonumber \\
		P_{1,13}(k) &\to& -\frac{2}{175\pi^2}P_{\rm lin}(k) \int_0^{\infty} dp p^2 P_{\rm lin}(k), \nonumber \\
		P_{2,13}(k) &\to&  \frac{16}{189\pi^2}P_{\rm lin}(k) \int_0^{\infty} dp p^2 P_{\rm lin}(k), \nonumber \\
		P_{3,13}(k) &\to&  \frac{12}{175\pi^2}P_{\rm lin}(k) \int_0^{\infty} dp p^2 P_{\rm lin}(k).
\end{eqnarray}
Similarly, for $r=p/k\gg1$,
Eq.~(\ref{13_k_small}) leads to
\begin{eqnarray}
		P_{0,13}(k) &\to& \frac{8k^2}{189\pi^2}  P_{\rm lin}(k) \int_0^{\infty} dp P_{\rm lin}(p) ,\nonumber \\
		P_{1,13}(k) &\to& -\frac{2k^2}{25\pi^2}  P_{\rm lin}(k) \int_0^{\infty} dp P_{\rm lin}(p) ,\nonumber \\
		P_{2,13}(k) &\to& \frac{16k^2}{189\pi^2} P_{\rm lin}(k) \int_0^{\infty} dp P_{\rm lin}(p) ,\nonumber \\
		P_{3,13}(k) &\to& \frac{4k^2}{175\pi^2}  P_{\rm lin}(k) \int_0^{\infty} dp P_{\rm lin}(p).
\end{eqnarray}
These show 
\begin{eqnarray}
		P_{13}(k)&=& \sum_{\ell=0}^3 P_{\ell,13}(k) + P_{13}|_{\rm ZA}(k) \to
		-\frac{61k^2}{630\pi^2} P_{\rm lin}(k) \int_0^{\infty} dp P_{\rm lin}(p) \quad \mbox{for $p/k\gg1$}.
\end{eqnarray}
\section{NON-LINEAR CORRECTIONS IN THE 2-LOOP SPT}
\label{ap:SPT_2loop}

The SPT 2-loop solution is described as $P_{\rm 2\mathchar`-loop} = P_{33a} + P_{33b} + P_{24} + P_{15}$.
The Zel'dovich approximation leads to
\begin{eqnarray}
		P_{\rm 2\mathchar`-loop}|_{\rm ZA}(k)
		&=& \frac{1}{3!} \int d^3q e^{-i\kk\cdot\qq} \left\{ \sum_{\ell=0}^2i^{\ell}\Sigma_{\ell,\rm lin}(k,q){\cal L}_{\ell}(\hat{k}\cdot\hat{q})
		- \bar{\Sigma}_{\rm lin}(k) \right\}^3 \nonumber \\
		&=& P_{33b}|_{\rm ZA}(k) - \bar{\Sigma}_{\rm lin}(k) P_{22}|_{\rm ZA}(k) 
		+ \frac{1}{2}\left( \bar{\Sigma}_{\rm lin}(k) \right)^2P_{\rm lin}(k),
\end{eqnarray}
where
\begin{eqnarray}
		P_{33b}|_{\rm ZA}(k) 
		&=&  \frac{1}{3!}
		\int d^3q e^{-i\kk\cdot\qq} \left\{ \sum_{\ell=0}^2i^{\ell}
		\Sigma_{\ell,\rm lin}(k,q){\cal L}_{\ell}(\hat{k}\cdot\hat{q}) \right\}^3\nonumber \\
		&=& 4\pi \int_0^{\infty} dq q^2 
		\Bigg\{j_0(kq) \frac{\left( \Sigma_{0, \rm lin}(k,q) \right)^3}{3!} 
		+ j_2(kq) \frac{\left( \Sigma_{0, \rm lin}(k,q) \right)^2}{2!} \left( \Sigma_{2, \rm lin}(k,q) \right) \nonumber \\
		&&\hspace{2cm} + 	\Bigg(\frac{18}{35}j_4(kq) - \frac{2}{7}j_2(kq) + \frac{1}{5}j_0(kq) \Bigg)
		\frac{\left( \Sigma_{2, \rm lin}(k,q) \right)^2}{2!} \left( \Sigma_{0, \rm lin}(k,q) \right)  \nonumber \\
		&&\hspace{2cm} + \Bigg( \frac{18}{77} j_6(kq) - \frac{108}{385}j_4(kq) + \frac{3}{7} j_2(kq) - \frac{2}{35}j_0(kq) \Bigg)
		\frac{\left( \Sigma_{2, \rm lin}(k,q) \right)^3}{3!} \Bigg\}.
\end{eqnarray}
The 1-loop LPT leads to
\begin{eqnarray}
		P_{\rm 2\mathchar`-loop}|_{\rm LPT,1\mathchar`-loop}(k)
		&=&  P_{\rm 2\mathchar`-loop}|_{\rm ZA}(k)  \nonumber \\
		&& + \int d^3q e^{-i\kk\cdot\qq}
		\Bigg\{ \left( \sum_{\ell=0}^3 i^{\ell}\left( \Sigma_{\ell,22}(k,q)+\Sigma_{\ell,13}(k,q) \right){\cal L}_{\ell}(\hat{k}\cdot\hat{q}) 
		- \left( \bar{\Sigma}_{22}(k) + \bar{\Sigma}_{13}(k) \right)\right) \nonumber \\
		&&\hspace{3cm}\times \left(  \sum_{\ell=0}^2 i^{\ell}\Sigma_{\ell,\rm lin}(k,q){\cal L}_{\ell}(\hat{k}\cdot\hat{q}) 
		- \bar{\Sigma}_{\rm lin}(k)\right)\Bigg\} \nonumber \\
		&=& P_{33b}|_{\rm LPT, 1\mathchar`-loop} + P_{24}|_{\rm LPT, 1\mathchar`-loop}  \nonumber \\
		&& 
		- \bar{\Sigma}_{\rm lin}(k) P_{13}(k) - \frac{1}{2}\left( \bar{\Sigma}_{\rm lin}(k) \right)^2 P_{\rm lin}(k)
		- \left( \bar{\Sigma}_{13}(k) + \bar{\Sigma}_{22}(k) \right)P_{\rm lin}(k),\nonumber \\
\end{eqnarray}
where
\begin{eqnarray}
		 P_{33b}|_{\rm LPT, 1\mathchar`-loop}(k) &=&  
		  \frac{1}{3!}
		\int d^3q e^{-i\kk\cdot\qq} \left\{ \sum_{\ell=0}^2i^{\ell}
		\Sigma_{\ell,\rm lin}(k,q){\cal L}_{\ell}(\hat{k}\cdot\hat{q}) \right\}^3\nonumber \\
		 && + \int d^3q e^{-i\kk\cdot\qq}
		\left\{ \left( \sum_{\ell=0}^3i^{\ell} \Sigma_{\ell,22}(k,q){\cal L}_{\ell}(\hat{k}\cdot\hat{q}) \right)
		\left(  \sum_{\ell=0}^2i^{\ell} \Sigma_{\ell,\rm lin}(k,q){\cal L}_{\ell}(\hat{k}\cdot\hat{q}) \right)\right\} \nonumber \\
		 &=& 
		 P_{33b}|_{\rm ZA}(k) \nonumber \\
		 &+& 4\pi\int_0^{\infty} dq q^2 \Bigg\{
		 j_0(kq) \Sigma_{0,\rm lin}(k,q) \Sigma_{0,22}(k,q)  \nonumber \\
		&& \hspace{2cm} +  j_1(kq) \Sigma_{1,22}(k,q)\Sigma_{0,\rm lin}(k,q) 
		   + j_2(kq) \Sigma_{2,22}(k,q)\Sigma_{0,\rm lin}(k,q) \nonumber \\
	 	&& \hspace{2cm} + j_3(kq) \Sigma_{3,22}(k,q) \Sigma_{0,\rm lin}(k,q)
	     + j_2(kq) \Sigma_{2,\rm lin}(k,q) \Sigma_{0, 22}(k,q) \nonumber \\
	 	&& \hspace{2cm} + \left( -\frac{2}{5} j_1(kq) + \frac{3}{5} j_3(kq) \right) 
		 \Sigma_{1, 22}(k,q)  \Sigma_{2, \rm lin}(k,q) \nonumber \\
	    && \hspace{2cm} + \left( \frac{1}{10}j_0(kq) - \frac{1}{7} j_2(kq) + \frac{9}{35}j_4(kq) \right) 
         \Sigma_{2, 22}(k,q)  \Sigma_{2, \rm lin}(k,q)  \nonumber \\
		&& \hspace{2cm} +  \left( \frac{9}{35}j_1(kq) - \frac{4}{15}j_3(kq) + \frac{10}{21}j_5(kq) \right) 
		\Sigma_{3, 22}(k,q)   \Sigma_{2, \rm lin}(k,q) \Bigg\},\nonumber \\
		 \label{P33_LPT}
\end{eqnarray}
and 
\begin{eqnarray}
		 P_{24}|_{\rm LPT, 1\mathchar`-loop}(k) &=&  -	\bar{\Sigma}_{\rm lin}(k) P_{22}(k) \nonumber \\
		 &+&  4\pi\int_{0}^{\infty} dq q^2 \Bigg\{j_0(kq)\Sigma_{0,\rm lin}(k,q) \Sigma_{0,13}(k,q) \nonumber \\
		 && \hspace{2cm} + j_1(kq)  \Sigma_{1,13}(k,q) \Sigma_{0,\rm lin}(k,q)
		   + j_2(kq) \Sigma_{2,13}(k,q) \Sigma_{0,\rm lin}(k,q) \nonumber \\
	     && \hspace{2cm} + j_3(kq)  \Sigma_{3,13}(k,q) \Sigma_{0,\rm lin}(k,q)
	     + j_2(kq) \Sigma_{2,\rm lin}(k,q)\Sigma_{0, 13}(k,q) \nonumber \\
		 && \hspace{2cm} + \Bigg\{ \left( -\frac{2}{5} j_1(kq) + \frac{3}{5} j_3(kq) \right) 
		 \Sigma_{1, 13}(k,q)  \Sigma_{2, \rm lin}(k,q)\nonumber \\
		 && \hspace{2cm} + \left( \frac{1}{10}j_0(kq) - \frac{1}{7} j_2(kq) + \frac{9}{35}j_4(kq) \right) 
        \Sigma_{2, 13}(k,q)  \Sigma_{2, \rm lin}(k,q) \nonumber \\
		 && \hspace{2cm} +  \left( \frac{9}{35}j_1(kq) - \frac{4}{15}j_3(kq) + \frac{10}{21}j_5(kq) \right) 
		 \Sigma_{3, 13}(k,q)  \Sigma_{2, \rm lin}(k,q) \Bigg\}. \nonumber \\
		 \label{P24_LPT}
\end{eqnarray}

\section{$J^{(n)}$}
\label{ap:JJ}
\subsection{Zel'dovich Approximation}
The specific forms of $J^{(n)}$ are given from $n=2$ to $n=4$ as follows
\begin{eqnarray}
		J^{(2)}(z,k,q) &= & \frac{\left( D^2 \Sigma_{2, \rm lin}(k,q) \right)^2}{2!} 
		\Bigg\{\frac{18}{35}j_4(kq) - \frac{2}{7}j_2(kq) + \frac{1}{5}j_0(kq) \Bigg\} , \nonumber \\
		J^{(3)}(z,k,q) &= &  \frac{\left( D^2 \Sigma_{2, \rm lin}(k,q) \right)^3}{3!} 
		\Bigg\{
		\frac{18}{77} j_6(kq) - \frac{108}{385}j_4(kq) + \frac{3}{7} j_2(kq) - \frac{2}{35}j_0(kq) \Bigg\}, \nonumber \\
		J^{(4)}(z,k,q) &= &  \frac{\left( D^2 \Sigma_{2, \rm lin}(k,q) \right)^4}{4!}
		\Bigg\{
		\frac{72}{715}j_8(kq) - \frac{72}{385}j_6(kq) + \frac{1836}{5005}j_4(kq) - \frac{20}{77}j_2(kq) + \frac{3}{35}j_0(kq) \Bigg\}.
		\label{J_Z}
\end{eqnarray}

\subsection{LPT at the One-loop order}
\begin{eqnarray}
		J^{(2)}(z,k,q)
		&=& \frac{1}{2!} 
		\Bigg\{ j_0(kq) \left( -\frac{1}{3} \Sigma_1^2(z,k,q)  + \frac{1}{5} \Sigma_2^2(z,k,q) - \frac{1}{7}\Sigma_3^2(z,k,q)\right) \nonumber \\
	    &&\ \ \  + j_1(kq) \left( -\frac{4}{5}\Sigma_1(z,k,q) \Sigma_2(z,k,q) + \frac{18}{35} \Sigma_2(z,k,q) \Sigma_3(z,k,q) \right) \nonumber \\
		&&\ \ \  + j_2(kq) \left( \frac{2}{3} \Sigma_1^2(z,k,q)  - \frac{2}{7} \Sigma_2^2(z,k,q) + \frac{4}{21}\Sigma_3^2(z,k,q)
		                   -\frac{6}{7}\Sigma_1(z,k,q) \Sigma_3(z,k,q) \right) \nonumber \\
   	    &&\ \ \  + j_3(kq) \left( \frac{6}{5}\Sigma_1(z,k,q) \Sigma_2(z,k,q) - \frac{8}{15}\Sigma_2(z,k,q) \Sigma_3(z,k,q) \right) \nonumber \\
		&&\ \ \ + j_4(kq) \left( \frac{18}{35} \Sigma_2^2(z,k,q)  - \frac{18}{77} \Sigma_3^2(z,k,q) 
		   + \frac{8}{7}\Sigma_1(z,k,q) \Sigma_3(z,k,q) \right) \nonumber \\
		   &&\ \ \  + j_5(kq) \left( \frac{20}{21} \Sigma_2(z,k,q) \Sigma_3(z,k,q) \right) 
		   + j_6(kq) \left( \frac{100}{231} \Sigma_3^2(z,k,q)\right) \Bigg\}.
		   \label{J_2}
\end{eqnarray}

\bibliographystyle{apj}
%\bibliography{ms_p}

\end{document}